\documentclass[aps,prd,twocolumn,showpacs,superscriptaddress, floatfix]{revtex4-2}  % 

\usepackage{lineno}
\usepackage{graphicx}  % needed for figures
\usepackage{dcolumn}   % needed for some tables
\usepackage{bm}        % for math
\usepackage{amssymb}   % for math
\usepackage[caption=false]{subfig}
\usepackage{xcolor}
\usepackage{amsmath}
\usepackage{placeins}

\usepackage{ulem}

\newcommand{\nuantinu}{(anti\nobreakdash)\hspace{0pt}neutrino}
\newcommand{\nuantinus}{(anti\nobreakdash)\hspace{0pt}neutrinos}

\begin{document}

\title{Measurement of electron neutrino and antineutrino cross sections at low momentum transfer}

%% List of institution addresses, in command form.
\newcommand{\Rutgers}{Rutgers, The State University of New Jersey, Piscataway, New Jersey 08854, USA}
\newcommand{\Hampton}{Hampton University, Dept. of Physics, Hampton, VA 23668, USA}
\newcommand{\Dortmund}{Institute of Physics, Dortmund University, 44221, Germany }
\newcommand{\Otterbein}{Department of Physics, Otterbein University, 1 South Grove Street, Westerville, OH, 43081 USA}
\newcommand{\JMU}{James Madison University, Harrisonburg, Virginia 22807, USA}
\newcommand{\Florida}{University of Florida, Department of Physics, Gainesville, FL 32611}
\newcommand{\UCIrvine}{Department of Physics and Astronomy, University of California, Irvine, Irvine, California 92697-4575, USA}
\newcommand{\CBPF}{Centro Brasileiro de Pesquisas F\'{i}sicas, Rua Dr. Xavier Sigaud 150, Urca, Rio de Janeiro, Rio de Janeiro, 22290-180, Brazil}
\newcommand{\PUCP}{Secci\'{o}n F\'{i}sica, Departamento de Ciencias, Pontificia Universidad Cat\'{o}lica del Per\'{u}, Apartado 1761, Lima, Per\'{u}}
\newcommand{\INRM}{Institute for Nuclear Research of the Russian Academy of Sciences, 117312 Moscow, Russia}
\newcommand{\Jlab}{Jefferson Lab, 12000 Jefferson Avenue, Newport News, VA 23606, USA}
\newcommand{\Pittsburgh}{Department of Physics and Astronomy, University of Pittsburgh, Pittsburgh, Pennsylvania 15260, USA}
\newcommand{\Guanajuato}{Campus Le\'{o}n y Campus Guanajuato, Universidad de Guanajuato, Lascurain de Retana No. 5, Colonia Centro, Guanajuato 36000, Guanajuato M\'{e}xico.}
\newcommand{\Athens}{Department of Physics, University of Athens, GR-15771 Athens, Greece}
\newcommand{\Tufts}{Physics Department, Tufts University, Medford, Massachusetts 02155, USA}
\newcommand{\WM}{Department of Physics, William \& Mary, Williamsburg, Virginia 23187, USA}
\newcommand{\FNAL}{Fermi National Accelerator Laboratory, Batavia, Illinois 60510, USA}
\newcommand{\Purdue}{Department of Chemistry and Physics, Purdue University Calumet, Hammond, Indiana 46323, USA}
\newcommand{\MCLA}{Massachusetts College of Liberal Arts, 375 Church Street, North Adams, MA 01247}
\newcommand{\UMD}{Department of Physics, University of Minnesota -- Duluth, Duluth, Minnesota 55812, USA}
\newcommand{\Northwestern}{Northwestern University, Evanston, Illinois 60208}
\newcommand{\UNI}{Facultad de Ciencias F\'{i}sicas, Universidad Nacional Mayor de San Marcos, CP 15081, Lima, Per\'{u}}
\newcommand{\Rochester}{Department of Physics and Astronomy, University of Rochester, Rochester, New York 14627 USA}
\newcommand{\Austin}{Department of Physics, University of Texas, 1 University Station, Austin, Texas 78712, USA}
\newcommand{\USM}{Departamento de F\'{i}sica, Universidad T\'{e}cnica Federico Santa Mar\'{i}a, Avenida Espa\~{n}a 1680 Casilla 110-V, Valpara\'{i}so, Chile}
\newcommand{\Geneva}{University of Geneva, 1211 Geneva 4, Switzerland}
\newcommand{\Chicago}{Enrico Fermi Institute, University of Chicago, Chicago, IL 60637 USA}
\newcommand{\hired}{}
\newcommand{\OregonState}{Department of Physics, Oregon State University, Corvallis, Oregon 97331, USA}
\newcommand{\oxford}{Oxford University, Department of Physics, Oxford, OX1 3PJ United Kingdom}
\newcommand{\umiss}{University of Mississippi, Oxford, Mississippi 38677, USA}
\newcommand{\upenn}{Department of Physics and Astronomy, University of Pennsylvania, Philadelphia, PA 19104}
\newcommand{\AMU}{Department of Physics, Aligarh Muslim University, Aligarh, Uttar Pradesh 202002, India}
\newcommand{\wroclaw}{University of Wroclaw, plac Uniwersytecki 1, 50-137 Wroa\l{}aw, Poland}
\newcommand{\Mohali}{Department of Physical Sciences, IISER Mohali, Knowledge City, SAS Nagar, Mohali - 140306, Punjab, India}
\newcommand{\CINVESTAV}{Departamento de Fisica Col. San Pedro Zacatenco, 07360 Mexico, DF, Av. Instituto PolitÃ©cnico Nacional, Mexico}
\newcommand{\york}{York University, Department of Physics and Astronomy, Toronto, Ontario, M3J 1P3 Canada}
\newcommand{\ND}{Department of Physics and Astronomy, University of Notre Dame, Notre Dame, Indiana 46556, USA}
\newcommand{\ICL}{The Blackett Laboratory,  Imperial College London,  London SW7 2BW, United Kingdom}
\newcommand{\warwick}{Department of Physics, University of Warwick, Coventry, CV4 7AL, UK}
\newcommand{\qmul}{G O Jones Building, Queen Mary University of London, 327 Mile End Road, London E1 4NS, UK}

\newcommand{\mascencioThanks}{Now at Iowa State University, Ames, IA 50011, USA}
\newcommand{\amitbashyalThanks}{Now at  High Energy Physics Department, Argonne National Lab, 9700 S Cass Ave, Lemont, IL 60439}
\newcommand{\finerThanks}{Now at Los Alamos National Laboratory, Los Alamos, New Mexico 87545, USA}
\newcommand{\kleykampThanks}{now at Department of Physics and Astronomy, University of Mississippi, Oxford, MS 38677}
\newcommand{\bamThanks}{Now at University of Minnesota, Minneapolis, Minnesota 55455, USA}
\newcommand{\adrianThanks}{Now at Department of Physics, Drexel University, Philadelphia, Pennsylvania 19104, USA}
\newcommand{\byaeggyThanks}{Now at Department of Physics, University of Cincinnati,  Cincinnati, Ohio 45221, USA}
\newcommand{\lazazuetareyesThanks}{now at Syracuse University, Syracuse, NY 13244, USA}

% 55 total signatories.

\author{S.~Henry}                         \affiliation{\Rochester}
\author{H.~Su}                            \affiliation{\Pittsburgh}
\author{S.~Akhter}                        \affiliation{\AMU}
\author{Z.~~Ahmad~Dar}                    \affiliation{\WM}  \affiliation{\AMU}
\author{V.~Ansari}                        \affiliation{\AMU}
\author{M.~V.~Ascencio}\thanks{\mascencioThanks}  \affiliation{\PUCP}
\author{M.~Sajjad~Athar}                  \affiliation{\AMU}
\author{A.~Bashyal}\thanks{\amitbashyalThanks}  \affiliation{\OregonState}
\author{M.~Betancourt}                    \affiliation{\FNAL}
\author{J.~L.~Bonilla}                    \affiliation{\Guanajuato}
\author{A.~Bravar}                        \affiliation{\Geneva}
\author{G.A.~D\'{i}az~}                   \affiliation{\FNAL}  \affiliation{\Rochester}
\author{J.~Felix}                         \affiliation{\Guanajuato}
\author{L.~Fields}                        \affiliation{\ND}
\author{R.~Fine}\thanks{\finerThanks}     \affiliation{\Rochester}
\author{P.K.Gaur}                         \affiliation{\AMU}
\author{S.M.~Gilligan}                    \affiliation{\OregonState}
\author{R.~Gran}                          \affiliation{\UMD}
\author{E.Granados}                       \affiliation{\Guanajuato}
\author{D.A.~Harris}                      \affiliation{\york}  \affiliation{\FNAL}
\author{A.L.~Hart}                        \affiliation{\qmul}
\author{J.~Kleykamp}\thanks{\kleykampThanks}  \affiliation{\Rochester}
\author{A.~Klustov\'{a}}                  \affiliation{\ICL}
\author{M.~Kordosky}                      \affiliation{\WM}
\author{D.~Last} \affiliation{\upenn}
\author{A.~Lozano}\thanks{\adrianThanks}  \affiliation{\CBPF}
\author{X.-G.~Lu}                         \affiliation{\warwick}  \affiliation{\oxford}
\author{S.~Manly}                         \affiliation{\Rochester}
\author{W.A.~Mann}                        \affiliation{\Tufts}
\author{C.~Mauger}                        \affiliation{\upenn}
\author{K.S.~McFarland}                   \affiliation{\Rochester}
\author{M.~Mehmood}                       \affiliation{\york}
\author{B.~Messerly}\thanks{\bamThanks}   \affiliation{\Pittsburgh}
\author{J.G.~Morf\'{i}n}                  \affiliation{\FNAL}
\author{D.~Naples}                        \affiliation{\Pittsburgh}
\author{J.K.~Nelson}                      \affiliation{\WM}
\author{C.~Nguyen}                        \affiliation{\Florida}
\author{A.~Olivier}                       \affiliation{\ND}  \affiliation{\Rochester}
\author{V.~Paolone}                       \affiliation{\Pittsburgh}
\author{G.N.~Perdue}                      \affiliation{\FNAL}  \affiliation{\Rochester}
\author{C.~Pernas}                        \affiliation{\WM}
\author{K.-J.~Plows}                      \affiliation{\oxford}
\author{M.A.~Ram\'{i}rez}                 \affiliation{\upenn}  \affiliation{\Guanajuato}
\author{R.D.~Ransome}                     \affiliation{\Rutgers}
\author{N.~Roy}                           \affiliation{\york}
\author{D.~Ruterbories}                   \affiliation{\Rochester}
\author{H.~Schellman}                     \affiliation{\OregonState}
\author{C.~J.~Solano~Salinas}             \affiliation{\UNI}
\author{V.S.~Syrotenko}                   \affiliation{\Tufts}
\author{E.~Valencia}                      \affiliation{\Guanajuato}  \affiliation{\WM}
\author{N.H.~Vaughan}                     \affiliation{\OregonState}
\author{A.V.~Waldron}                     \affiliation{\qmul}  \affiliation{\ICL}
\author{C.~Wret}                          \affiliation{\Rochester}
\author{B.~Yaeggy}\thanks{\byaeggyThanks}  \affiliation{\USM}
\author{L.~Zazueta}\thanks{\lazazuetareyesThanks}  \affiliation{\WM}

\collaboration{The MINERvA Collaboration}
\noaffiliation

\date{\today}
\begin{abstract}
Accelerator based neutrino oscillation experiments seek to measure the relative number of electron and muon \nuantinus\ at different $L/E$ values. However high statistics studies of neutrino interactions are almost exclusively measured using muon \nuantinus\ since the dominant flavor of neutrinos produced by accelerator based beams are of the muon type.  This work reports new measurements of electron \nuantinu\ interactions in hydrocarbon, obtained by strongly suppressing backgrounds initiated by muon flavor \nuantinus.  Double differential cross sections as a function of visible energy transfer, $E_\text{avail}$, and transverse momentum transfer, $p_T$, or three momentum transfer, $q_3$ are presented.
\end{abstract}

\maketitle

\section{Introduction}
Predictions of interactions of GeV energy \nuantinus\ with nuclear targets present challenges for experiments seeking to precisely measure neutrino flavor oscillations.  Both the DUNE~\cite{Acciarri:2015uup} and Hyper-Kamiokande~\cite{Abe:2011ts} experiments are designed to measure muon to electron neutrino flavor transitions with uncertainties on the order one percent. Consequently accurate predictions of detector efficiencies, backgrounds, energy reconstruction, and cross sections, which make use of the measurements presented here, are needed. Compounding the problem for DUNE and Hyper-Kamiokande is that their near detectors, which are used for studying neutrino interactions, see primarily a flux of muon neutrinos and only a small fraction of electron neutrinos.  Therefore constraints that are solely derived from measurements of muon neutrino interactions will need to be theoretically corrected for electron neutrinos.  

Electron-neutrino and muon-neutrino charged current cross sections differ for two reasons.  First, the tensor structure of the hadronic current and its contraction with the lepton current yields terms that depend explicitly on the square of the lepton mass compared to combinations of the target mass, neutrino energy, and energy transfer~\cite{LlewellynSmith:1971zm}.  These terms are largely negligible for electron neutrinos at accelerator energies, however they may provide non-negligible subleading corrections for muon neutrinos at low neutrino energies or energy transfers.   Secondly, momentum and energy transfer limits for electron and muon neutrino interactions differ because the kinematic limits in momentum and energy transfer are a function of lepton mass.  If the energy and three-momentum transfer from the incoming neutrino to the final-state lepton are denoted $q_0$ and $q_3$ respectively, conservation of energy and momentum requires
%(~\refcomComment{"require" - "requires"}) 
that 
\begin{eqnarray}
 E_\nu-\sqrt{E_\nu^2-2 E_\nu q_0-m_l^2+q_0^2}<q_3 \nonumber\\ < E_\nu+\sqrt{E_\nu^2-2 E_\nu q_0-m_l^2+q_0^2},
    \label{eqn:q0-q3-conservationOfFourMomentum}
\end{eqnarray}
where $E_\nu$ is the energy of the incoming neutrino and $m_l$ is the final-state lepton mass.  A second constraint comes from the relationship between initial and final-state target invariant masses.  If we denote the initial invariant mass $M$ and the final mass $M+\Delta$, then this implies a maximum three momentum transfer
\begin{eqnarray}
    q_3&\leq&\frac{1}{2 M (2 E_\nu+M)}\times\left( 2 E_\nu^2 M  -\Delta ^2 E_\nu \right. \nonumber\\
    &-&\left. 2 \Delta  E_\nu M+E_\nu m_l^2 +(E_\nu+M)\sqrt{\eta}\right) ,~\text{where} \nonumber\\
    \eta & \equiv & 4 E_\nu^2 M^2-4 E_\nu M \left(\Delta  (\Delta +2 M)+m_l^2\right) \nonumber\\
    & + & (m_l^2-\Delta^2 ) 
    \left(m_l^2-(\Delta +2 M)^2\right).
    \label{eqn:q3-invariantMassLimit}
\end{eqnarray}
Expanding in $m_l$, we see
\begin{eqnarray}
   q_3&\leq&\left(E_\nu-\frac{\Delta  (\Delta +2 M)}{2 M}\right) \nonumber\\ &-&m_l^2 \left(\frac{E_\nu+M}{2 E_\nu M-\Delta  (\Delta +2 M)}-\frac{1}{2 M}\right)+{\cal O}\left(m_l^4\right),
    \label{eqn:q3-invariantMassLimit-expand}
\end{eqnarray}
A simple case in~\eqref{eqn:q3-invariantMassLimit-expand} is where $\Delta=0$, which is approximately correct for elastic and quasielastic scattering from nucleons, gives a limit of $q_3<E_\nu-\frac{m_l^2}{2E_\nu}$. Both~\eqref{eqn:q0-q3-conservationOfFourMomentum} and \eqref{eqn:q3-invariantMassLimit-expand} show that increasing $m_l$ eliminates regions of allowed energy and momentum transfer.  Since it is the reactions of muon neutrinos that are studied with high statistics in near detectors, the extrapolation into certain regions of energy and momentum transfer are not well-explored experimentally. 

Another effect recently discussed in the literature concerns %(\refcomComment{"is that of" - "concerns"}) 
radiative corrections which have a strong dependence on the mass of the final-state lepton~\cite{Tomalak:2021hec,Tomalak:2022xup}.  This cited work concludes that the effect of radiative corrections can be precisely predicted, although those predictions are not currently implemented
%(\refcomComment{added "implemented"})
in neutrino interaction models used by experiments.

The effects of electron and muon neutrino interaction differences have been studied within specific models of neutrino interaction cross sections on nucleons and nuclei~\cite{Day:2012gb,Ankowski:2017yvm,Nieves:2017lij,Nikolakopoulos:2019qcr,Dolan:2021rdd}.  Such studies illuminate possible differences between electron and muon neutrino interactions but are not exhaustive. 

With sufficient statistics, and with strong rejection and control of backgrounds, it is possible to directly measure the interactions of electron neutrinos and antineutrinos.  This paper describes such a measurement with the MINERvA detector~\cite{Aliaga:2013uqz} using the broadband NuMI~\cite{Adamson:2015dkw} beam located at Fermi National Accelerator Laboratory.  To keep backgrounds low and well-controlled, the measurement is performed at energy and momentum transfers less than $\sim 1$~GeV, much less than the incoming neutrino energy.  

A complication in these measurements is that energy transfer cannot be directly measured in a broadband neutrino beam, but instead must be inferred from the visible recoil products in the detector.  To approximate what is measured calorimetrically, we employ a proxy used in previous MINERvA measurements~\cite{Rodrigues:2015hik,MINERvA:2021wjs}, $i.e.$ replace $q_0$ in the measurement by the quantity $E_{\text{avail}}$, defined as
\begin{equation}
   E_{\text{avail}}\equiv \sum_{\text{protons}} T_p +  \sum_{\pi^\pm} T_{\pi^\pm} + \sum_{\pi^0} E_{\pi^0},
   \label{eq:Eavail}
\end{equation}
where the sums are over final-state particles, and $T_X$ indicates the kinetic, rather than total energy $E_X$ of a final-state particle $X$.  The weak decay products of strange, or heavier quark, baryons are included 
%(\refcomComment{removed "to the sum"})
by adding their total energies to the sum, and by subtracting (or adding) a nucleon mass in the case of baryons (antibaryons).   For scattering from nuclei, this quantity differs from $q_0$ in that it does not have the kinetic energy of final-state neutrons nor the rest mass of charged pions, and ignores any additional excitation energy or mass differences in the final-state nuclear system.  

\subsection{Past results on electron neutrino interactions at GeV energies}

Because of the relatively small number of electron neutrinos in GeV energy accelerator beams and high backgrounds, previous measurements are few and are often statistics and background limited.  Qualitatively, previous measurements fall into several categories.  Some measurements have measured only flux integrated total cross sections, or total cross sections as a function of derived neutrino energy, with relatively low statistics, from tens to a few hundreds of events~\cite{Gargamelle:1977pmg,T2K:2015ydf,ArgoNeuT:2020kir,MicroBooNE:2021gfj}.   These low statistics measurements are unlikely to constrain electron neutrino interaction models to the level needed by future appearance oscillation experiments.  The T2K and MicroBooNE experiments have produced measurements of flux-integrated lepton kinematics for samples of order a hundred or few hundred events~\cite{T2K:2014lbi,T2K:2020lrr,MicroBooNE:2021ppm} that can be compared to models and could be sensitive to large deviations, $>10$\%.

Several measurements have additional capabilities to test models.  The NOvA experiment has made a high statistics measurement of cross sections as a function of lepton kinematics~\cite{NOvA:2022see}, with nearly $10^4$ signal events with good purity, but makes its measurements in relatively wide bins of lepton energy and angle.  MINERvA~\cite{MINERvA:2015jih} and MicroBooNE~\cite{MicroBooNE:2022tdd} have measured events without final-state pions, a sample presumably dominated by single and multinucleon knockout, with good purity, and with samples of order $10^3$ and $10^2$ events, respectively.  These samples complement the measurements reported here because of their sensitivity to this exclusive and important reaction channel.

The measurement reported in this article, by contrast to these previous results, is high statistics with tens of thousands of signal events in each of the $\nu_e$ and $\bar{\nu}_e$ samples. In addition it measures correlations between lepton kinematics (as measured primarily by $p_T$ and the derived three momentum transfer, $q_3$) and a variable which is visible energy transfer to the final-state.  It is inclusive, although in the lower visible energy and momentum transfer regions. It is dominated by single and multinucleon knockout events, and reports over a wide range of momentum transfer, up to $1.6$~GeV in transverse momentum transfer, $p_T$ or $1.2$~GeV in inferred $q_3$.
 
\section{MINERvA experiment and NuMI beam line}

The MINERvA detector is a fine-grained tracking calorimeter with a fully active solid-scintillator tracker forming the bulk of the inner detector (ID). Upstream of the tracker is an area of nuclear targets -- carbon, water, iron, and lead, interleaved with tracking planes. The downstream part of the ID contains Electromagnetic CALorimeter (ECAL) and Hadronic CALorimeter (HCAL).  The ID is surrounded by side ECAL and side HCAL.  Downstream, the MINOS near detector served as a muon spectrometer for MINERvA. Muon charge and momentum measurements were provided for muons with momenta above $\sim$1.5 GeV/c. 

The active detector elements are solid-scintillator strips of triangular cross section, with a 3.3 cm base, 1.7 cm high, arranged in planes where neighboring strips alternate orientation with respect to the beam. Charge sharing between neighboring strips provides a spatial resolution of $\sim$3 mm.  Scintillation light due to a charged particle traversing the scintillator is collected by a wavelength shifting fiber located at the center of each strip and routed through clear optical fibers to M64 Hamamatsu photomultiplier tubes (PMT). The electrical signals from front end boards mounted on top of each PMT box are readout via the data acquisition system.  The detector consists of hexagonal modules containing one or two active planes mounted on a steel frame. The orientation of strips in the planes can be vertical (X), +60$^\circ$ (U), or -60$^\circ$ (V). Four types of modules were built: (i) tracker modules with strip orientations X+U, X+V; (ii) ECAL with 0.2cm lead sheets, plus planes with strip orientations X+U or X+V; (iii) HCAL with a 2.54cm thick iron plate, and plane with strip orientation X, U, or V; and (iv) Target modules with passive carbon, water, iron or lead targets. 

\begin{figure}
    \centering    \includegraphics[scale=0.22]{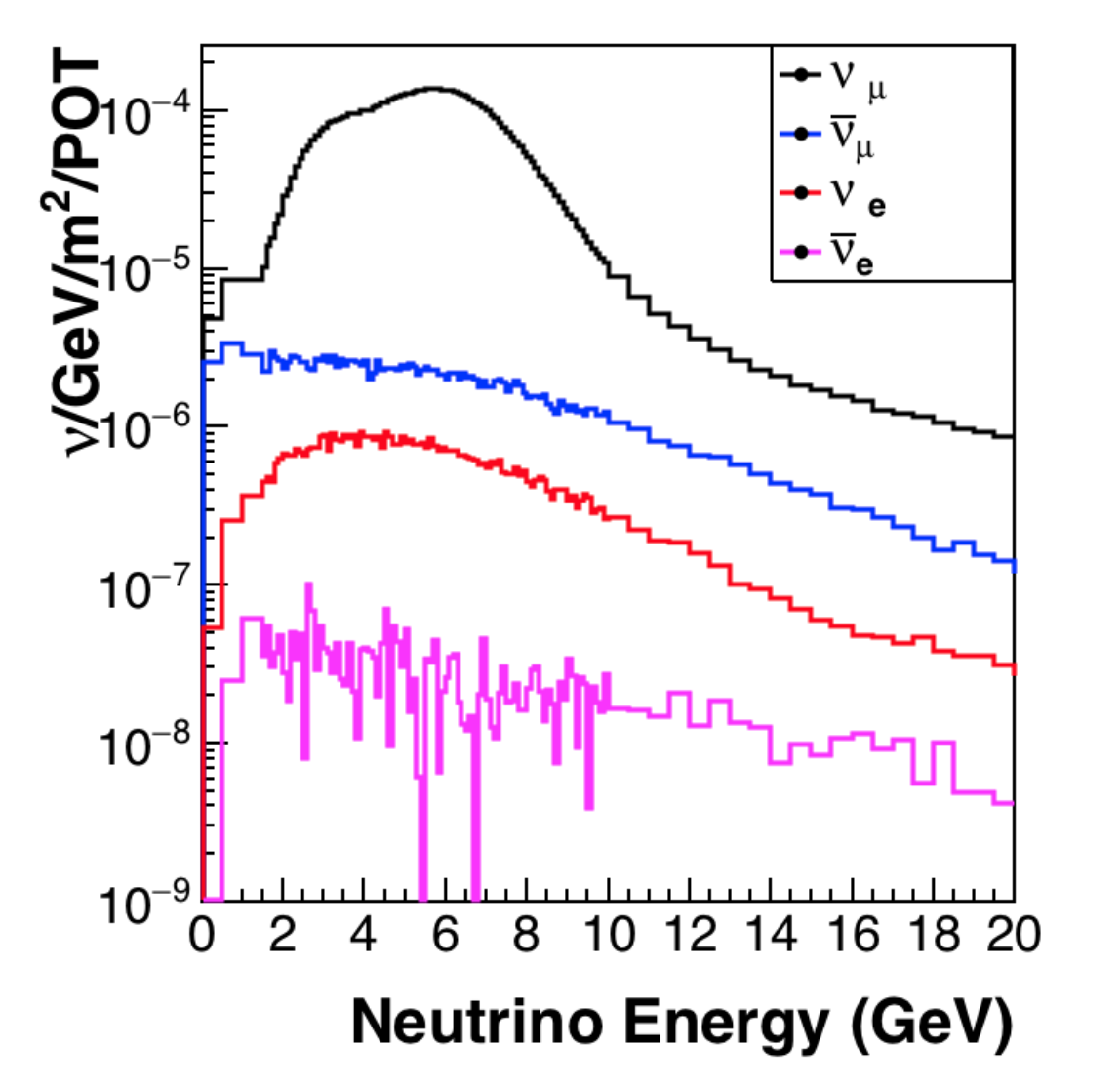}\includegraphics[scale=0.225]{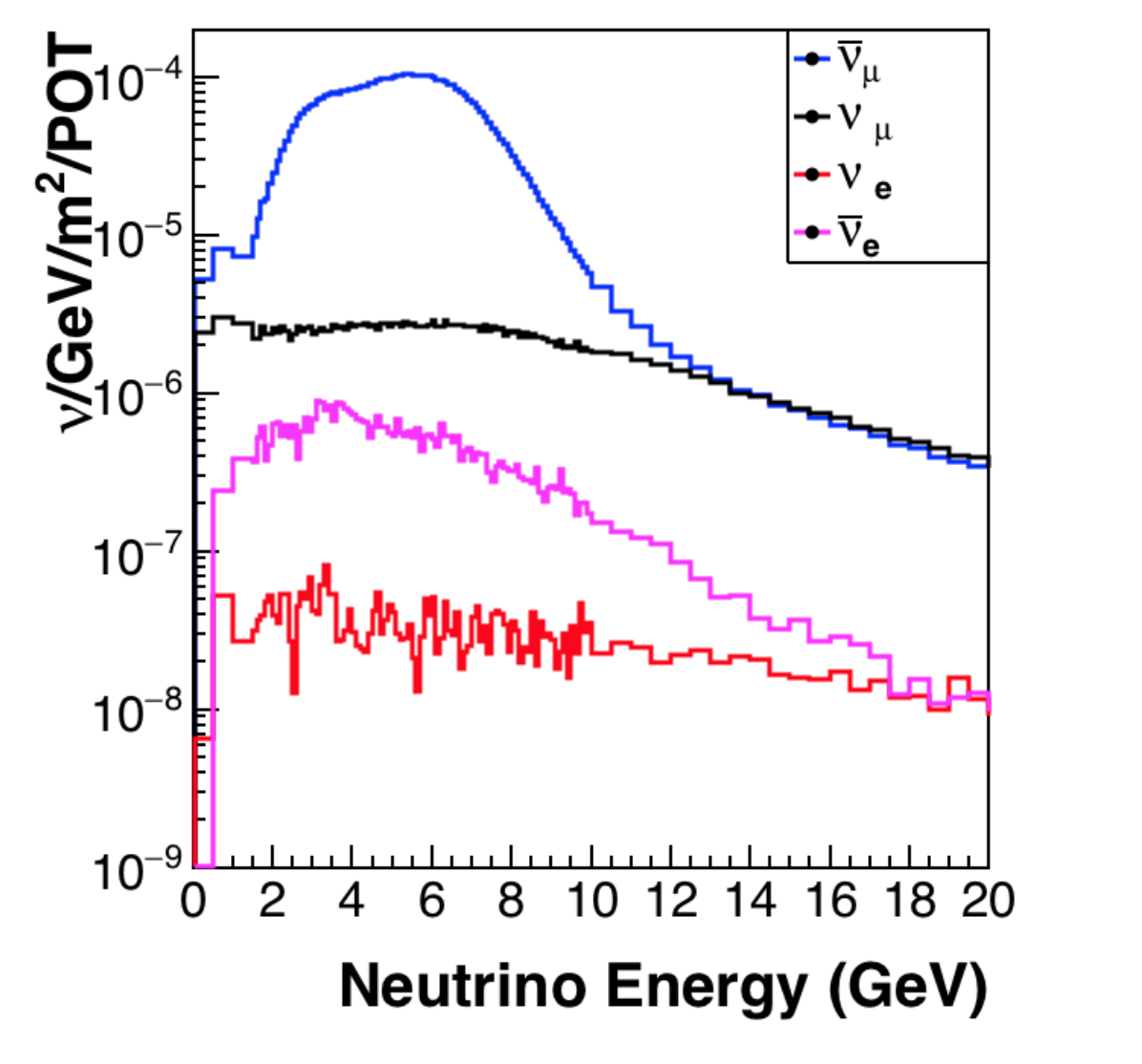}
    \caption{Flux predictions for FHC mode (Left) and RHC mode (Right). The contributions from all neutrino types are shown for each beam mode.}
    \label{fig:MIN_flux_FRHC}
\end{figure}

MINER$\nu$A utilizes the intense broadband NuMI (Neutrinos at the Main Injector) beam running at FNAL. FNAL's Main Injector accelerates protons up to 120 GeV which are directed to a carbon target.   The pions produced by the proton interactions on the carbon target are focused by two horns and allowed to decay in a 675 meter decay pipe. Undecayed pions are absorbed in the hadron absorber just downstream of the decay region and the decay muons are absorbed in the following 240  meters of rock before reaching the detector hall.  Different energy tunes are available by varying the location of target and horns. The two main configurations are known as the low-energy (LE) and medium-energy (ME) beam tunes. The results presented here are based on the ME configuration.  NuMI also allows for neutrino or antineutrino beam running by sign selecting pions and kaons by setting the magnetic horn current direction. The forward horn current (FHC) polarity produces predominantly muon neutrinos. The reverse horn current (RHC) polarity produces predominantly muon antineutrinos. However both FHC and RHC contain antineutrino and neutrino contamination, respectively, on the order of a few percent. This cross-contamination results from kaon decay producing neutrinos at the higher end of the energy spectrum and from muon decay that creates neutrinos at the lower end. The neutrino flux prediction (see Fig~\ref{fig:MIN_flux_FRHC}) used by MINERvA is derived from a Geant4 simulation of the NuMI beamline which is constrained by measurements from neutrino-electron elastic scattering\cite{Valencia:2019mkf,MINERvA:2022vmb}.

\section{Neutrino Interaction Simulation}
Neutrino-nucleus interactions are simulated using GENIE v2.12.6. GENIE\cite{Andreopoulos:2009rq} is a Monte Carlo (MC) neutrino interaction event generator which simulates multiple neutrino-nucleus interaction channels exclusively, including the three primary channels -- charged current quasielastic (CCQE), resonant pion production (RES), and deep inelastic scattering (DIS) - as well as subdominant channels such as charm and coherent pion production.  The quasielastic interactions are simulated using the Llewellyn-Smith formalism\cite{LlewellynSmith:1971zm} with BBBA05 vector form factor modeling\cite{FORMFACTOR}, where the axial form factor uses the dipole form with an axial mass of $M_A = 0.99$ GeV. The Rein-Sehgal model\cite{REIN198179} with axial mass of $M_A^{RES} = 1.12$ GeV is employed to simulate resonance productions. DIS interactions are simulated using the leading order model with the Bodek-Yang prescription\cite{DIS}. In addition, "two particle two hole" (2p2h) interactions are simulated using the Valencia model\cite{PhysRevC.83.045501,PhysRevD.88.113007,schwehr2017genie} and coherent pion production is simulated by the other Rein-Sehgal model\cite{REIN198329}. The nucleon initial states are simulated using the relativistic Fermi gas model\cite{Smith:1972xh} with additional Bodek-Ritchie tail\cite{BRtail} while the FSI is simulated using the INTRANUKE-hA package\cite{FSI}, which is a hadronic cascade model. To better describe MINERvA data, there are tunes applied to the prediction of CCQE, RES, and 2p2h interactions, collectively referred to as MINERvA tune v1, and described in Ref.~\cite{MINERvA:2021owq}.

The coherent channel of GENIE does not simulate coherent scattering off hydrogen atoms, e.g., diffractive pion production. However, MINERvA data from its low-energy beam showed that the contribution of the neutral current (NC) diffractive process is sizable\cite{NCDIF}. In order to simulate this process, the charged current (CC) diffractive model in GENIE, which is an implementation of the work by Rein\cite{REIN198661}, is used with two modifications to turn the CC model into an NC model by producing a neutrino in the final-state, reducing the cross section by a factor of two for the expected CC/NC ratio.

\section{Electron Reconstruction and Identification} 

The reconstruction method employed in this analysis is a combination of the MINERvA electron neutrino CCQE measurement preformed with the LE dataset\cite{MINERvA:2015jih} and the muon neutrino low-recoil analyses\cite{Rodrigues:2015hik}. The electron candidates are reconstructed using the same method as used in the LE analysis but updated with retuned algorithms to cluster hits from neutrino interactions in time and with tracking improvements. 

An inclusive electron neutrino charged-current interaction sample is selected using the following four cuts. First, events with any MINOS matched tracks are rejected. Second, an electron candidate is constructed for each track that originated from the most upstream vertex and is contained in the MINERvA detector. The hits are considered as part of electron candidate if they are inside a 7.5 degree cone region with an apex at the event vertex and axis along the track direction or a cylindrical region of 50 mm radius extending from the event vertex along the track direction. If there is more than a three radiation length separation between a hit and the next downstream hit, this upstream hit will be tagged as the most downstream hit considered part of the electron candidate. Third, the collection of hits considered as the electron candidate is tested by a k-nearest-neighbor classifier using three variables: mean $dE/dx$, the fraction of energy deposited at the downstream end, and the median shower width \cite{Hastie:2009itz}. The classifier is trained to distinguish electromagnetic showers from track like particles using simulated single particle samples including electrons, muons, photons, changed pions, and protons. Events are selected if there is at least one electron candidate having a kNN score greater than 0.7. Lastly, the energy of the electron candidate is measured by employing a calorimetric sum of hit energies, corrected for passive materials. We choose the most energetic electron candidate as the primary candidate if multiple candidates pass the threshold.

\subsection{Electron and photon separation}
Additional selections are necessary because the kNN classifier is not optimized to distinguish between electrons and photons. We use the minimal energy deposition in a 100 mm sliding window from 25 mm to 500 mm downstream of the event vertex (measured along track direction) of the electron candidate as the discriminator and require the $dE/dx$ in the minimal window less than 2.4 MeV/cm (see Fig. \ref{fig:frontdedx}) to be considered as an electron. In addition, we reject events with multiple vertices since they are more likely to be a neutral-current interaction.

\subsection{Reconstruction of visible calorimetric energy} \label{sec:visibleEnergy}
The visible calorimetric energy is calculated as described in the MINERvA muon neutrino low-recoil measurement\cite{Rodrigues:2015hik}. We assume that hits that are not included in the electron candidate are the result of energy deposited by the hadronic system. These hits are summed using the calibrated visible energy in each subdetector and corrected for passive material. We construct two hadronic energy estimators using the visible energy in each subdetector to estimate $q_0$ and $E_\text{avail}$ separately. The energy transfer $q_0$ is estimated by summing the visible energies in all subdetectors and applying a spline correction to offset the bias observed by a MC study.  $E_\text{avail}$ is estimated (see Eq. \ref{eq:Eavail}) by summing up visible energies in the tracker and ECAL and applying a constant scale factor independent of visible energy. 
%(\refcomComment{replaced "flat" with "independent of visible energy"})
The spline correction applied to extract $q_0$ is model dependent, since it attempts to correct for energy that is not calorimetrically visible in the detector, such as kinetic energy of neutrons or rest masses of charged pions. For this reason, cross sections are reported as a function of $E_\text{avail}$, and $q_0$ is only used as a subleading input to construct $q_3$ as described below. 
%(\refcomComment{We rewrote the previous sentence from the ground up to make it more clear}) 
In addition, we estimate that 0.8\% of the EM shower energy leaks out of the electron candidate on average using a simulation study and we therefore correct the EM shower energy leakage from the reconstructed $E_\text{avail}$ and $q_0$. 

Finally, we reconstruct $q_3$ using lepton kinematics and reconstructed $q_0$:
\begin{align}
    q_3 &= \sqrt{Q^2 + q_0^2}  \nonumber \\
    Q^2 &= 2 (E_{l} + q_0) (E_{l}-|\vec{p_{l}}|\cos(\theta_{l})) - m_{l}^2
    \label{eqn:reconstructed-q3}
\end{align}

\section{Backgrounds and Control Samples} 

\begin{figure}
    \centering
    \includegraphics[width=8cm]{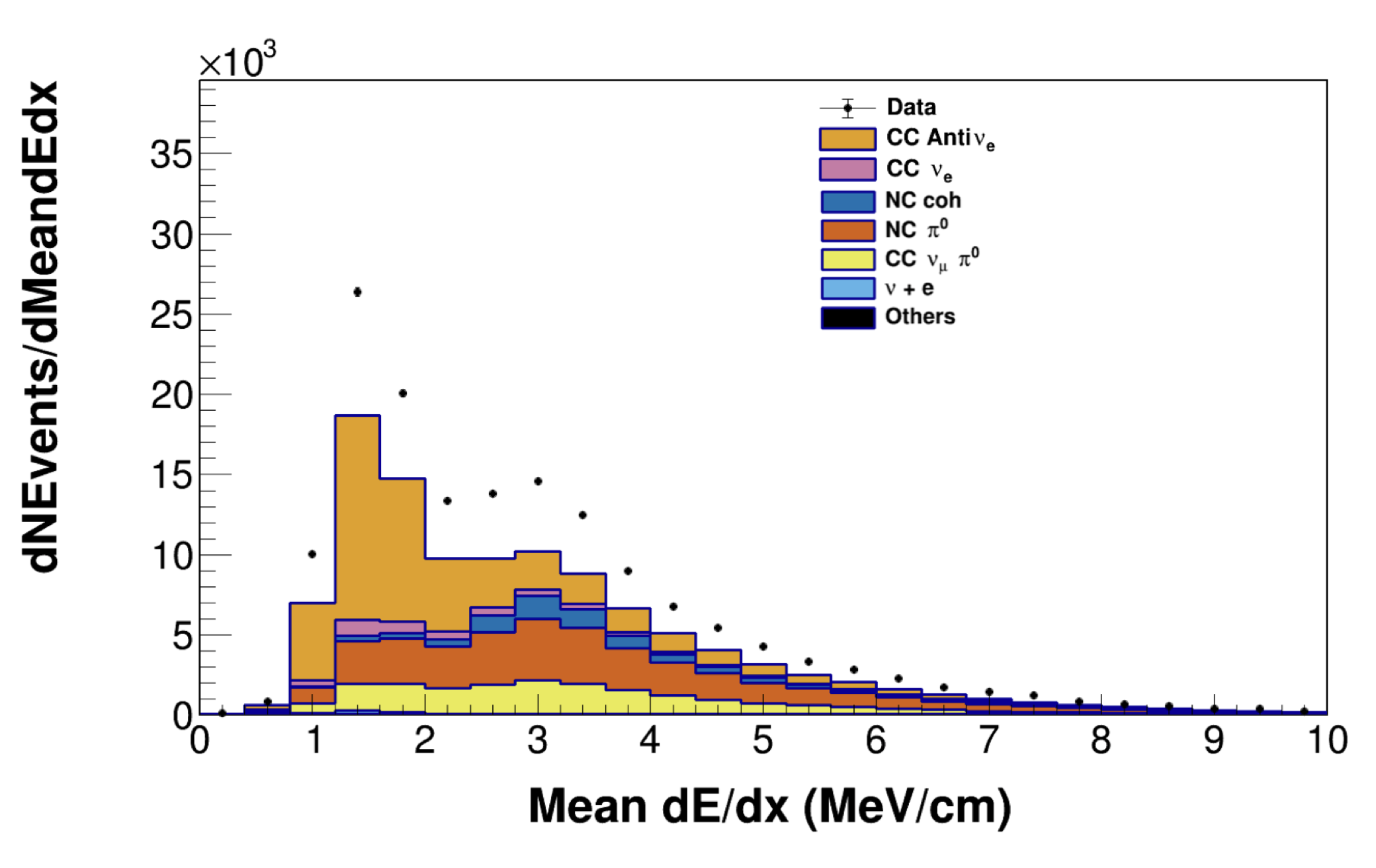}
    \caption{RHC mean dE/dX distribution.}
    \label{fig:frontdedx}
\end{figure}

Figure \ref{fig:frontdedx} shows the distribution of the mean \textit{dE/dx} quantity described above for both data and simulation. There is a large excess of data events in the background dominated region with mean \textit{dE/dx} $>$ 2.4 MeV/cm. 

This excess is similar to what was reported in MINERvA's LE data\cite{MINERvA:2015jih} and with the conclusion that it may be explained through diffractive pion production. NC diffractive $\pi^0$ production is similar to NC coherent $\pi^0$ production in that both are inelastic processes where a lepton and a pion are produced in the forward direction while leaving the struck nucleus in the ground state. The square of the four-momentum transfer, $\vert t \vert$, must be small to preserve the initial state of the nucleus. Since MINERvA’s tracker material is a CH-based hydrocarbon, there is a possibility a neutrino interaction will occur on a free proton, referred to as diffractive pion production.  Since the proton is much less massive than a carbon nucleus, the proton recoils visibly from the momentum imparted to the target proton in the MINERvA detector. The recoiling proton deposits its energy upstream of the ``vertex'' where the $\pi^0$ is identified as an electron candidate. There exists a model for an NC Diffractive scattering process in GENIE, from the work of Rein \cite{REIN198661}, that is valid for $W > 2.0$ GeV but with an underestimated cross section
%(\refcomComment{added "but with an underestimated cross section"}). 
GENIE's implementation of the Rein model is used to predict this background contribution.

We divide the background processes into two cases. The first case is when a $\pi^0$ is the only particle produced in the neutrino interaction which is inclusive of coherent NC pion production and NC diffractive pion production. The second is when photons are produced from a $\pi^0$ and additional particles are also produced simultaneously including NC incoherent pion production and nonelectron neutrino CC pion production. Given the 2.5 GeV energy requirement of the electron, the NC background typically consists of high hadronic invariant mass $W^2$ events defined as
\begin{equation}
    W^2 = M^2 + 2 M \nu,
\end{equation}
\noindent where $M$ is the nucleon mass.
The exception are the NC diffractive $\pi^0$ and NC coherent $\pi^0$ cases. Neutrino-electron elastic scattering also contributes to the signal sample at values of zero $E_{avail}$.

\begin{figure}
    \centering
    \includegraphics[width=8cm]{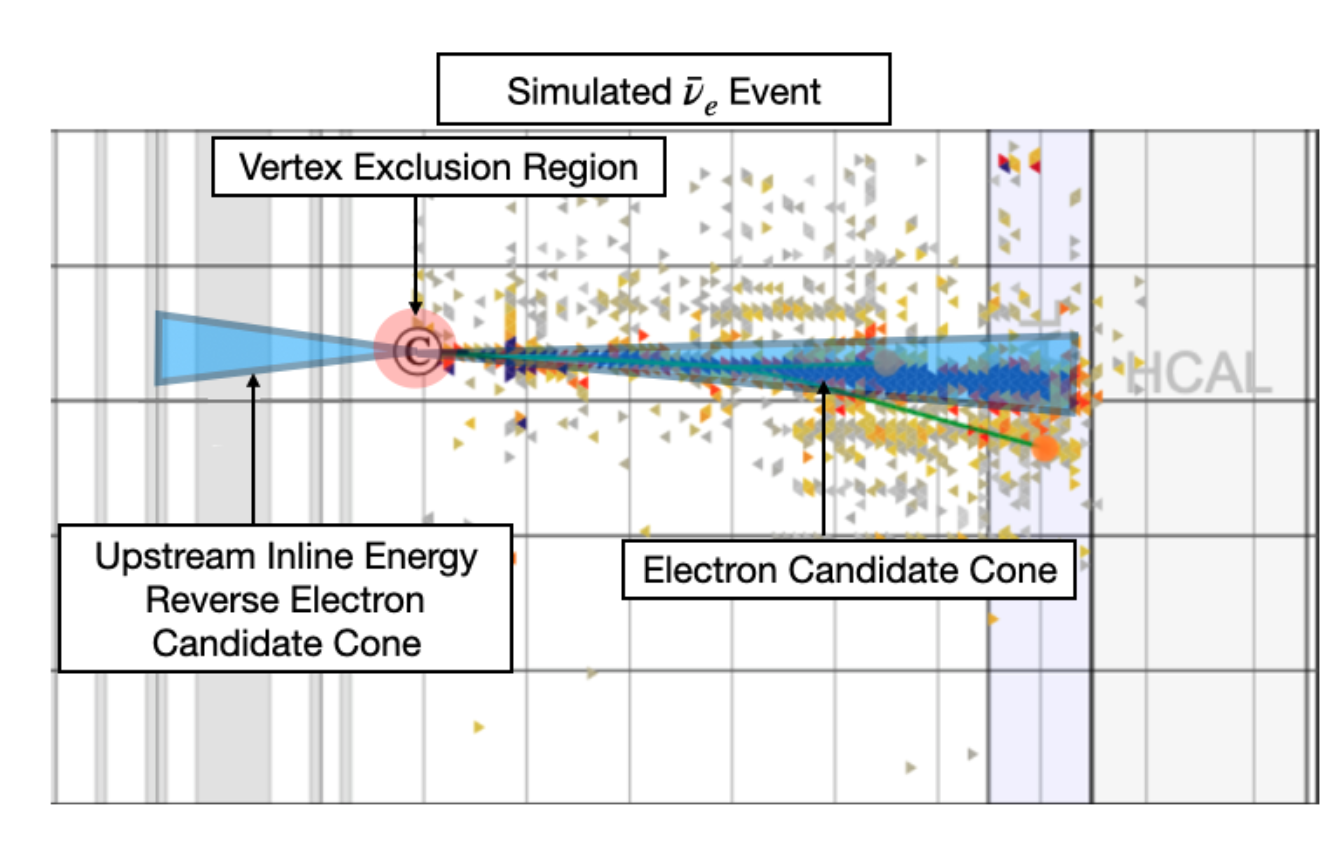}
    \caption{Graphic description of upstream inline energy. Upstream inline energy is defined as the energy deposited inside a backward oriented $7.5 ^\circ$ cone with its apex emanating from the interaction vertex.}
    \label{fig:iuecone}
\end{figure}

\FloatBarrier

\subsection{Background constraints}

\begin{figure}
    \centering
    \includegraphics[width=8cm]{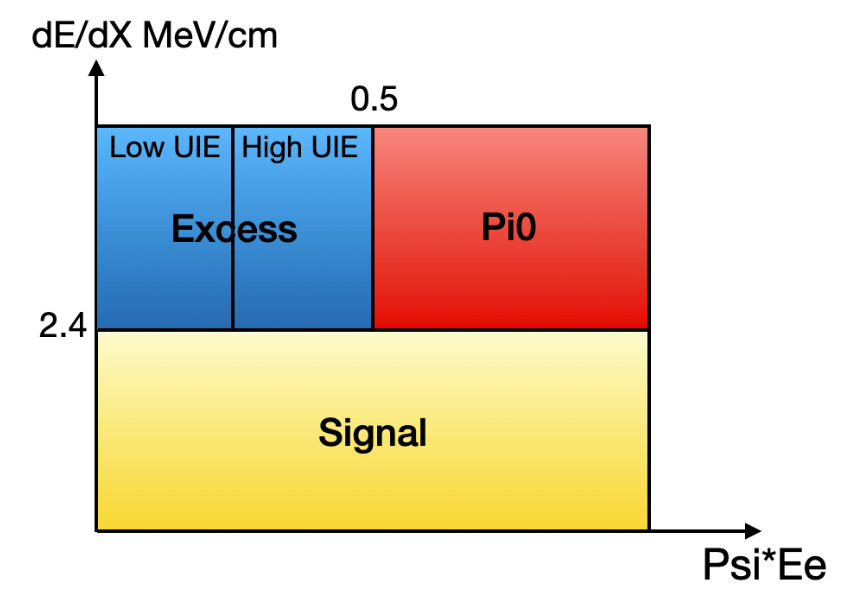}
    \caption{Representation of signal and sideband regions with respective cuts. The ``excess" sideband has been subdivided into low and high upstream inline energy.}
    \label{fig:sideband}
\end{figure}

The backgrounds are constrained by examining sidebands in the high dE/dx region which is further subdivided into three separate regions used to separate $\pi^0$ production channels. We use two variables to define the sidebands: upstream inline energy $E_{\rm UIE}$ and extra energy $\Psi E_{EM}$. Upstream inline energy is defined as the energy depositions inside of a reversed $7.5 ^\circ$ cone region, as shown in Fig. \ref{fig:iuecone}. $E_{\rm UIE}$ is the best discriminator between NC $\pi^0$ coherent and NC $\pi^0$ diffractive events, allowing for the capture of recoiling proton energy upstream of the event vertex. $\Psi E_{EM}$ is defined by the ratio of visible energy outside of the electron cone to energy inside:
\begin{equation}
    \Psi = \frac{E_{extra} + E_{\rm UIE}}{E_{EM}}
\end{equation}
The division of the sidebands regions are shown in Fig. \ref{fig:sideband} and  summarized as the incoherent $\pi^0$ region defined by $\psi E_{EM}$ $>$ 0.5 GeV and dE/dx $>$ 2.4 MeV/cm, the coherent region defined by $\psi E_{EM}$ $<$ 0.5 GeV, $E_{\rm UIE}$ $<$ 10 MeV and dE/dx $>$ 2.4 MeV/cm, and the diffractive region defined by $\psi E_{EM}$ $<$ 0.5 GeV, $E_{\rm UIE}$ $>$ 10 MeV and dE/dx $>$ 2.4 MeV/cm. 
%(\refcomComment{used incoherent, coherent, and diffractive $\pi^0$ definitions in figures 5-13  })

%(\refcomComment{resized font in table})

%begin{widetext}
\begin{table}[h!]
\fontsize{6pt}{6pt}
%\tiny
\centering
\sffamily
\begin{tabular}{cccccccc}
    \hline
    $E_{EM}$ (GeV)&[2.5,5)&[5,7.5)&[7.5,10)&[10,12.5)&[12.5,15)&[15,20) &\\
    Diffractive $\pi^0$ & 3.385 &7.413&9.535& 15.95 &23.21&9.807&\\
    Coherent $\pi^0$ & 1.970 & 2.258 & 2.936 &  2.614 & 2.018 & 5.363& \\
    \hline
    $P_{lep}^{t}$ (GeV)&[0,0.2)&[0.2,0.4)&[0.4,0.6)&[0.6,0.8)& [0.8,1.0)& [1.0,1.2)&[1.2,1.6)\\
    Noncoherent $\pi^0$ & 0.6897 &0.6945&0.7659&0.8151&0.9229 &1.014 &1.151\\
    \hline
\end{tabular}

\caption{FHC scale factors applied to $\pi^0$ production processes}
\label{tab:FHC_Background_SF}
\end{table}
\begin{table}[h!]
\fontsize{6pt}{6pt}\selectfont
%\tiny
\centering
\sffamily
\begin{tabular}{cccccccc}
    \hline
    $E_{EM}$ (GeV)&[2.5,5)&[5,7.5)&[7.5,10)&[10,12.5)&[12.5,15)&[15,20) &\\
    Diffractive $\pi^0$ &  5.03 & 7.868 & 7.095 & 10.114  &  10.767 & 4.134 &\\
    Coherent $\pi^0$ & 1.911 & 2.000 & 2.363 & 1.894 & 1.318 & 3.693 &\\
    \hline
    $P_{lep}^{t}$ (GeV)&[0,0.2)&[0.2,0.4)&[0.4,0.6)&[0.6,0.8)&[0.8,1.0)& [1.0,1.2)&[1.2,1.6)\\
    Noncoherent $\pi^0$ & 1.156 & 1.074 & 1.044 & 1.083 & 1.072  & 1.198 & 1.336\\
    \hline
\end{tabular}
\caption{RHC scale factors applied to $\pi^0$ production processes}
\label{tab:RHC_Background_SF}
\end{table}
%\end{widetext}

\begin{figure}
    \centering
    \includegraphics[width=8cm]{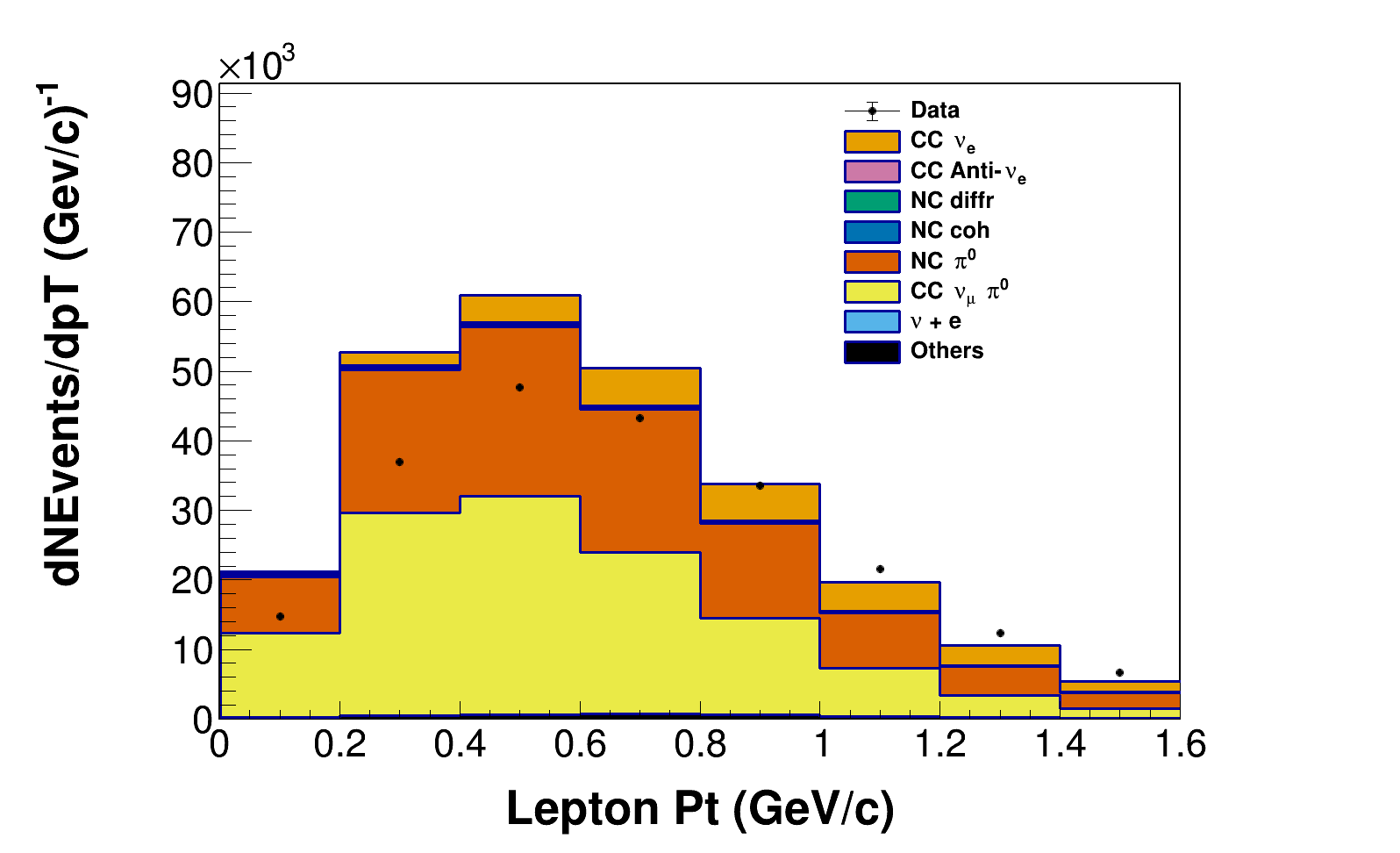}
    \includegraphics[width=8cm]{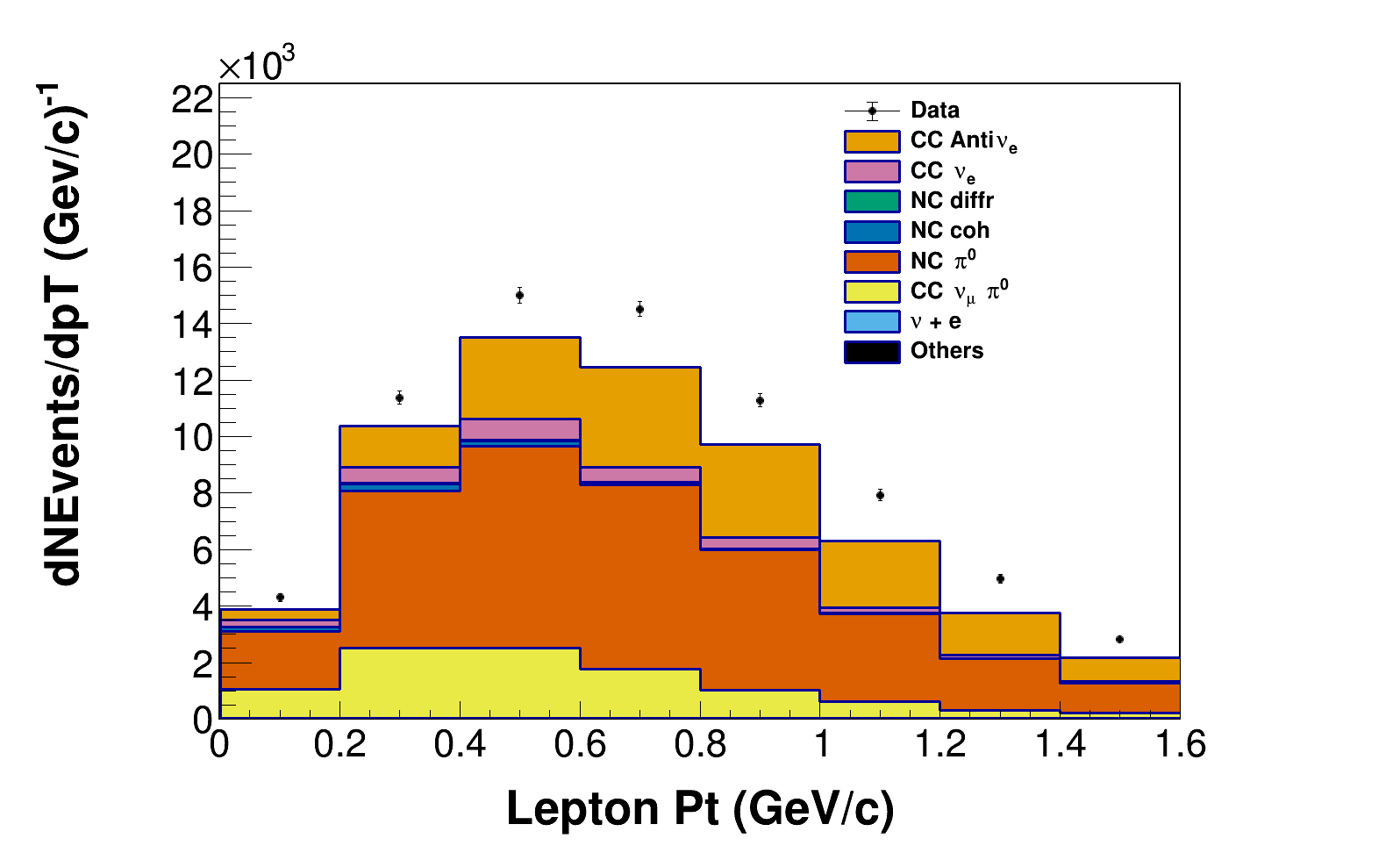}
    \caption{Prebackground tuned FHC (top) and RHC (bottom) $p_T$ distribution for the incoherent $\pi^0$ sideband ($dE/dx>$ 2.4 MeV/cm, $\psi*E_e>$ 0.5 GeV.)}
    \label{fig:unscaled_pt_pi0}
\end{figure}

\begin{figure}
    \centering
    \includegraphics[width=8cm]{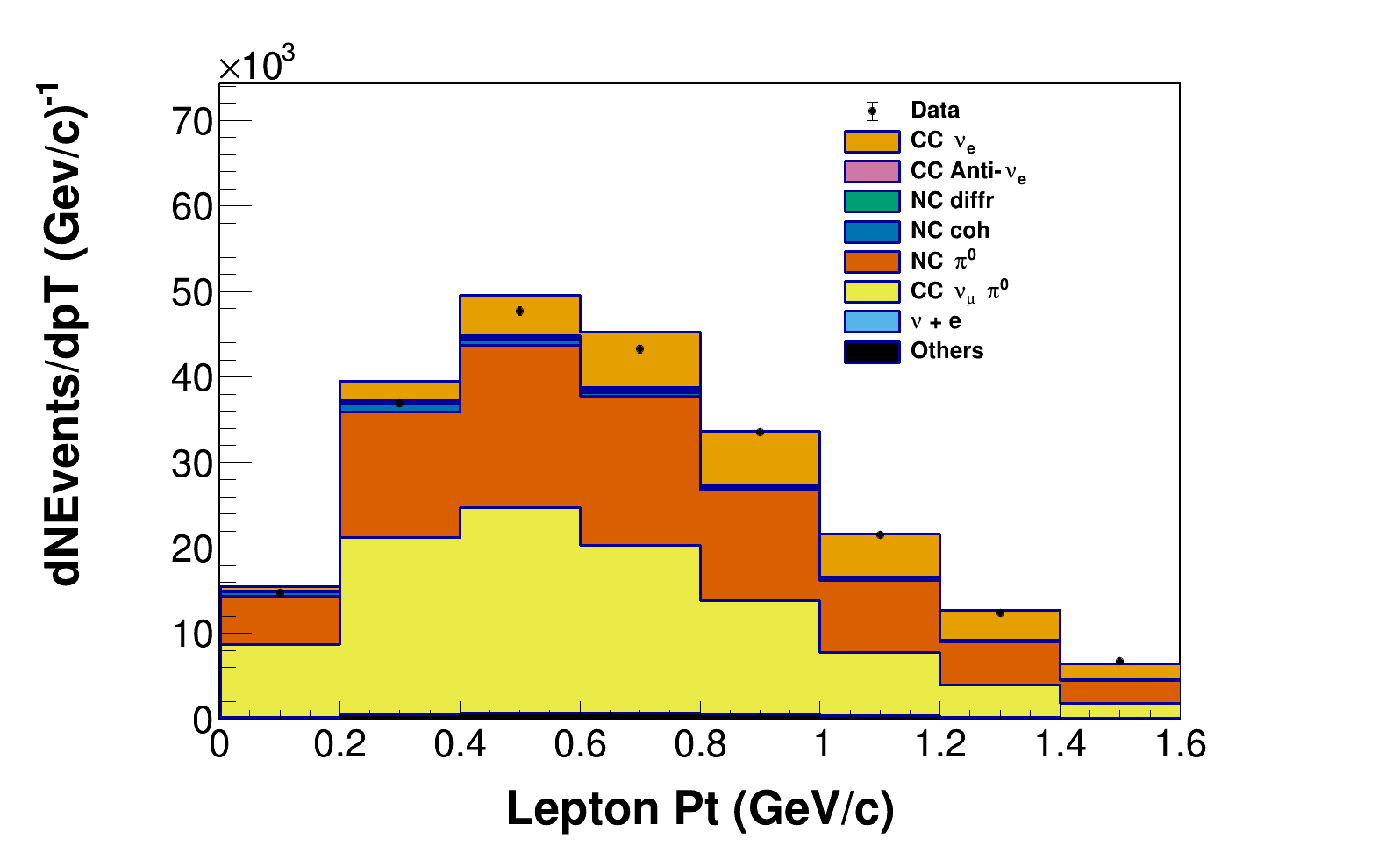}
    \includegraphics[width=8cm]{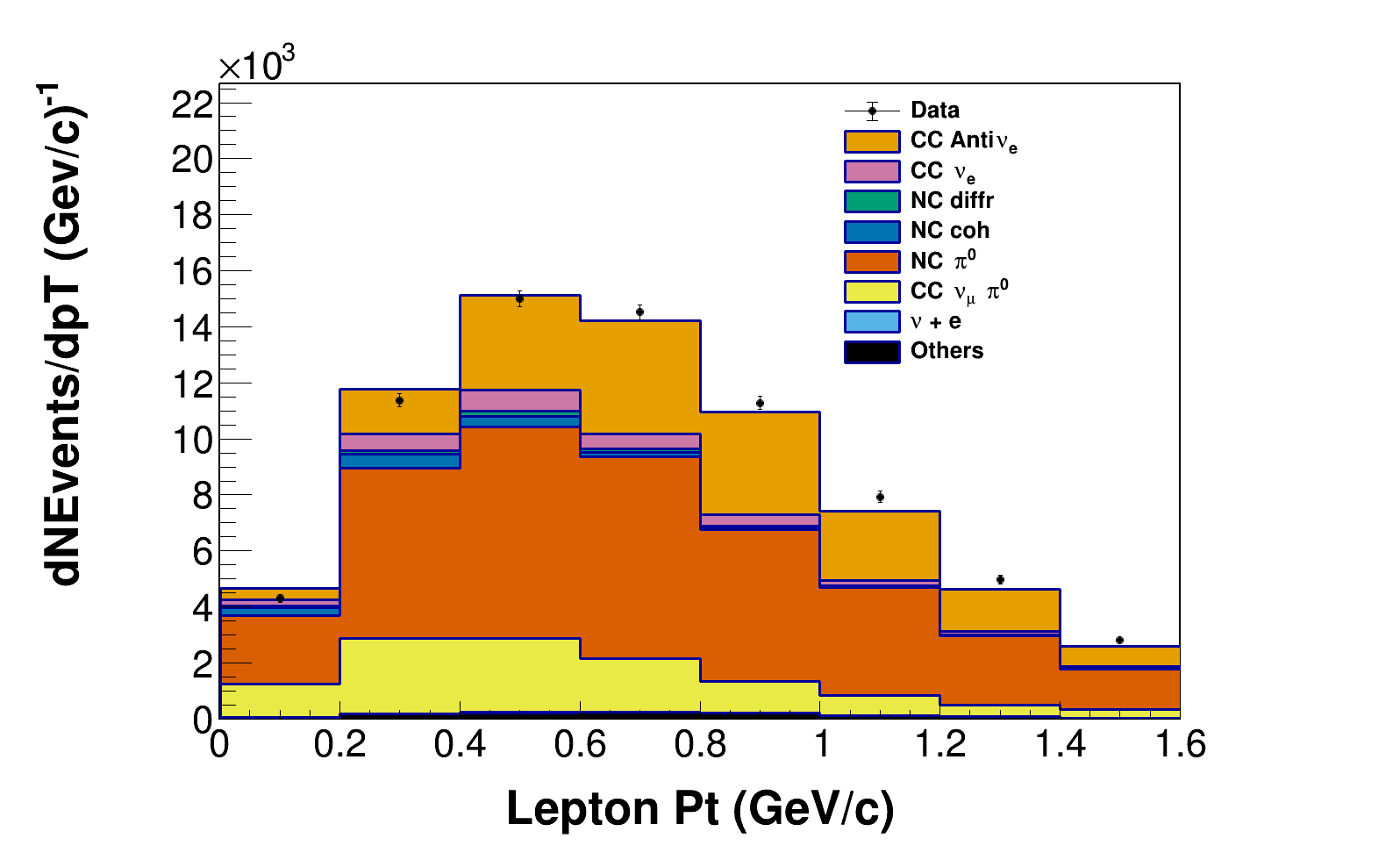}
    \caption{Postbackground tuned $p_T$ FHC (top) and RHC (bottom) distribution for the incoherent $\pi^0$ sideband  ($dE/dx>$ 2.4 MeV/cm, $\psi*E_e>$ 0.5 GeV.)}
    \label{fig:scaled_pt_pi0}
\end{figure}

\begin{figure}
    \centering
     \includegraphics[width=8cm]{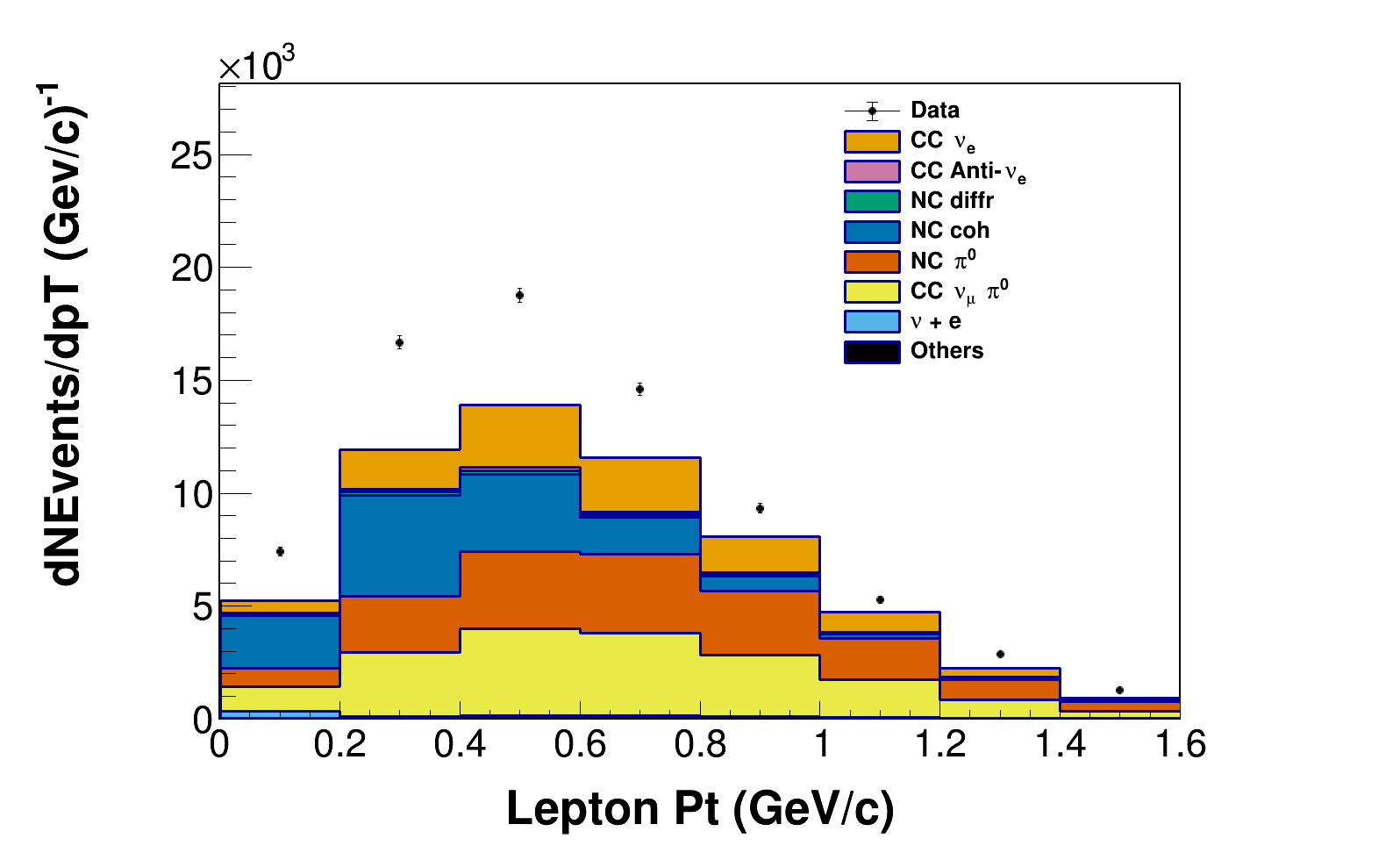}
    \includegraphics[width=8cm]{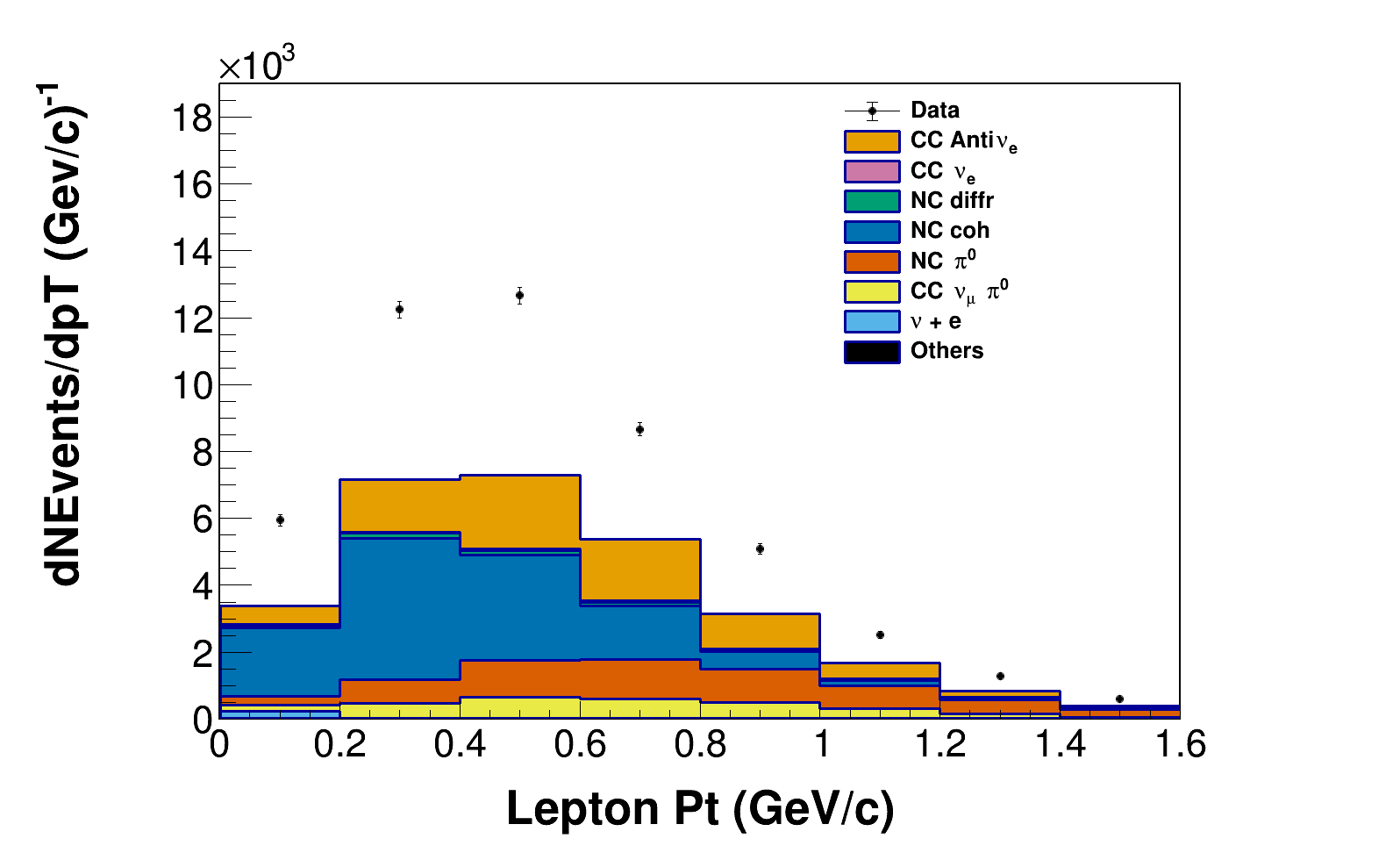}
    \caption{Prebackground tuned $p_T$ FHC (top) and RHC (bottom) $p_T$ distribution for the coherent $\pi^0$ (low UIE)  sideband ($dE/dx>$ 2.4 MeV/cm, $\psi*E_e<$ 0.5 GeV, $E_{uie}<$ 10 MeV.)}
    \label{fig:unscaled_pt_lowuie}
\end{figure}

\begin{figure}
    \centering
    \includegraphics[width=8cm]{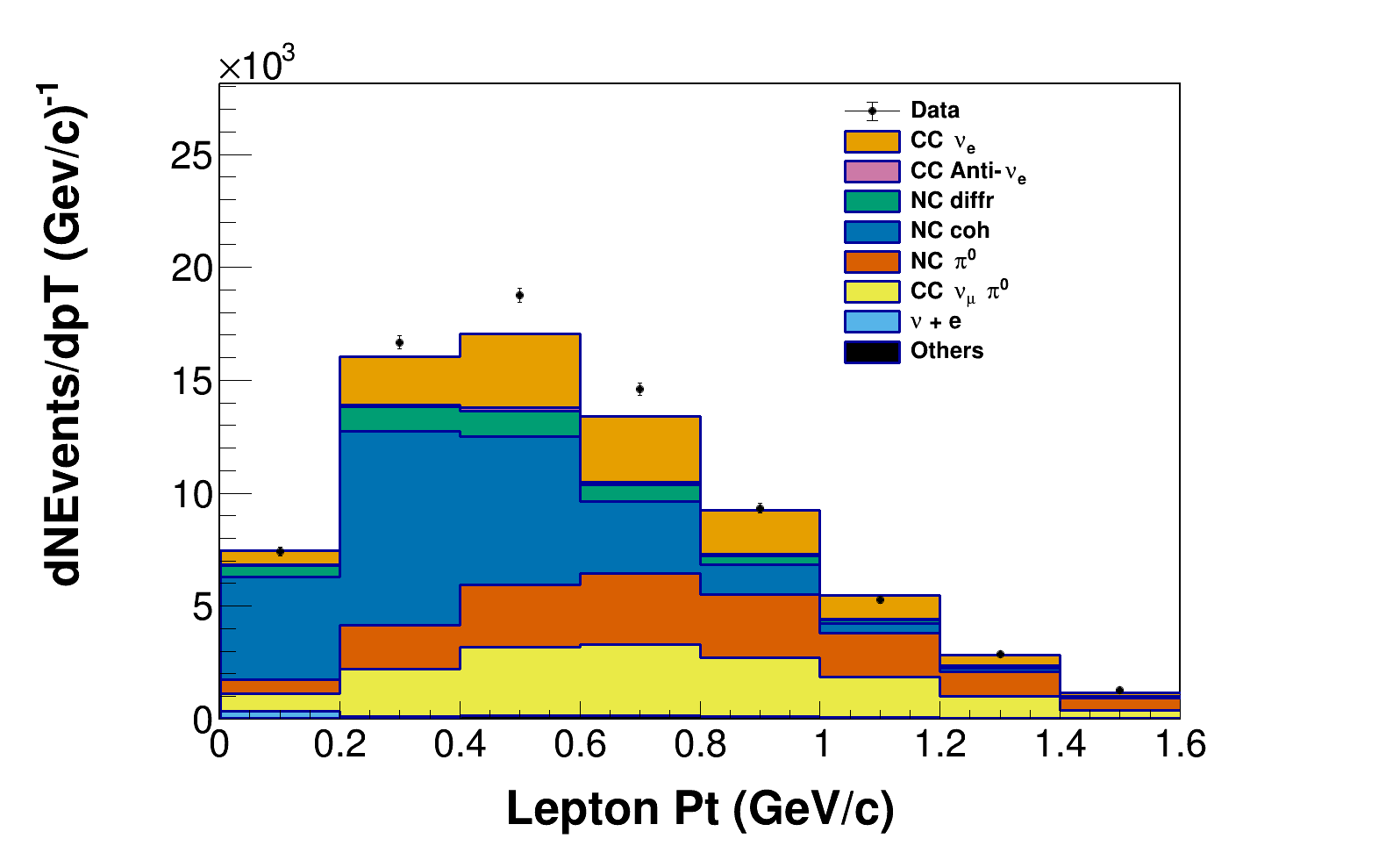}
    \includegraphics[width=8cm]{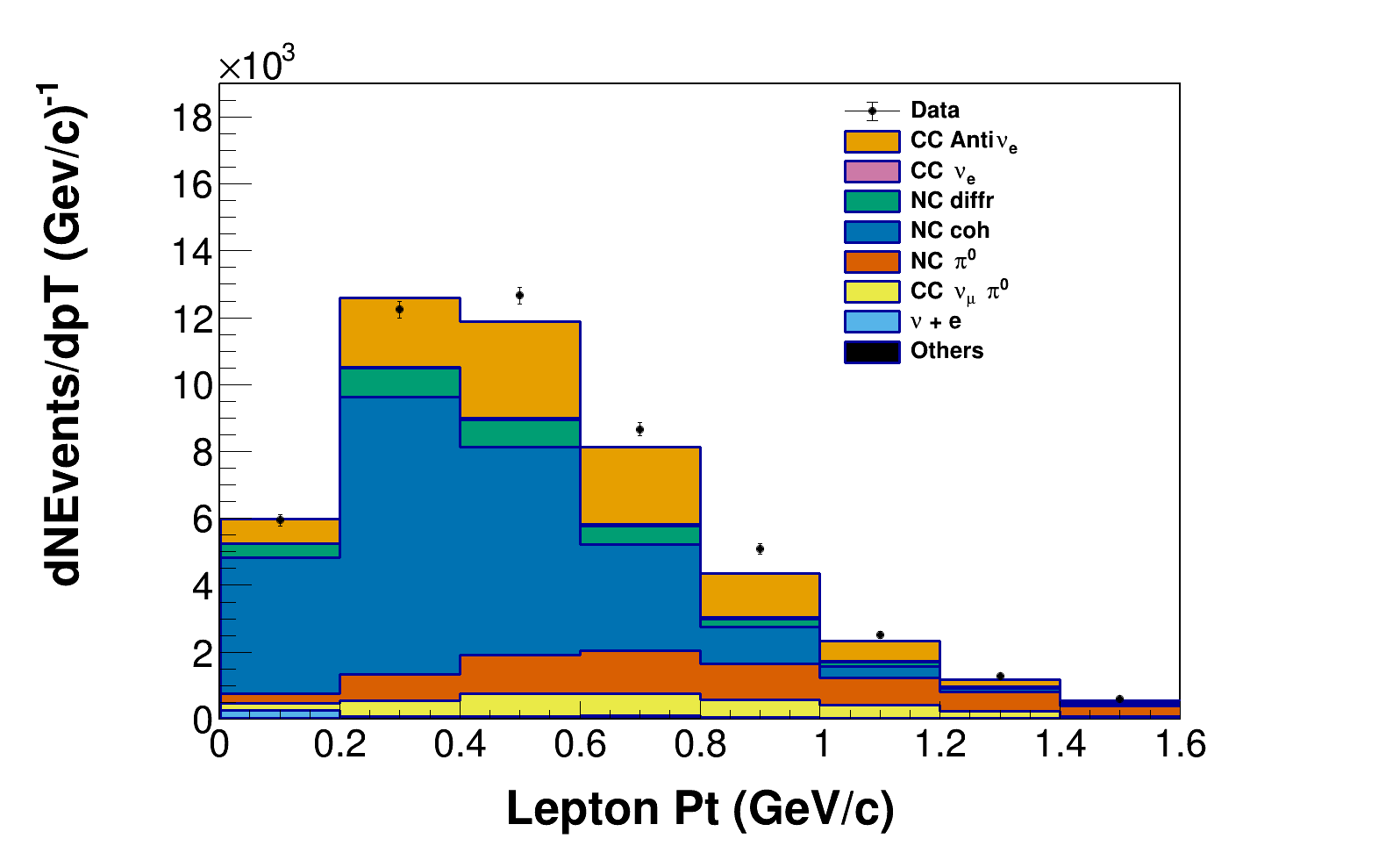}
    \caption{Postbackground tuned $p_T$ FHC (top) and RHC (bottom) $p_T$ distribution for the coherent $\pi^0$ (low UIE) sideband ($dE/dx>$ 2.4 MeV/cm, $\psi*E_e<$ 0.5 GeV, $E_{uie}<$ 10 MeV.)}
    \label{fig:scaled_pt_lowuie}
\end{figure}

\begin{figure}
    \centering
    \includegraphics[width=8cm]{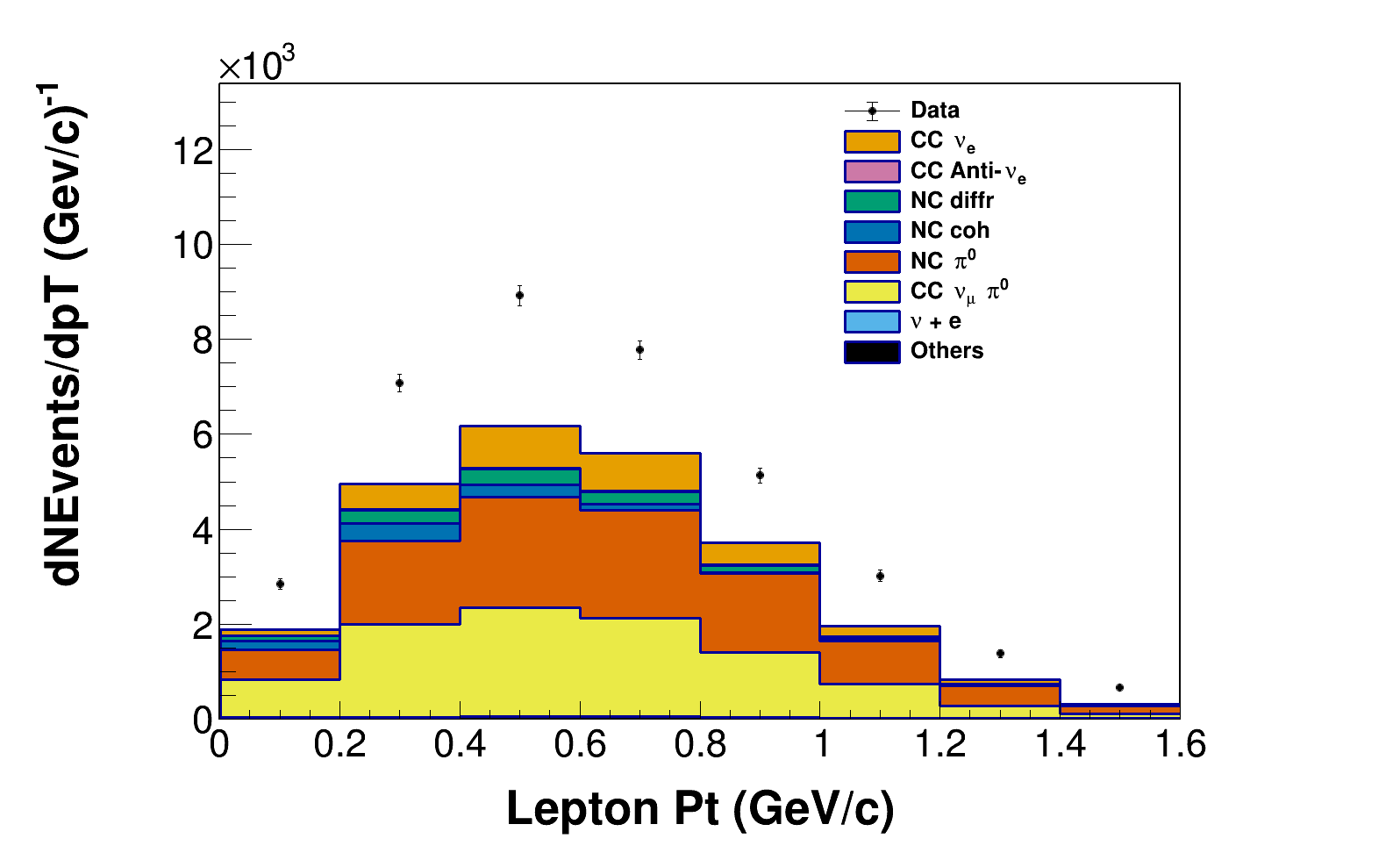}
    \includegraphics[width=8cm]{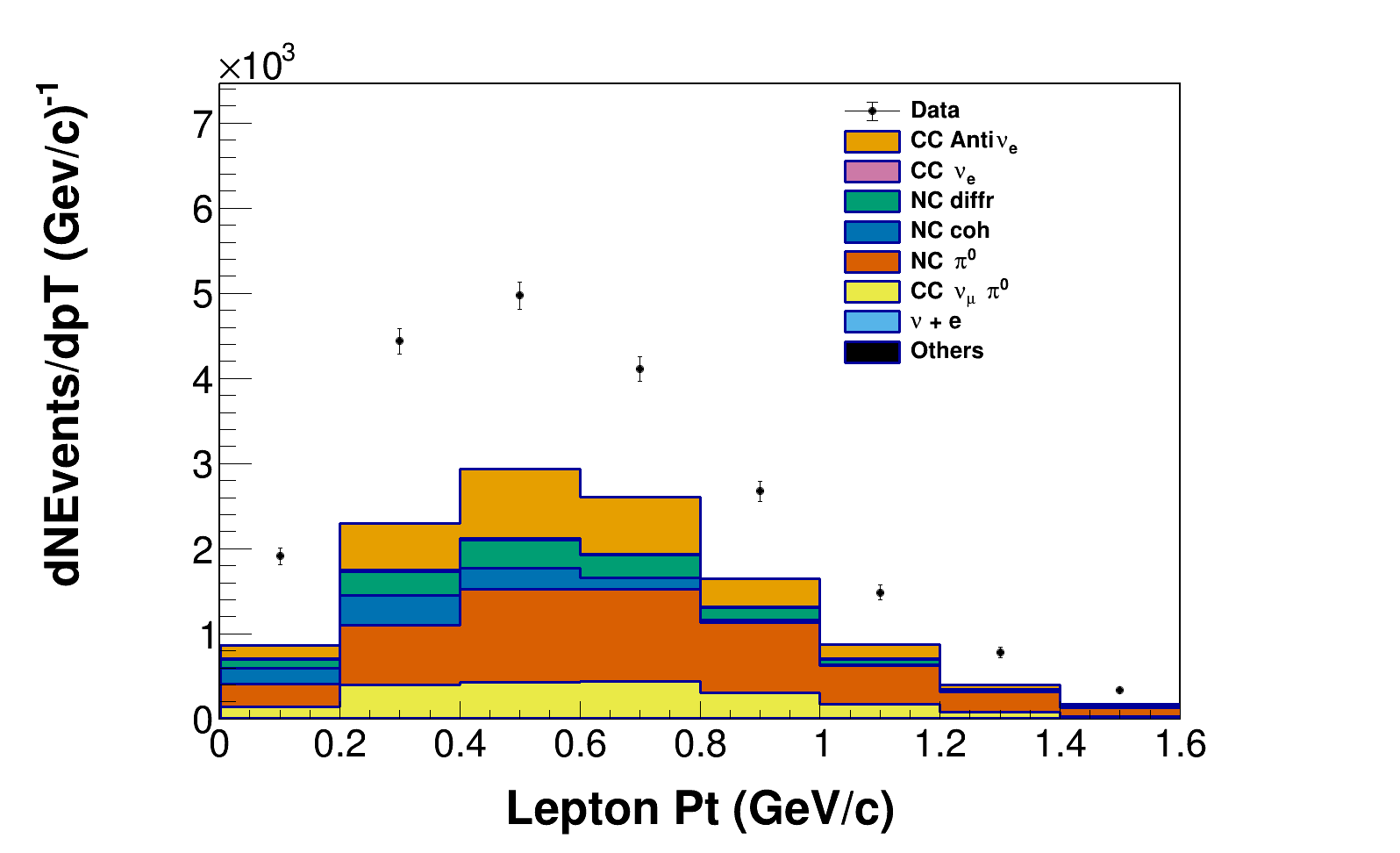}
    \caption{Prebackground tuned FHC (top) and RHC (bottom) $p_T$ distribution for the diffractive $\pi^0$ (high UIE) sideband ($dE/dx>$ 2.4 MeV/cm, $\psi*E_e<$ 0.5 GeV, $E_{uie}>$ 10 MeV.)}
    \label{fig:unscaled_pt_highuie}
\end{figure}

\begin{figure}
    \centering
    \includegraphics[width=8cm]{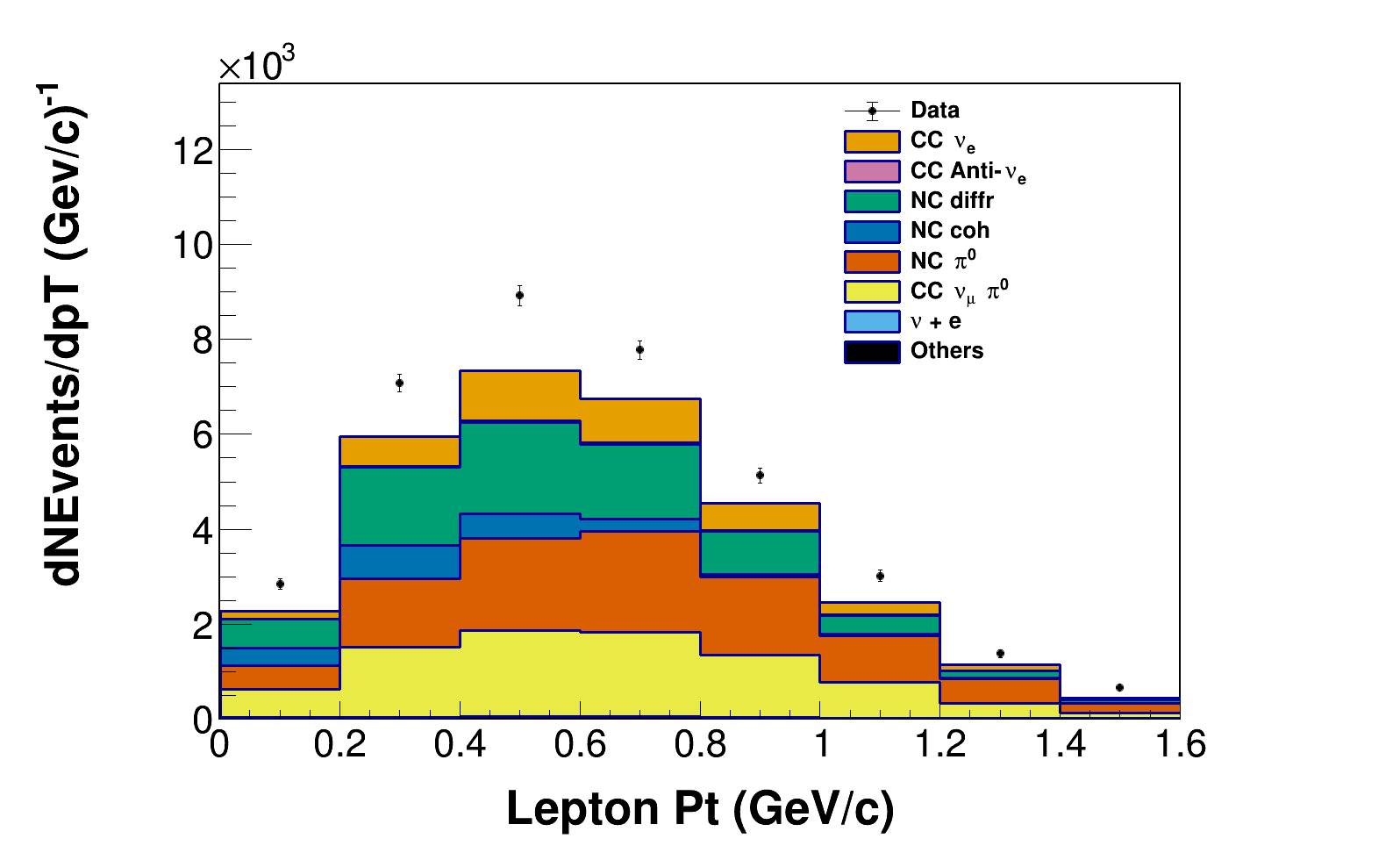}
    \includegraphics[width=8cm]{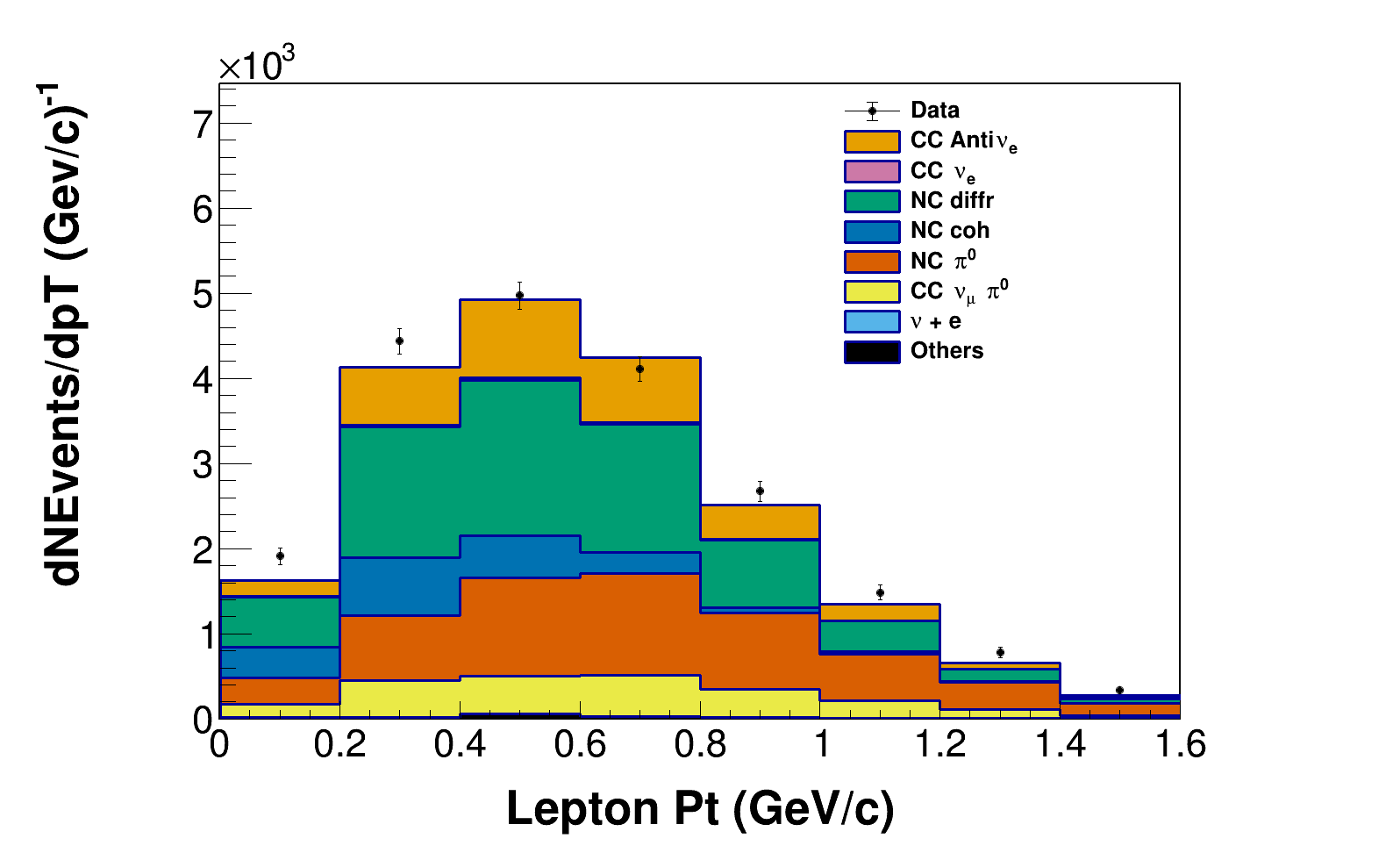}
    \caption{Postbackground tuned FHC (top) and RHC (bottom) $p_T$ distribution for the diffractive $\pi^0$ (high UIE) sideband  ($dE/dx>$ 2.4 MeV/cm, $\psi*E_e<$ 0.5 GeV, $E_{uie}>$ 10 MeV.)  The disagreement in FHC, as a result of tension between this region and the incoherent $\pi^0$ sideband, is discussed in the text.}
    \label{fig:scaled_pt_highuie}
\end{figure}

The normalization of the $\pi^0$ backgrounds are each fitted using distributions in both bins of $E_{avail}$ vs $q_3$ and $E_{avail}$ vs $p_T$ to obtain scale factors that represent the best estimate of the normalization of data compared to the GENIE prediction. The signal contribution is also tuned during this global fitting due to its non-negligible contribution in the sideband regions; however, this tune to the signal model is not applied to the signal model after the determination of the background from sidebands.  
%(\refcomComment{clarified this clause}) 
The fitting process, which is done through minimizing the negative log-likelihood assuming Poisson distribution, is done in two steps.  The first global fit is done with RHC data in $\Psi E_{EM}$ vs $p_T$ in bins of $E_{EM}$ which optimizes the NC coherent and diffractive processes. The background predictions of the coherent and diffractive $\pi^0$ processes are updated by applying scale factors on an event-by-event basis. The second global fit is done in $E_{avail}$ vs $p_T$ in bins of $p_T$ which optimizes noncoherent $\pi^0$ and signal processes separately for the respective FHC and RHC samples. The applied scale factors for the FHC and RHC analyses are found in Tables \ref{tab:FHC_Background_SF} and \ref{tab:RHC_Background_SF} respectively. Pretune and post-tune distributions in the sideband regions can be found in Figs. \ref{fig:unscaled_pt_pi0} - \ref{fig:scaled_pt_highuie} for RHC and FHC respectively. 

\FloatBarrier
\begin{figure}
    \centering
    \includegraphics[width=8cm]{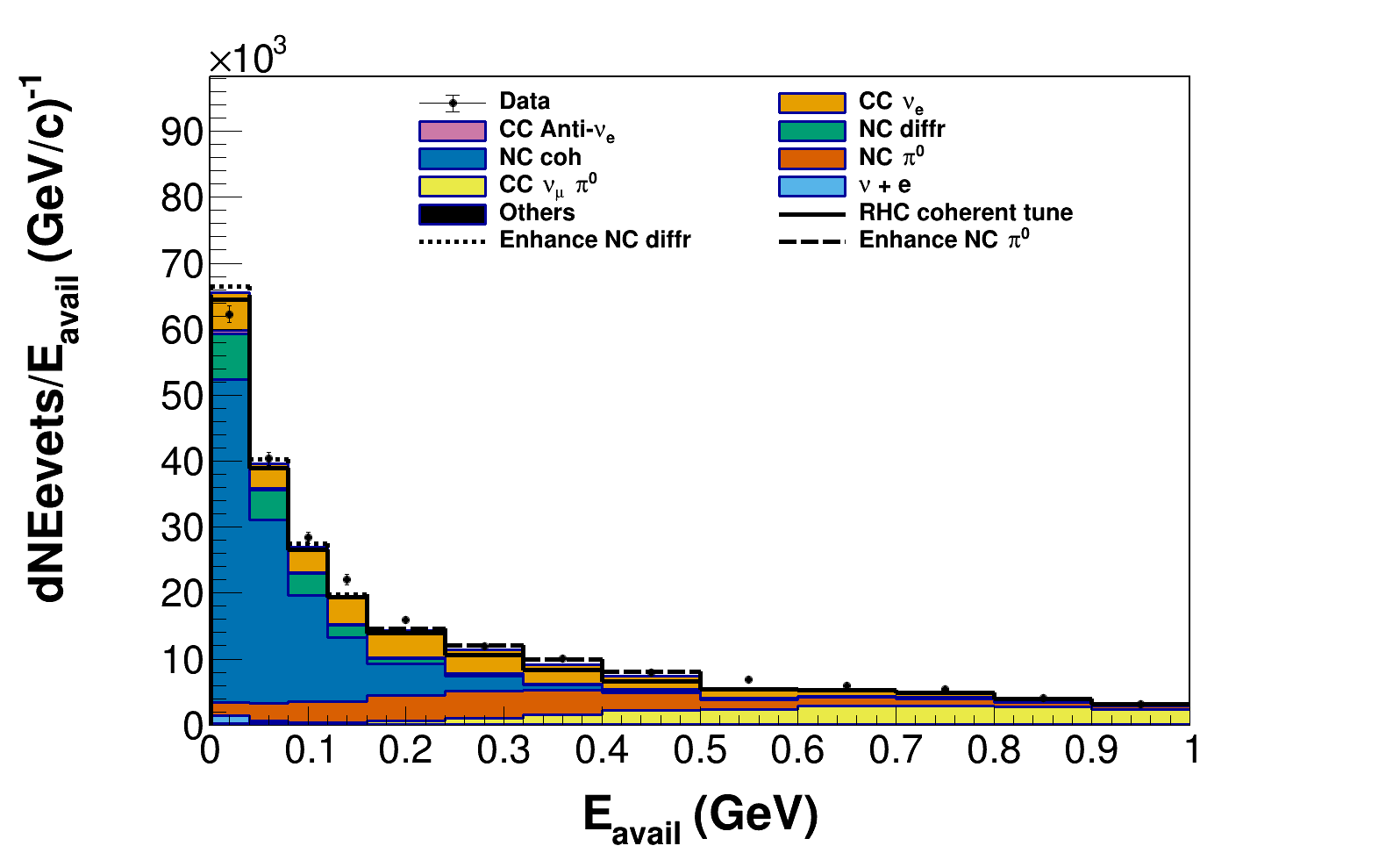}
     \includegraphics[width=8cm]{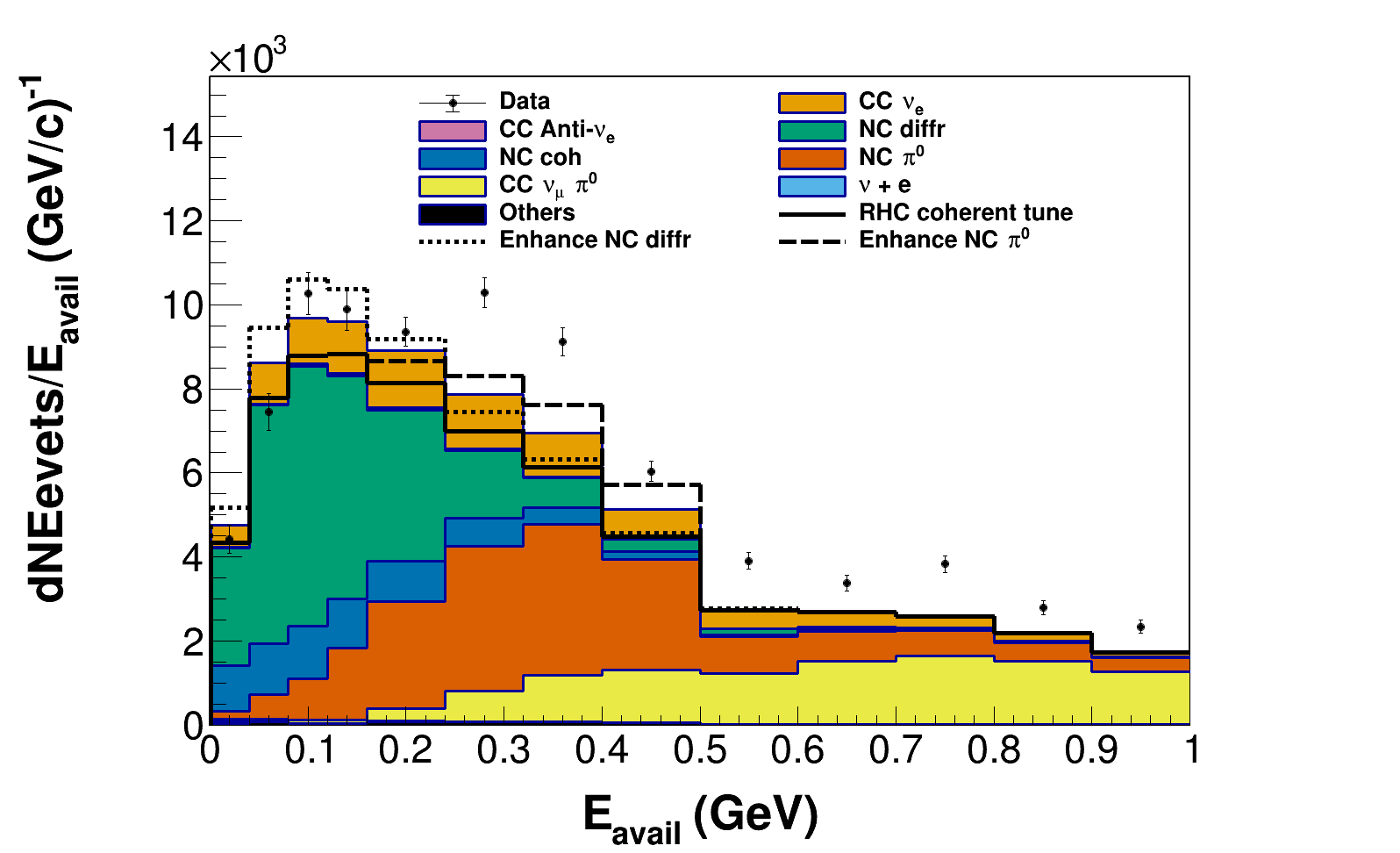}
      \includegraphics[width=8cm]{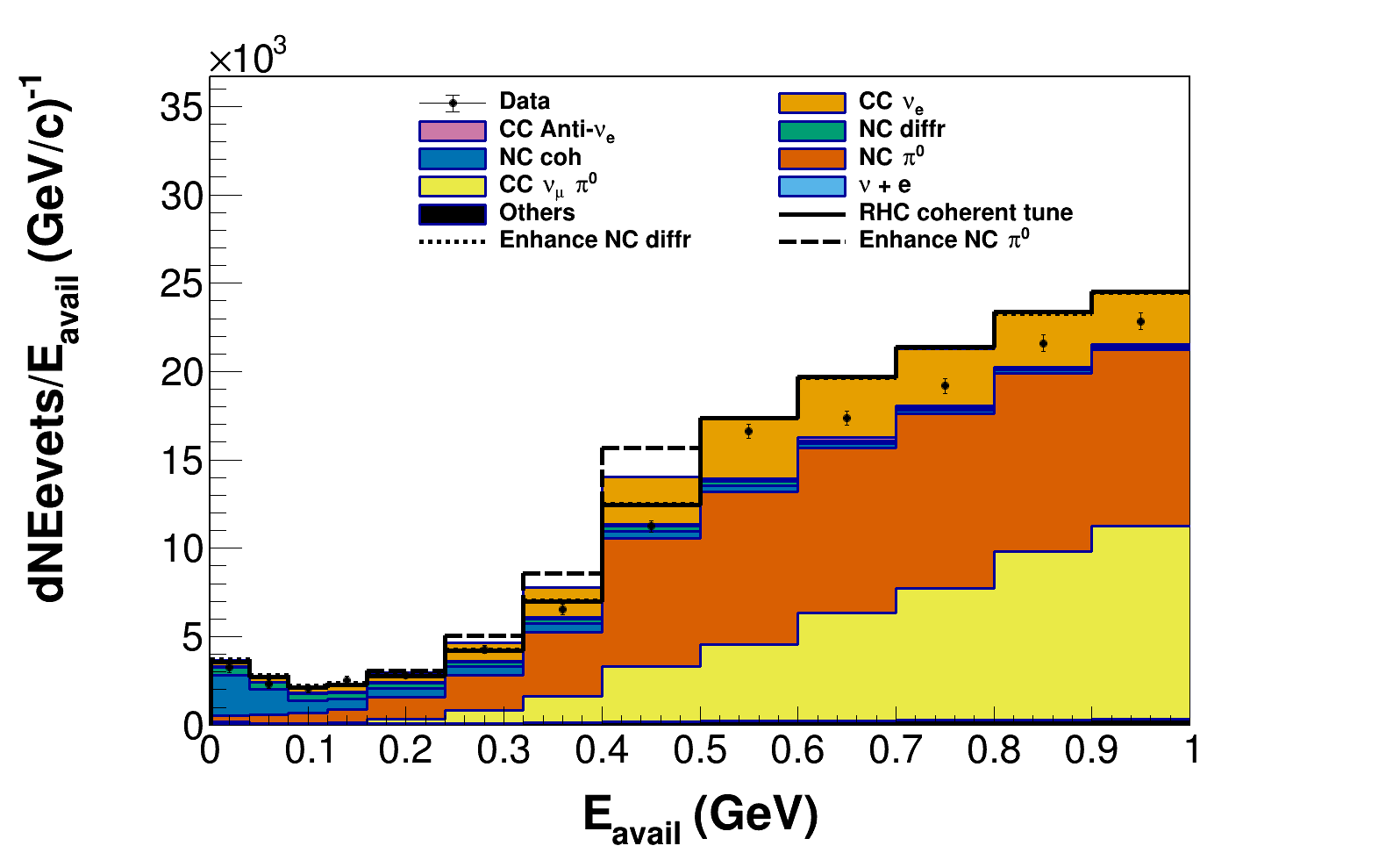}
    \caption{The high UIE sideband in FHC: demonstration of the tension in the tuning, and the alternate scenarios considered for (top) the coherent $\pi^0$ (low UIE) sideband, (middle) the diffractive $\pi^0$ (high UIE) sideband, and (bottom) the incoherent $\pi^0$ sideband.}
    \label{fig:scaled_pt_highuie_alternate}
\end{figure}

\subsubsection{Tensions in the FHC constraints}
As noted above, the diffractive and NC coherent backgrounds are estimated using the RHC samples because those processes are a larger fraction of the low UIE and high UIE sidebands, respectively in the RHC beam.  However, in the high UIE and incoherent $\pi^0$ FHC sideband, this fit is unable to reproduce the shape of the data as a function of $E_{avail}$, as shown in Fig.~\ref{fig:scaled_pt_highuie_alternate}.  Additional tunes and a systematic uncertainty on those tunes were developed to address this disagreement.  We considered two alternate hypotheses, neither of which describes the data well across all of the sidebands.  In the first, NC coherent and diffractive processes are allowed to have an additional normalization in the second global FHC fit.  The rationale is that the high-energy neutrino components in the two beams, above the focusing peak, are different. This could affect the relative event rates in the FHC and RHC samples if the cross section has a poorly modeled rate as a function of neutrino energy.  In the second hypothesis, a subset of the noncoherent $\pi^0$ processes that dominates the region with the observed high UIE disagreement ($0.2$ GeV $< E_{avail} < 0.5$ GeV and $ P_{lep}^{t} < 1$ GeV) are enhanced independently from other noncoherent $\pi^0$ processes with a separate scale factor.  We use the average of these two fits as our base background prediction, and take the difference between the two as an assessment of the systematic uncertainty in this procedure.  Note that the background comes from all of these contributions together, and so the systematic underprediction of the FHC high UIE sideband coupled with the systematic underprediction of the incoherent $\pi^0$ sideband do not indicate that the background is poorly estimated.

\FloatBarrier

\subsection{Interpretation of the Coherent and Diffractive Contributions}
As shown in Tables \ref{tab:FHC_Background_SF} and \ref{tab:RHC_Background_SF}, the scale factors for the NC diffractive $\pi^0$ background process as well as the NC coherent $\pi^0$ process are large. We believe the explanation for these large-scale factors is that the diffractive and coherent $\pi^0$ processes are not well-modeled by our reference GENIE model.  We will discuss in turn the evidence that these events are in fact single $\pi^0$ in nature, the relative strength of the coherent and diffractive processes, and the reliability of the GENIE model prediction as a function of $E_{\pi^0}$. 

\begin{figure}[tp]
    \centering
    \includegraphics[width=8.5cm]{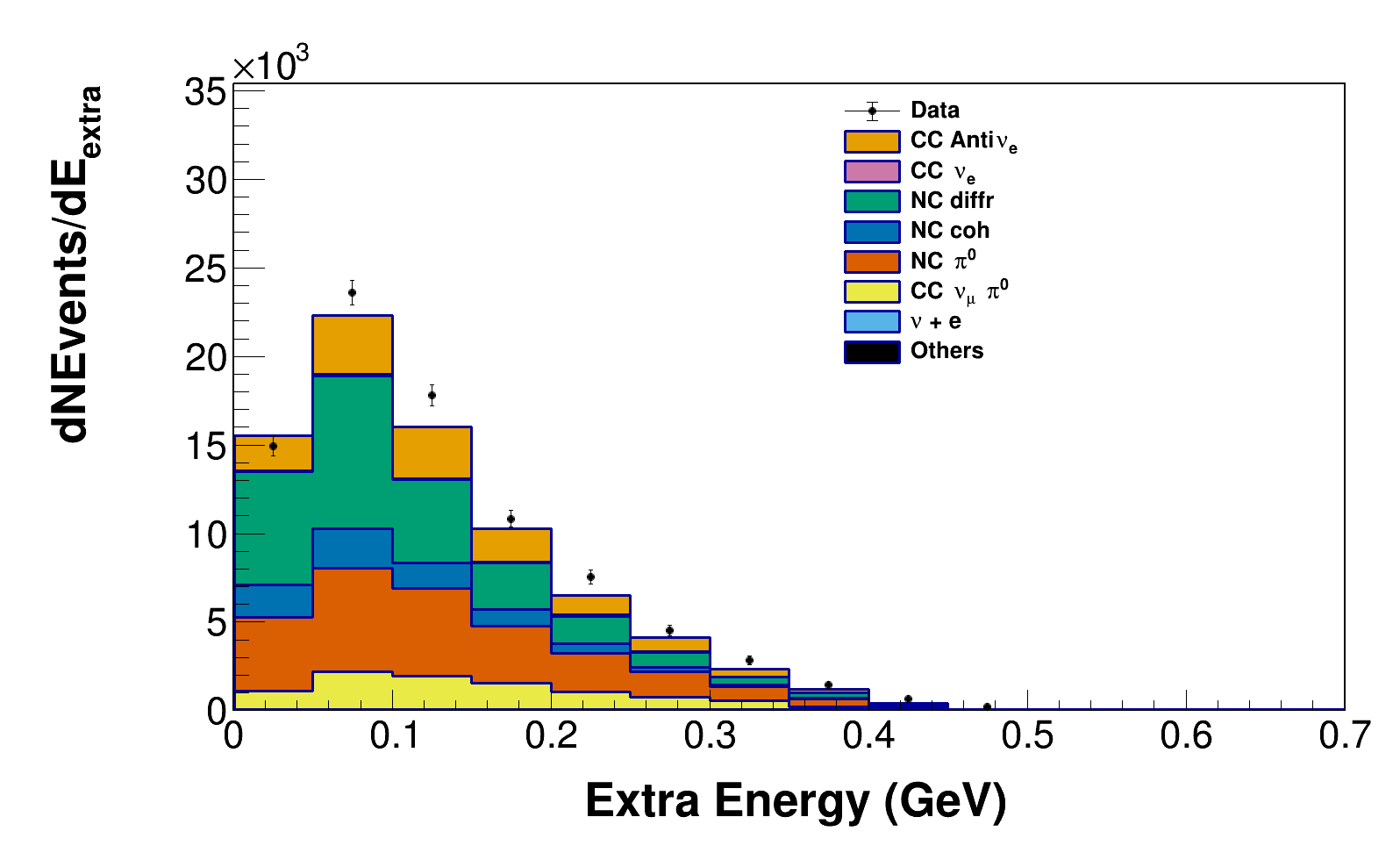}
    \caption{RHC energy outside of electron candidate cone and vertex region in the diffractive $\pi^0$  (high UIE) sideband region of dE/dx $>$ 2.4 MeV/cm, $\psi*E_e < 0.5 GeV$, $E_{uie} >$ 10 MeV.}
    \label{fig:extraenergy}
\end{figure}

Measurements of events from the sidebands targeting coherent and diffractive $\pi^0$ production, the low UIE and high UIE sidebands  respectively, do support the hypothesis that these events have a high energy $\pi^0$. Since electromagnetic cascades spread out transversely to the direction of propagation, there is a range of energy where single-photon showers can be distinguished from multiphoton showers based on transverse size. Median shower width, or median transverse width, provides the extent to which an electromagnetic cascade spreads transversely to its direction of propagation.

\begin{figure}[tp]
    \centering
    \includegraphics[width=8.5cm]{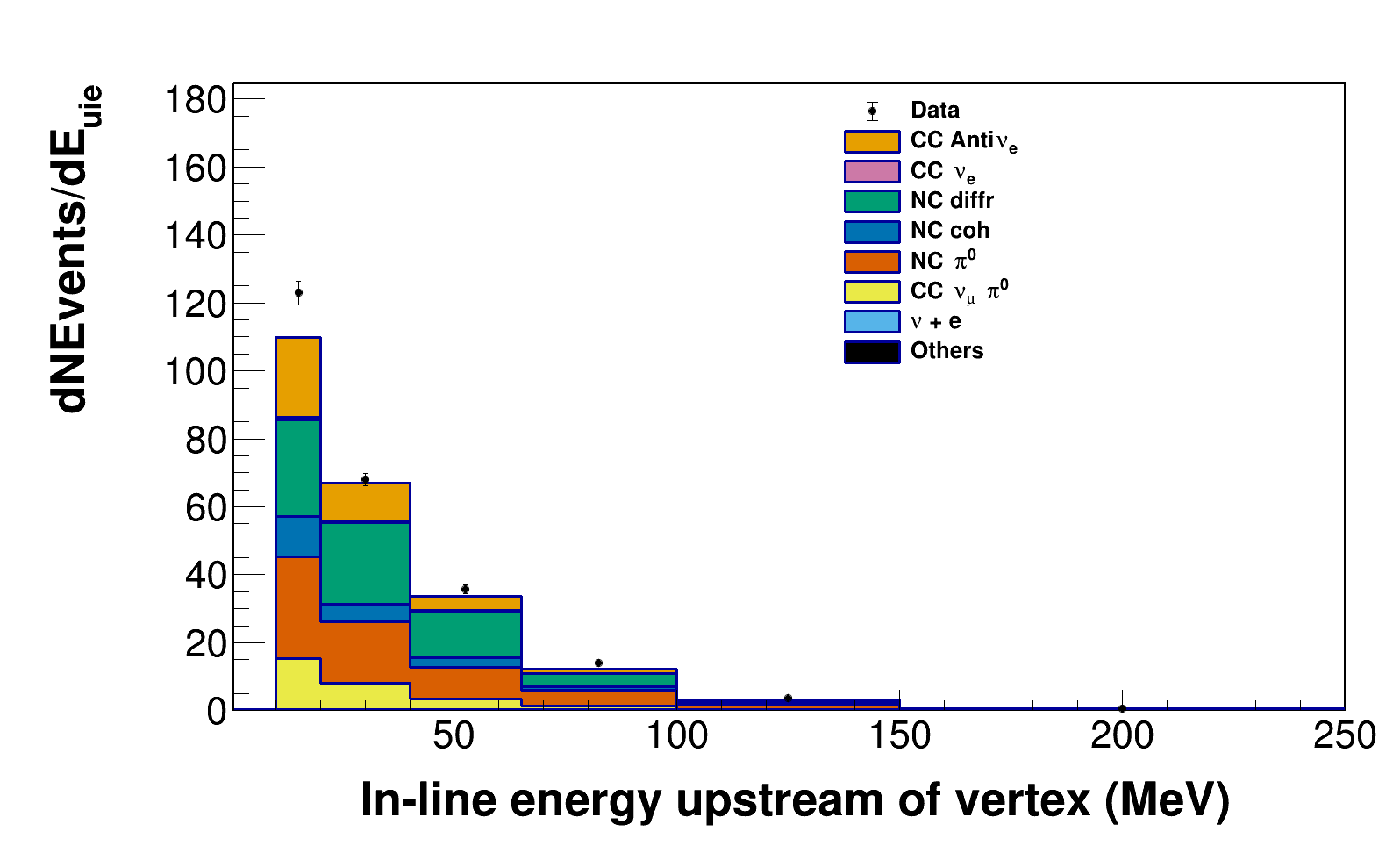}
    \caption{RHC upstream inline energy in the diffractive $\pi^0$  (high UIE) sideband region of dE/dx $>$ 2.4 MeV/cm, $\psi*E_e < 0.5 GeV$, $E_{uie} >$ 10 MeV.}
    \label{fig:inlineupstream}
\end{figure}

Figure \ref{fig:extraenergy} shows the post background tuned energy outside the electron candidate cone and the vertex region, referred to as the extra energy ($E_{extra}$). Most events from the high and low UIE sidebands populate the first few bins and are well-described. From these distributions, it is apparent that the event has little nonshower activity.  Additionally, the post background tuned inline-upstream energy cone distribution, shown in Fig. \ref{fig:inlineupstream}, indicates that the shape of the diffractive and coherent processes agree with what we would expect from energy upstream of the event vertex.
We note that the relative rate of the diffractive reaction, with high upstream inline energy, and the coherent events, with low upstream inline energy is consistent with naive scaling arguments which would suggest a dependence on the atomic number, $A$, somewhere between $A^{1/3}$ and $A^{2/3}$ between carbon and hydrogen.  
%This suggests that the large difference between the two in the GENIE model, approximately a factor of $30$, is incorrect. 
%(\refcomComment{reworded following sentence to clarify issue with GENIE})
This suggests that the large
difference (approximately a factor of $30$) between the two energy regions as implemented in the GENIE model, is incorrect.

On the subject of the $\pi^0$ energy dependence of the scale factors, a separate MINERvA analysis studying neutrino-induced coherent $\pi^+$ production on different targets~\cite{MINERvA:2022esg}  concluded that the Rein-Sehgal and PCAC-based Belkov-Kopeliovich (B-K) models do not accurately describe the angular dependence on $\theta_\pi$, the energy-dependence on $E_\pi$, or the A-dependence. The fact that this is also seen in the charged current analog reaction makes the energy dependent scale factors needed in this analysis more plausible.

\begin{figure}[tp]
    \centering
    \includegraphics[width=8cm]{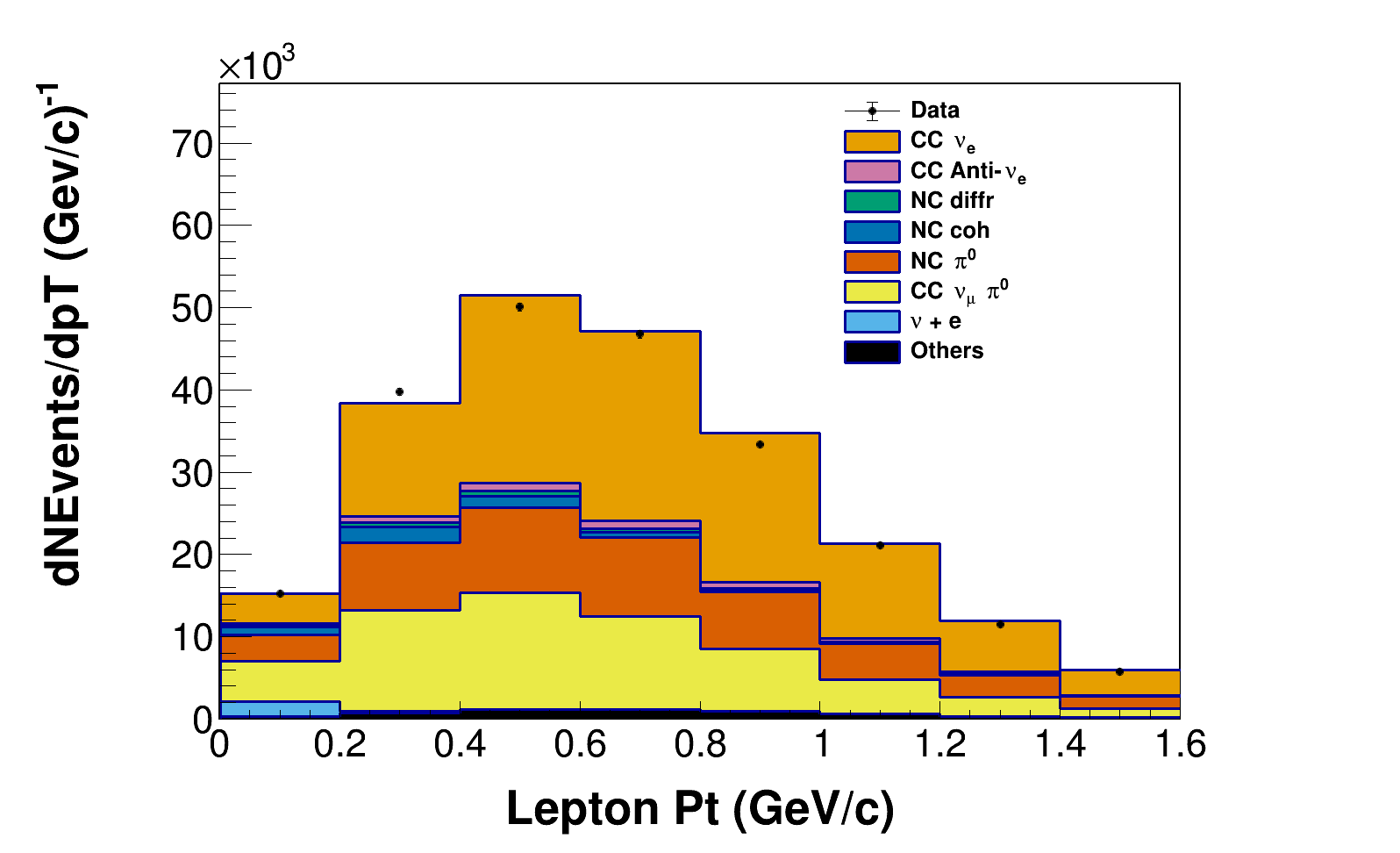}
    \includegraphics[width=8cm]{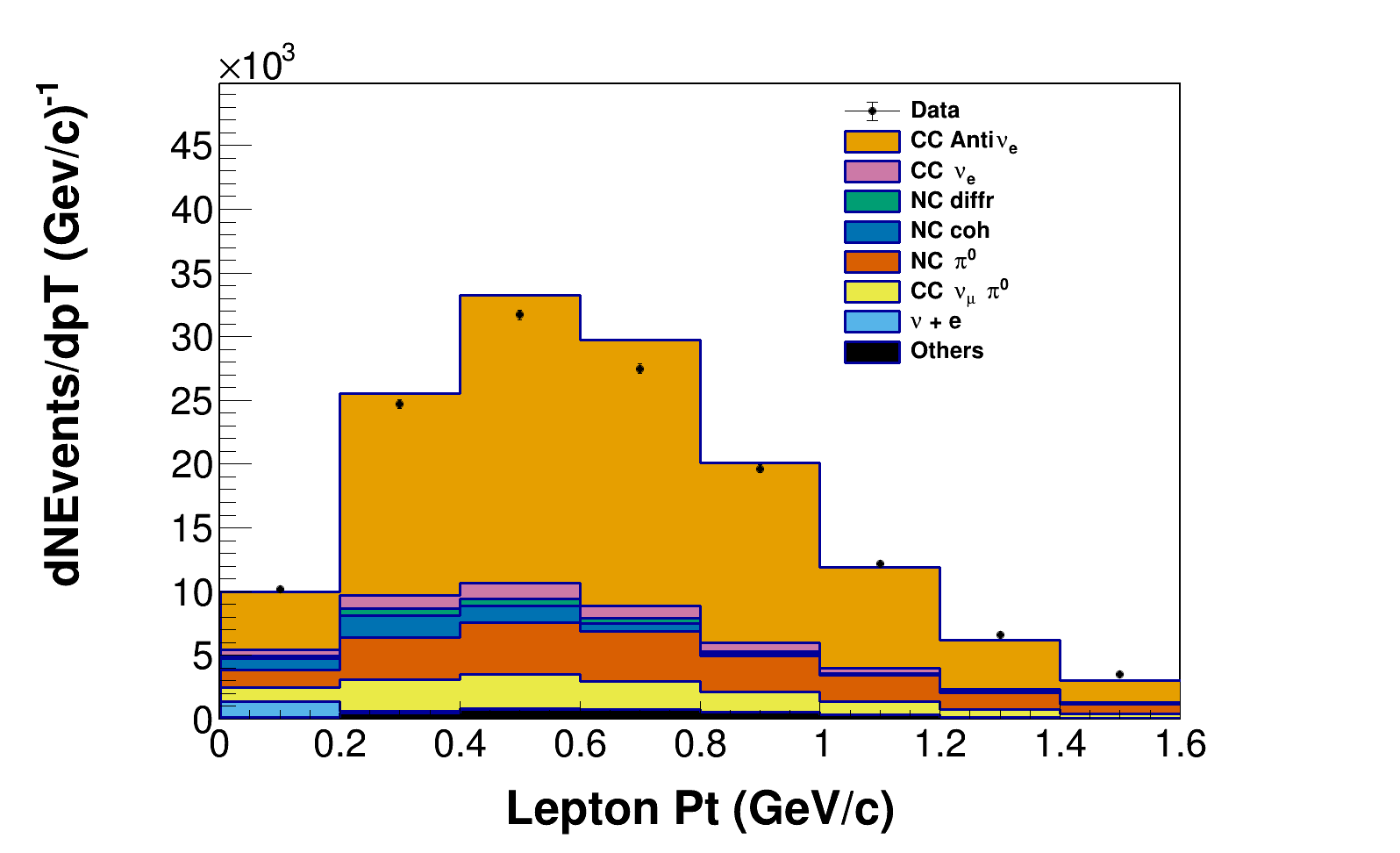}
    \caption{Postbackground tuned FHC (top) and RHC (bottom) $p_T$ distribution for the signal region of $dE/dx<$ 2.4 MeV/cm.}
    \label{fig:scaled_pt_signal}
\end{figure}
\subsection{Background subtracted signal distributions}
Figure \ref{fig:scaled_pt_signal} shows the lepton $p_T$ distribution for the signal region for both FHC and RHC samples. The FHC sample has approximately 46,700 selected events with a total estimated background of 24,600 events. The RHC sample has approximately 28,300 selected events and a total estimated background of 8,000 events.

%\ksmComment{This section should probably conclude with a summary of the sizes of the samples, and the total estimated background.  Could do with an equivalent to Fig.~10 for the signal region instead.}

\section{Separation of $\nu_e$ and $\bar{\nu}_e$ Events}

While $\bar{\nu}_e$ and $\nu_e$ events are indistinguishable in data, the MC simulation provides a prediction for the $\bar{\nu}_e$ contribution to the FHC sample and the $\nu_e$ contribution to the RHC sample.  To correct for the contamination from these events in the respective samples, we form an estimator based on FHC data and the MC simulation that gives a prediction of the $\nu_e$ background found in the RHC sample and vice versa. The procedure is identical for the two measurements, so we will describe the procedure to correct the RHC measurement.

\begin{figure}[tp]
    \centering
    \includegraphics[width=8cm]{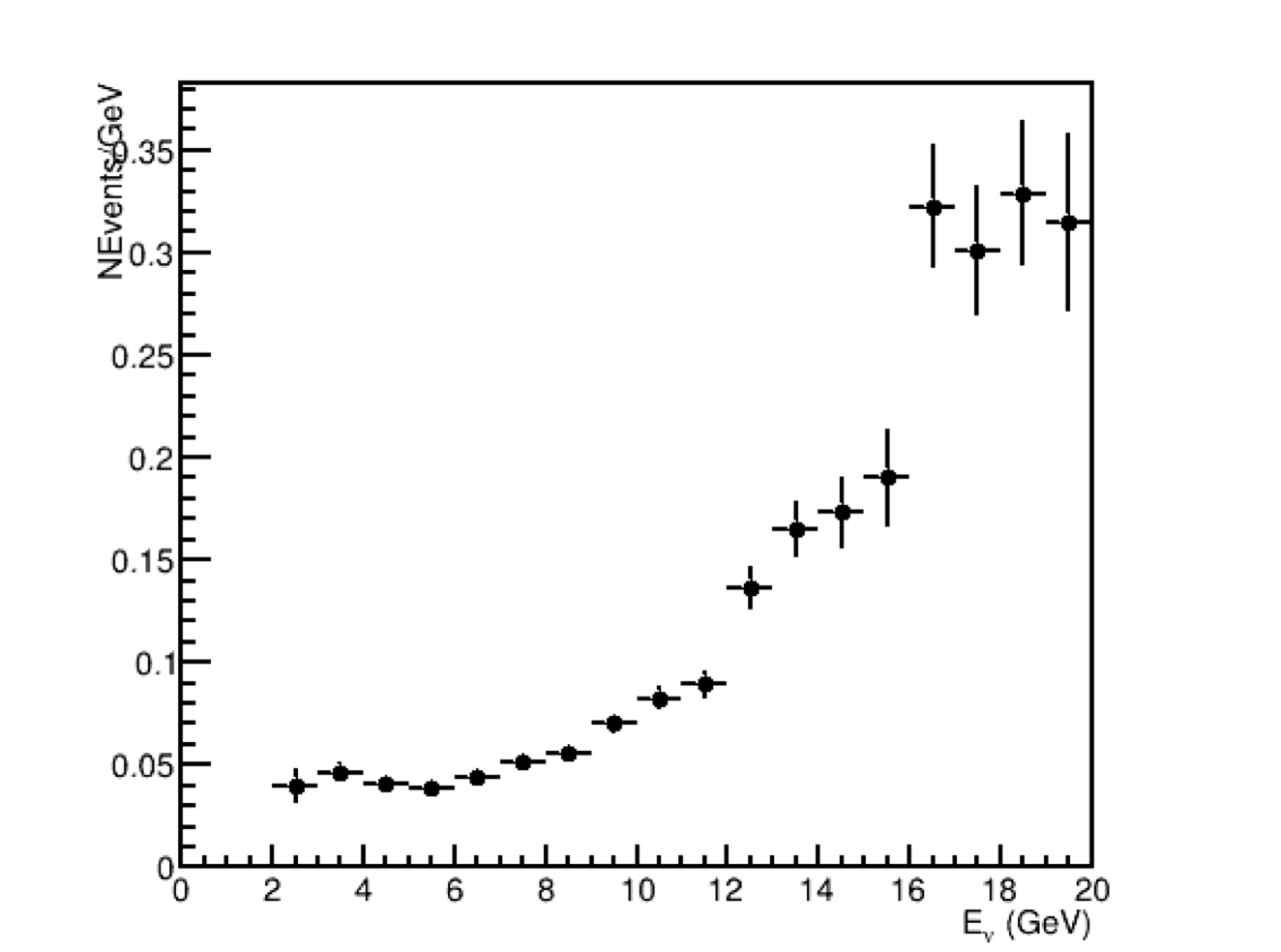}
    \caption{Ratio of RHC/FHC $\nu_e$ in true neutrino energy}
    \label{fig:RHCFHC}
\end{figure}

\begin{figure}
    \centering
    \includegraphics[width=8cm]{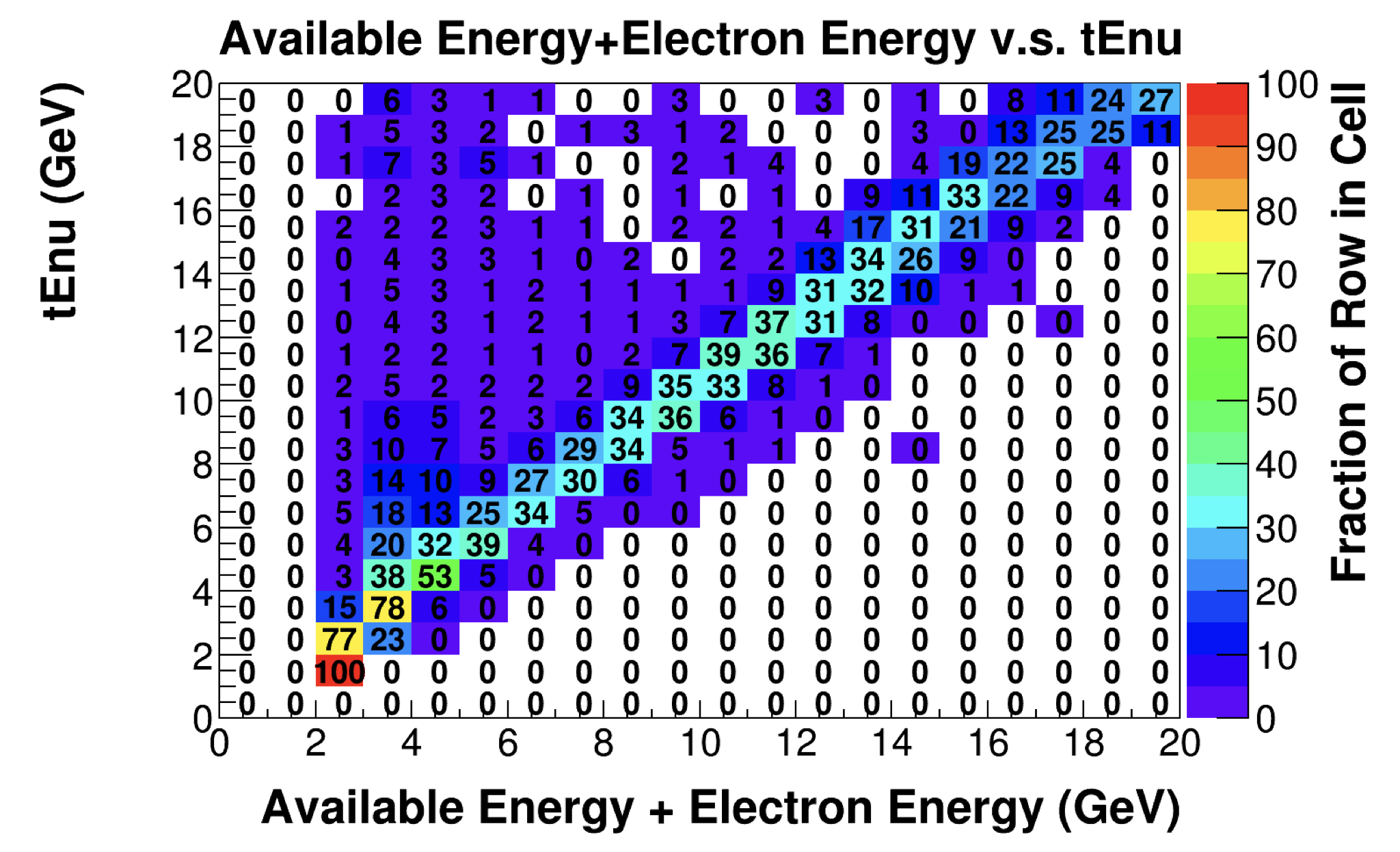}
    \caption{Neutrino energy estimator vs true neutrino energy in RHC.}
    \label{fig:NueEst}
\end{figure}

In this procedure, a corrected FHC sample is used to replace the MC prediction for the $\nu_e$ event rate in the RHC sample.  First a ratio of RHC/FHC $\nu_e$ events distributed in true neutrino energy is formed, seen in Fig.~\ref{fig:RHCFHC} $\nu_e$, as a function of true neutrino energy. 
This ratio must be applied to correct the FHC $\nu_e$ events to make a prediction for them in RHC. 
 To apply the correction to the data, a neutrino energy estimator is developed out of the reconstructed available energy and the reconstructed electron energy, 
\begin{align}
    E_{est} = E_{e} + E_{avail}
\end{align}
The accuracy of the formed energy estimator to predict the true neutrino energy for the samples in this analysis is shown in Fig. \ref{fig:NueEst}. 
The energy estimator value is then used to correct events on an event-by-event based by the ratio of the fluxes in the two beams as shown in Fig.~\ref{fig:RHCFHC}.
%(\refcomComment{"fo"->of, added "the" and "as"})

\begin{figure}[tp]
    \centering  \includegraphics[trim={40 0 0 0},clip=true,width=0.46\textwidth]{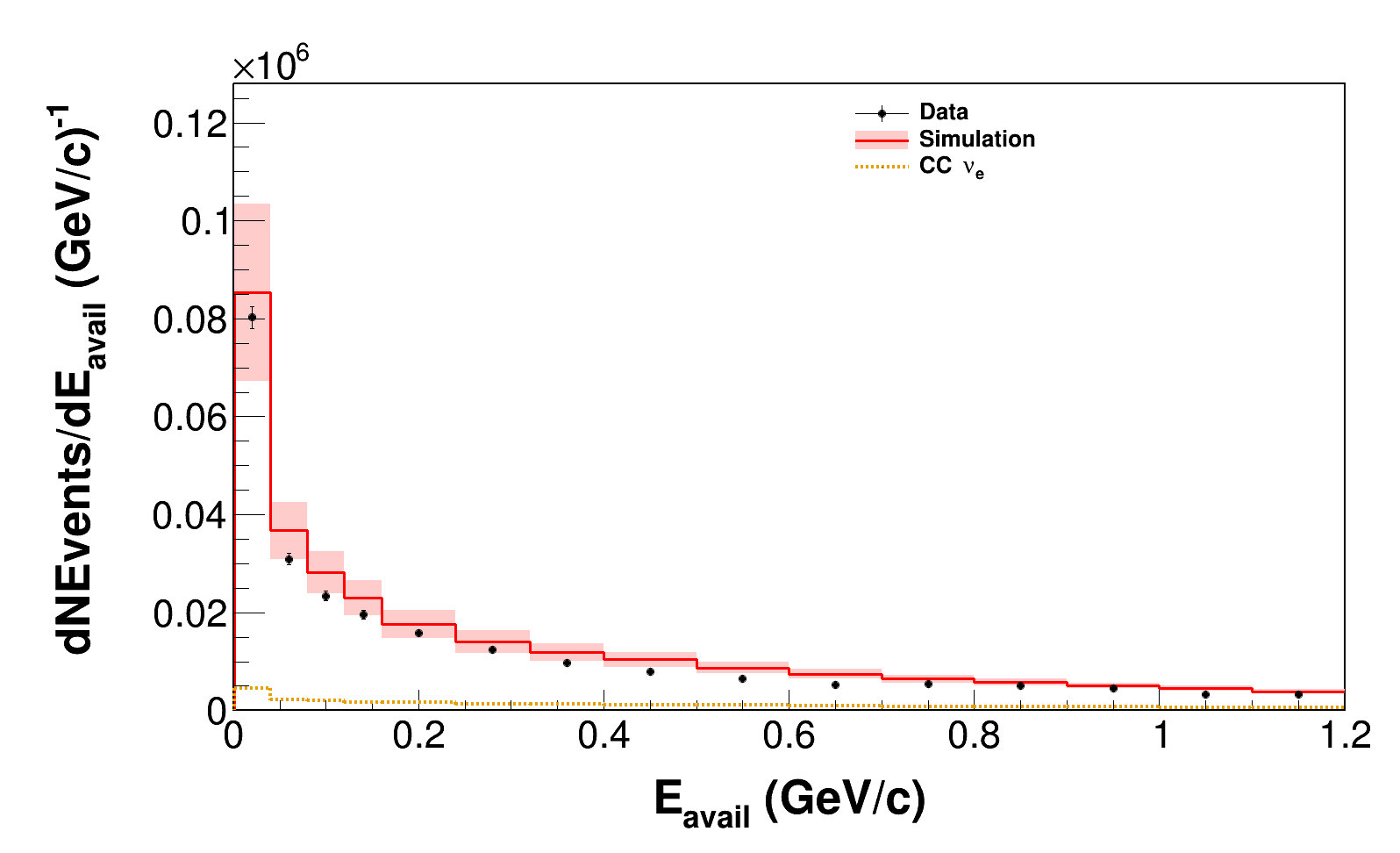}
    \includegraphics[width=0.54\textwidth]{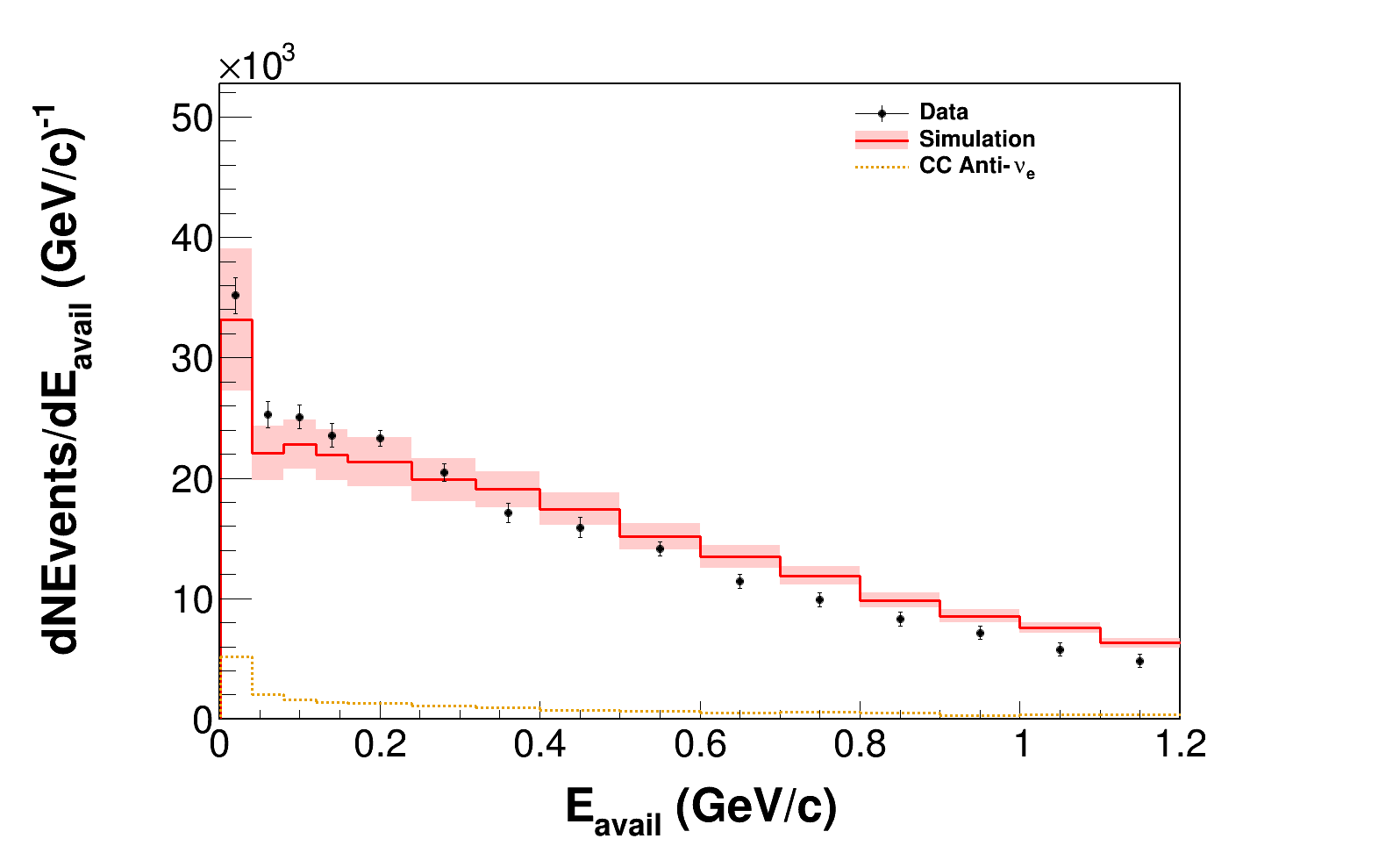}
    \caption{The top (bottom) plots show the scaled RHC (FHC) prediction for the electron antineutrino background to the FHC sample (electron neutrino background to the RHC sample) compared to the prediction for $p_T<1.6$~GeV/c.}
    \label{fig:wrongSignScaledPrediction}
\end{figure}

A complication is that in the data, there are background contributions to the RHC samples, as well as contributions from $\bar{\nu}_e$. The simulation is used to predict the initial $\bar{\nu}_e$ background to the RHC sample, and the other backgrounds are predicted as described above with the tunes to the control samples.  Each of these contributions is weighted on an event-by-event basis by the energy estimator from the reconstruction, whether the source is data or simulation.  After this weighting, the RHC $\nu_e$ prediction is formed by taking the corrected FHC data, and subtracting the corrected $\nu_e$ background prediction and the corrected other sources of background.  Because the flux correction is made event-by-event, these samples can be used to predict the background in the measured reconstucted variables.  This procedure is iterated once, replacing the initial MC prediction of $\bar{\nu}_e$ events with the data corrected version from this procedure.  The resulting background estimations, shown in Figure~\ref{fig:wrongSignScaledPrediction}, are less than a few percent in most bins, and largest in high $p_T$ bins with high $E_{avail}$ in $\bar{\nu}_e$ sample or low $E_{avail}$ in $\nu_e$ sample.  

\section{Measurement of Differential Cross Sections}
Calculation of the flux-integrated differential cross section per nucleon for kinematic variable $x$, in bins of i, is measured by the following equation:
\begin{equation}
  \label{eq:xsec_formula}
  (\frac{d\sigma}{dx})_i = \frac{\sum_jU_{ij}(N_j^{data}-N_j^{bkg})}{\epsilon_i T \Phi (\Delta x)_i} 
\end{equation}
where $(\frac{d\sigma}{dx})_i$ is the differential cross section as function of $x$ at bin $i$, $U_{ij}$ is the unfolding matrix, $N_j^{data}$ is the measured number of events in bin $j$ of reconstructed variable $x$, $N_j^{bkg}$ is the predicted number of background events in bin $j$, $\epsilon_i$ is estimated acceptance at bin $i$, $T$ is number of nucleon targets, $\Phi$ is integrated neutrino flux (or integrated antineutrino flux), and $(\Delta x)_i$ is bin width normalization of bin $i$.

The double differential cross sections $d^2 \sigma/dE_{avail}dq_3$ and $d^2 \sigma/dE_{avail}dp_T$ are calculated using the selected number of events and subtracting the number of background events predicted by the simulation. An iterative unfolding approach using the D’Agostini\cite{D'Agostini:1994zf,DAgostini:2010xxxxx}  method as implemented in RooUnfold\cite{Adye:2011gm}  was used to correct the background-subtracted distributions for resolution effects. Several unfolding studies were carried out using randomly thrown pseudodata samples generated from alternate physics models. The number of unfolding iterations is determined through $\chi^2$ values calculated by comparing the unfolded pseudodata with the truth pseudodata. The alternate physics models used in the studies include modification to the 2p2h enhancement to include a reweight giving nn/pp or np pairs additonal strength and a modification of RPA suppression affecting the low $Q^2$ or high $Q^2$ regions. The result of the unfolding studies for both the $\bar{\nu}_e$ (measured in the RHC beam) and $\nu_e$ (measured in the FHC beam) analyses indicate that a different number of unfolding iterations are required for the two different distributions based on the different minimum $\chi^2$ values. It is decided that $E_{avail}$ vs $p_T$ will be unfolded with 10 iterations and $E_\text{avail}$ vs $q_3$ unfolded with 15 iterations.

\begin{figure*}[p]
    \centering
    \includegraphics[width=0.85\textwidth]{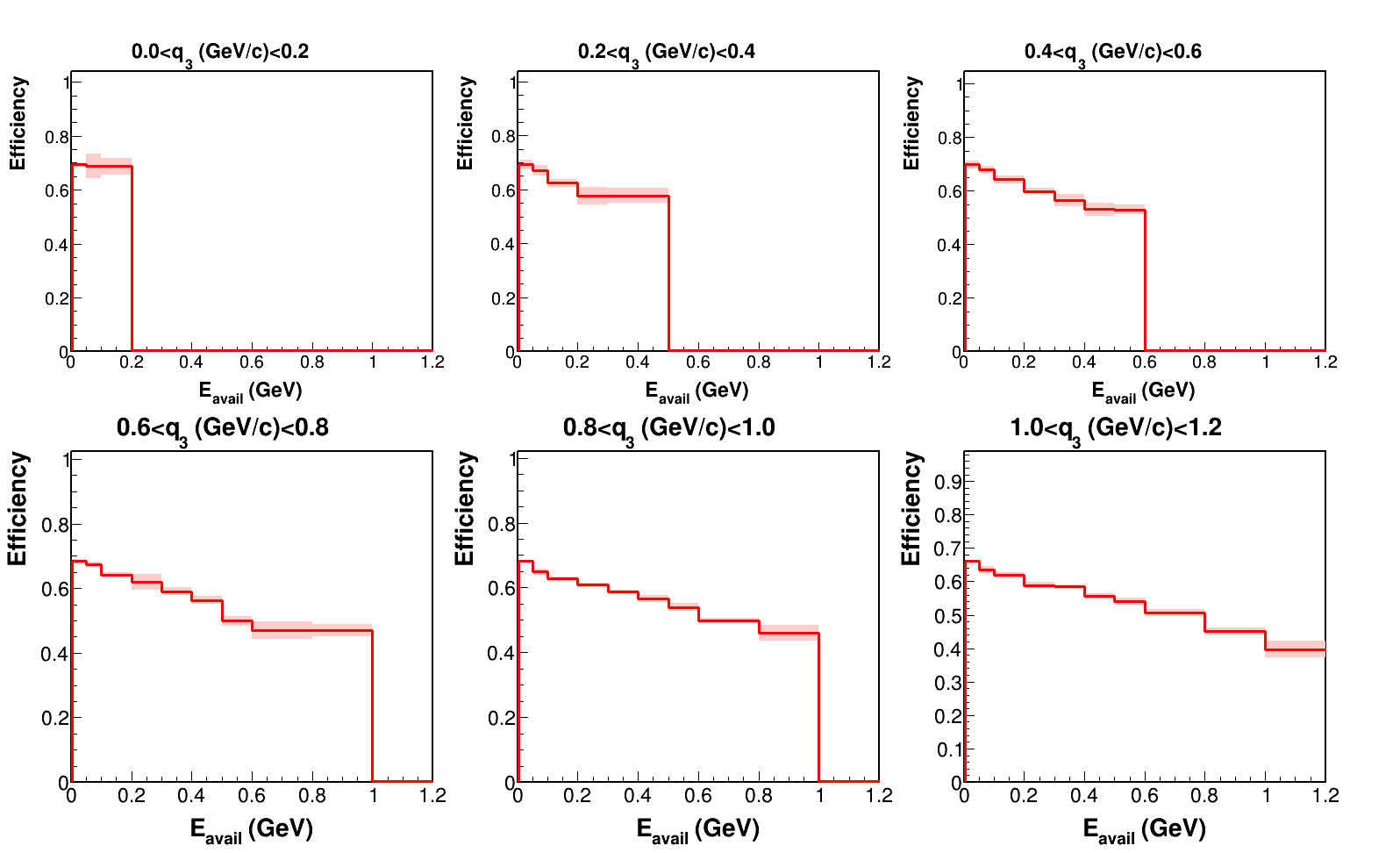}
    \caption{$\bar{\nu}_e$ efficiency for $E_{avail}$ vs $q_3$ distribution.}
    \label{fig:rhc_eff_q3}
\end{figure*}

\begin{figure*}[p]
    \centering
    \includegraphics[width=0.8\textwidth]{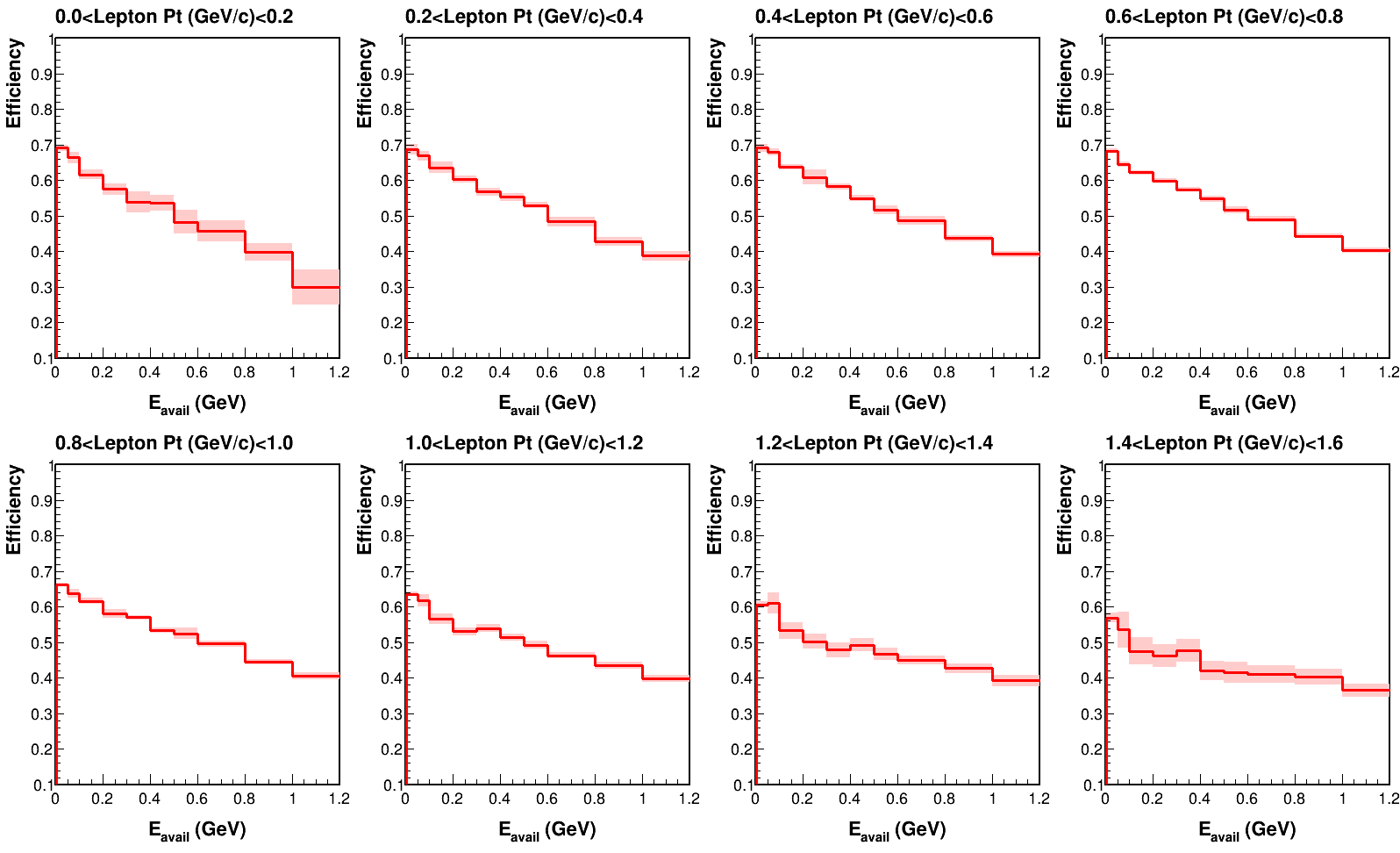}
    \caption{$\bar{\nu}_e$ efficiency for $E_{avail}$ vs $p_T$ distribution.}
    \label{fig:rhc_eff_pt}
\end{figure*}

\begin{figure*}[p]
    \centering
    \includegraphics[width=0.85\textwidth]{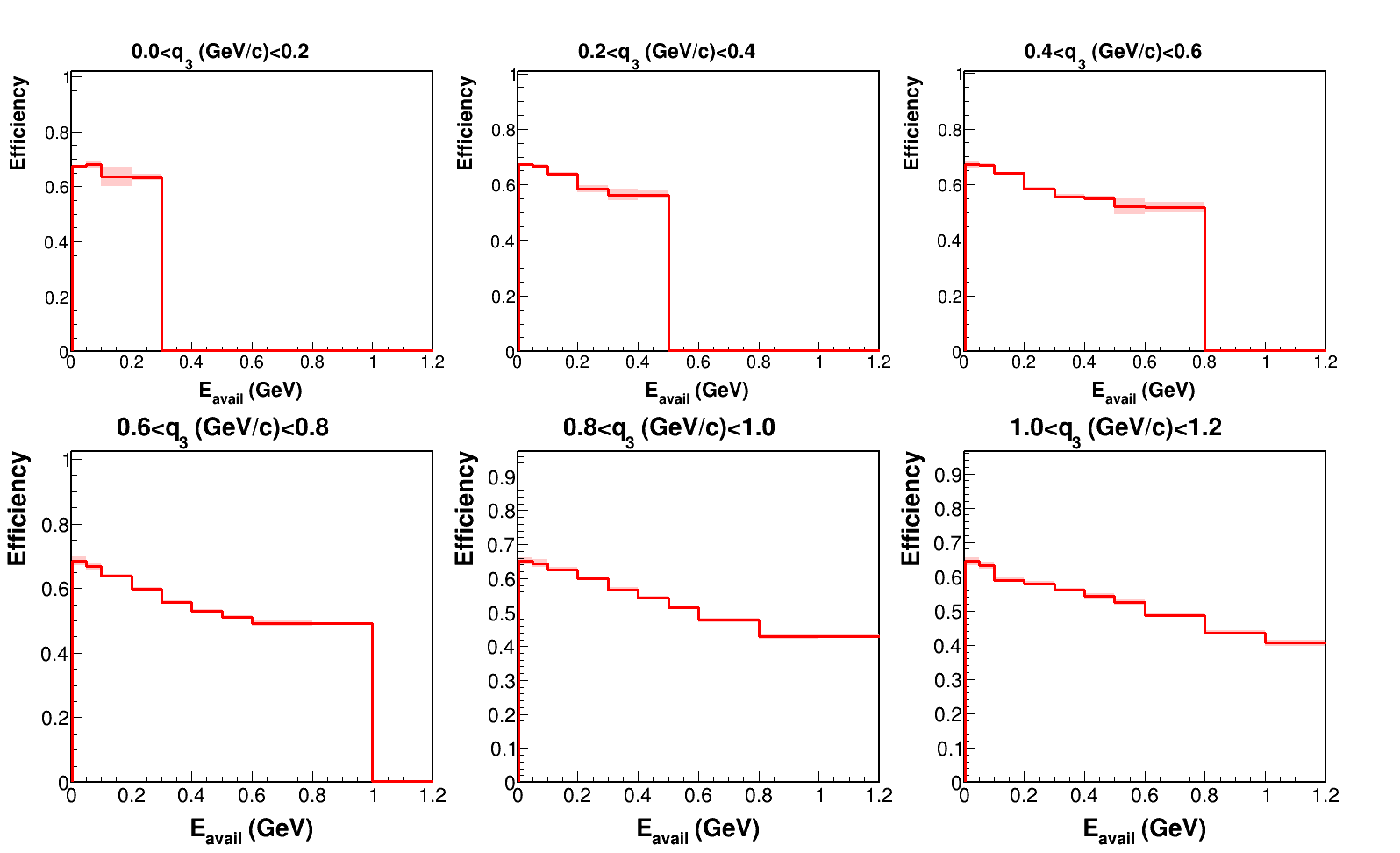}
    \caption{$\nu_e$ efficiency for $E_{avail}$ vs $q_3$ distribution.}
    \label{fig:fhc_eff_q3}
\end{figure*}

\begin{figure*}[p]
    \centering
    \includegraphics[width=0.8\textwidth]{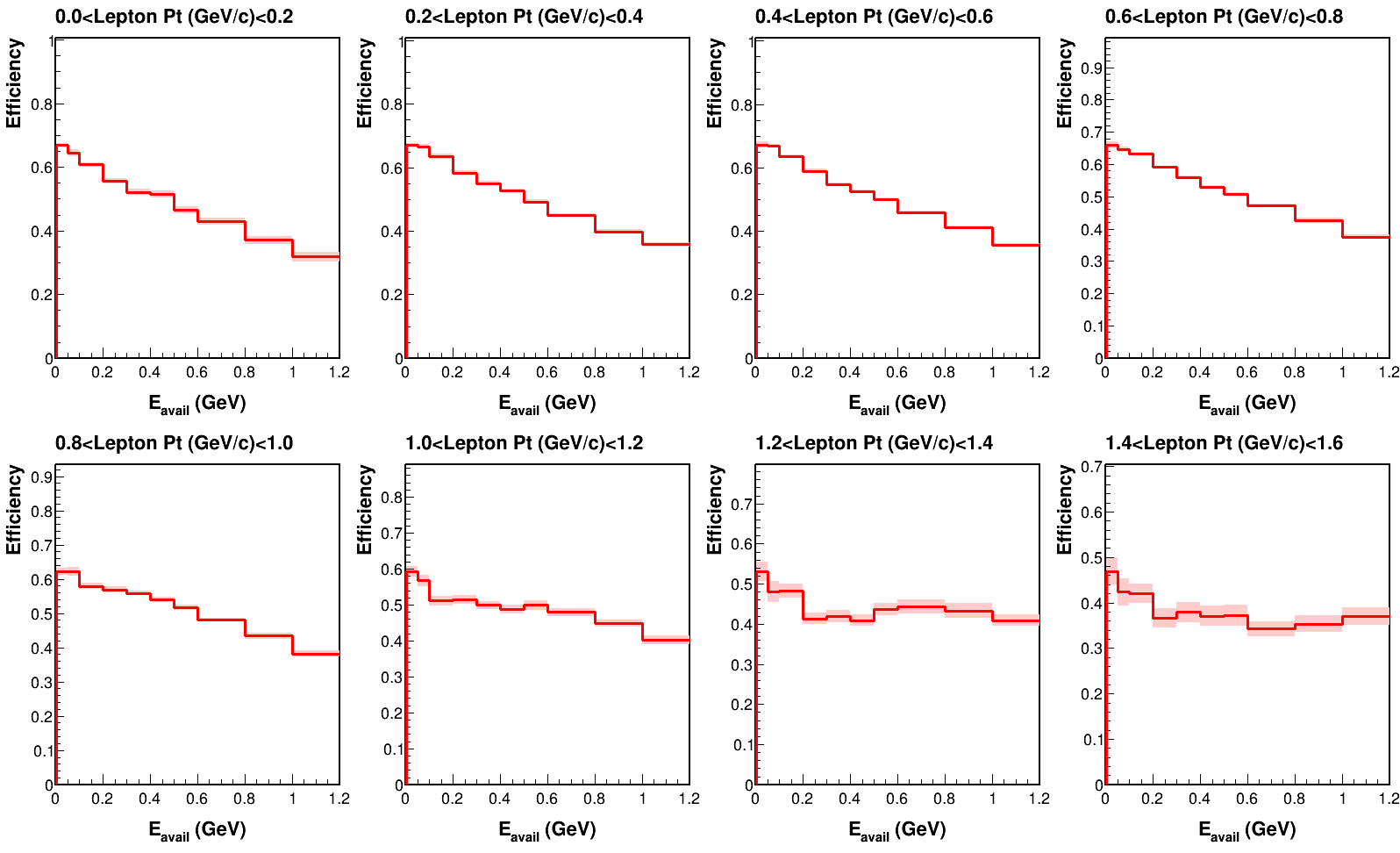}
    \caption{$\nu_e$ efficiency for $E_{avail}$ vs $p_T$ distribution.}
    \label{fig:fhc_eff_pt}
\end{figure*}

The number of events after unfolding is then divided by the efficiency. The $\bar{\nu}_e$ efficiency is found in Figs. \ref{fig:rhc_eff_q3} and \ref{fig:rhc_eff_pt} for the $E_{avail}$ vs $q_3$ and $E_{avail}$ vs $p_T$ distributions respectively.  In both cases, the efficiency decreases at higher $E_{avail}$ values. This is most likely because it is more difficult for the tracking algorithm to reconstruct a proper electron candidate track at higher $E_{avail}$ due to the greater amount of hadronic activity overlapping with EM showers. The equivalent $\nu_e$ efficiencies are found in Figs. \ref{fig:fhc_eff_q3} and \ref{fig:fhc_eff_pt}.  The inefficiency for high $E_{avail}$ events is due to the overlapping of EM showers and hadronic activity.  A few bins near the limit of $E_{avail}$ for a given $q_3$ contain very low statistics, making the evaluation of the efficiency difficult.  The efficiency in these low statistics bins are estimated by the average of adjacent bins because the efficiency is shown in nearby bins to be slowly varying.  These bins also have very high statistical uncertainties in the final cross section results. (
%\refcomComment{Moved and expanded the description of the edge of phase space efficiency calculations.})

\begin{figure*}[p]
    \centering
    \includegraphics[width=0.8\textwidth]{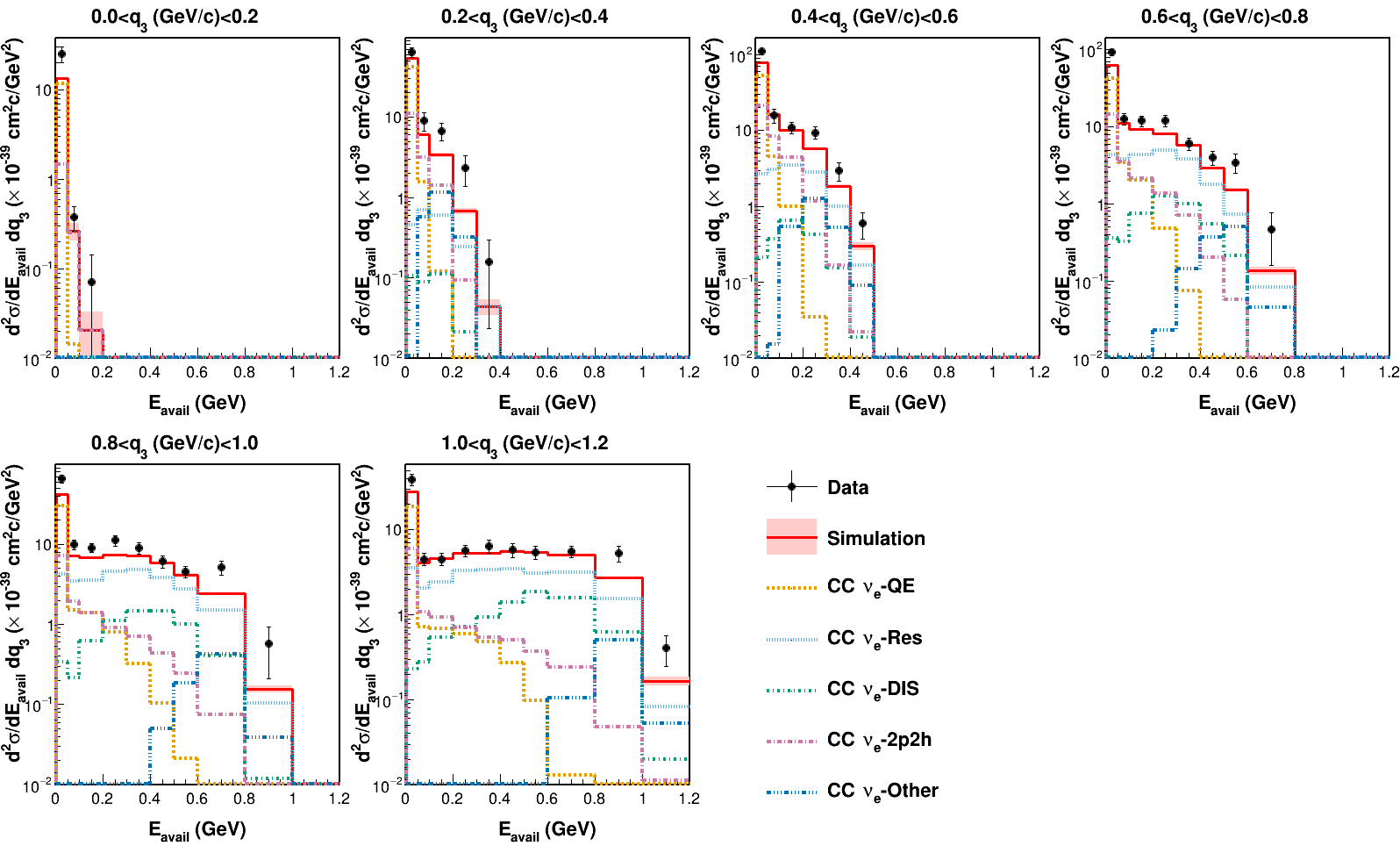}
    \caption{A decomposition of the $\bar{\nu}_e$ cross section result into contributing interaction types in $E_{avail}$ vs $q_3$ on a y log scale. The y axis is on a log scale truncated at $10^{-2}$ to enable a better view of the tail end of the cross section.}
    \label{fig:RHCQ3SigdepLogXSec}
\end{figure*}

\begin{figure*}[p]
    \centering
    \includegraphics[width=0.8\textwidth]{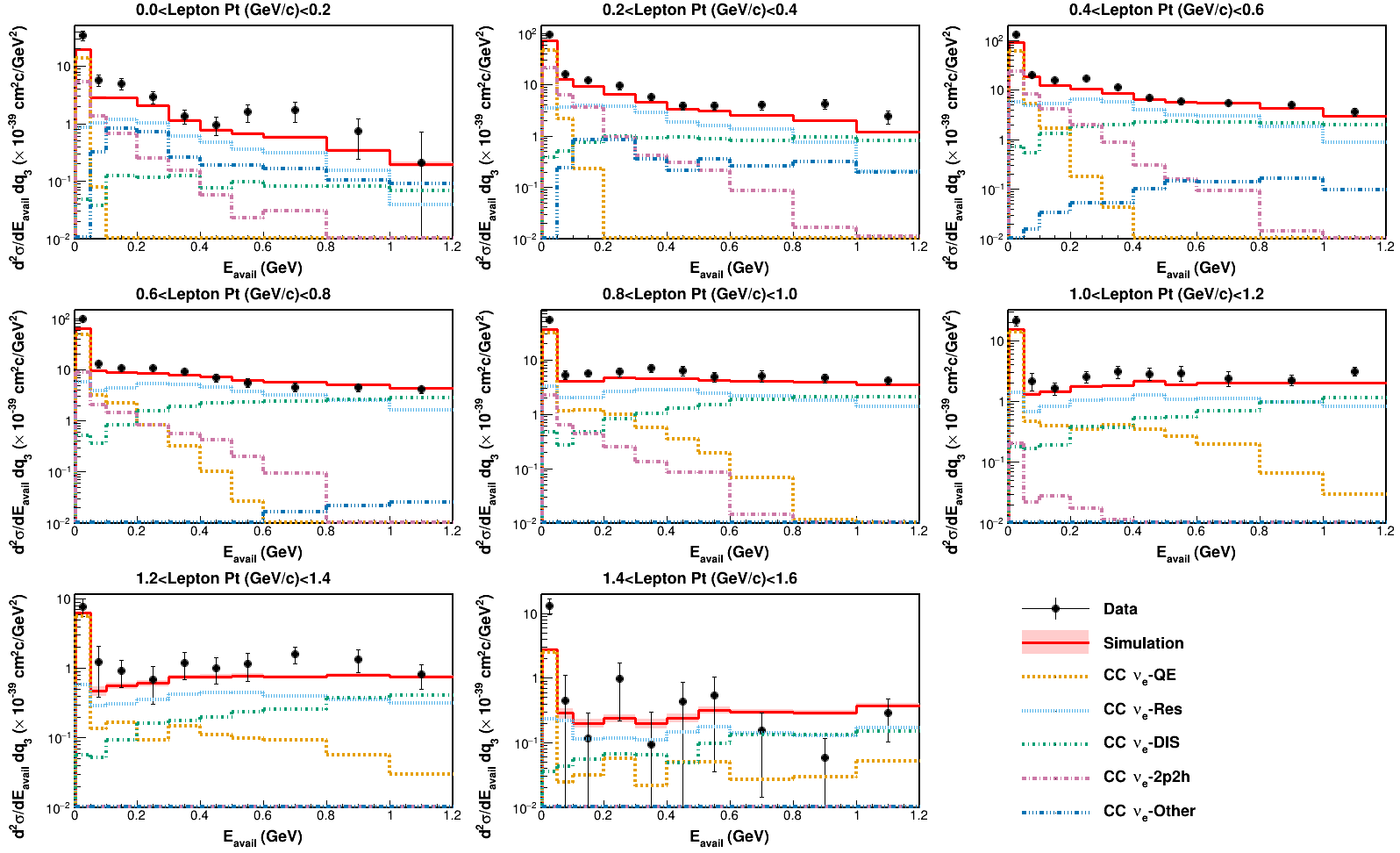}
    \caption{A decomposition of the $\bar{\nu}_e$ cross section result into contributing interaction types in $E_{avail}$ vs $p_T$ on a y log scale. The y axis is on a log scale truncated at $10^{-2}$ to enable a better view of the tail end of the cross section.}
    \label{fig:RHCPtSigdepLogXSec}
\end{figure*}

\begin{figure*}[p]
    \centering
    \includegraphics[width=0.8\textwidth]{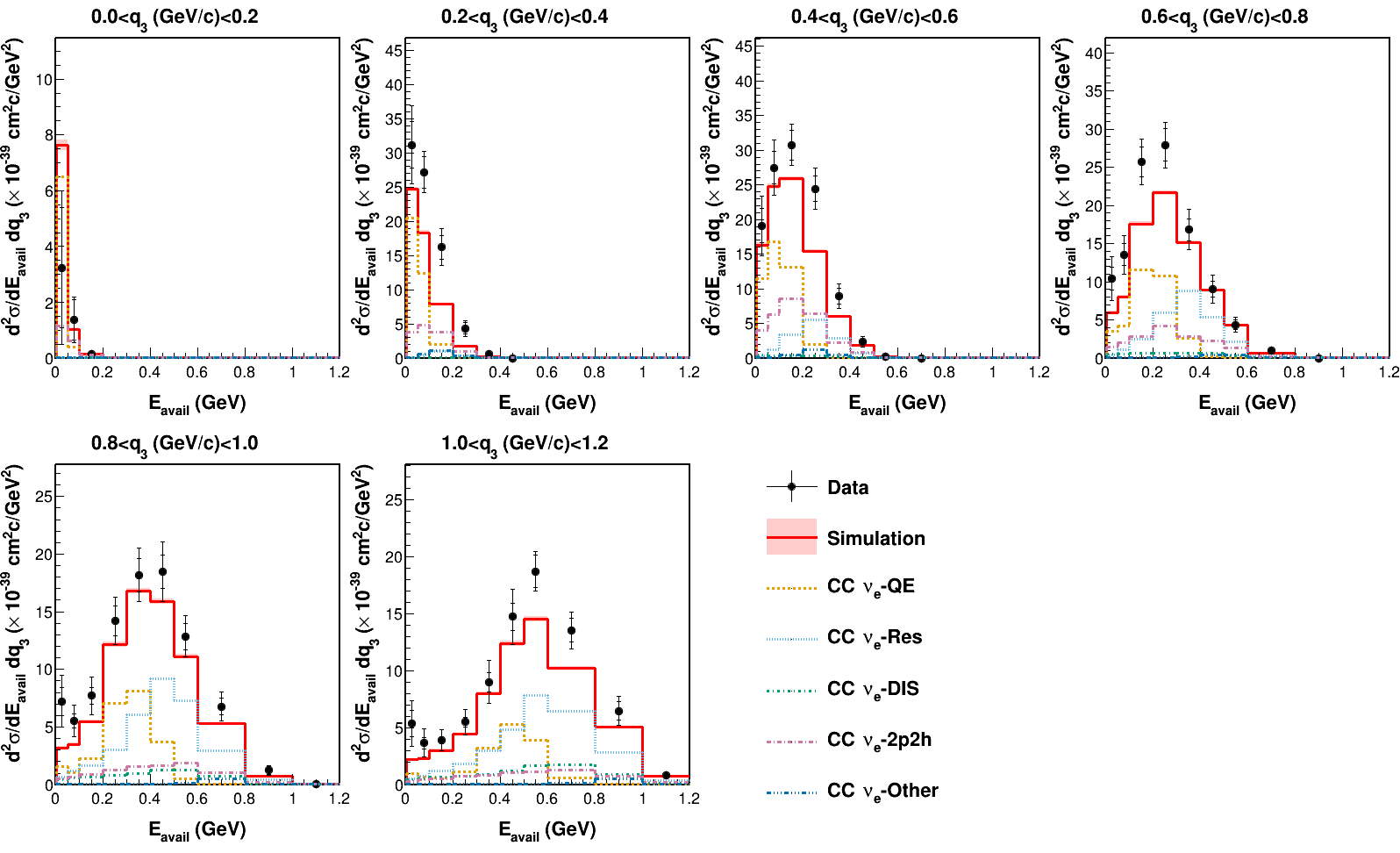}
    \caption{A decomposition of the $nu_e$ cross section result into contributing interaction types in $E_{avail}$ vs $q_3$.}
    \label{fig:FHCQ3SigdepXSec}
\end{figure*}

\begin{figure*}[p]
    \centering
    \includegraphics[width=0.8\textwidth]{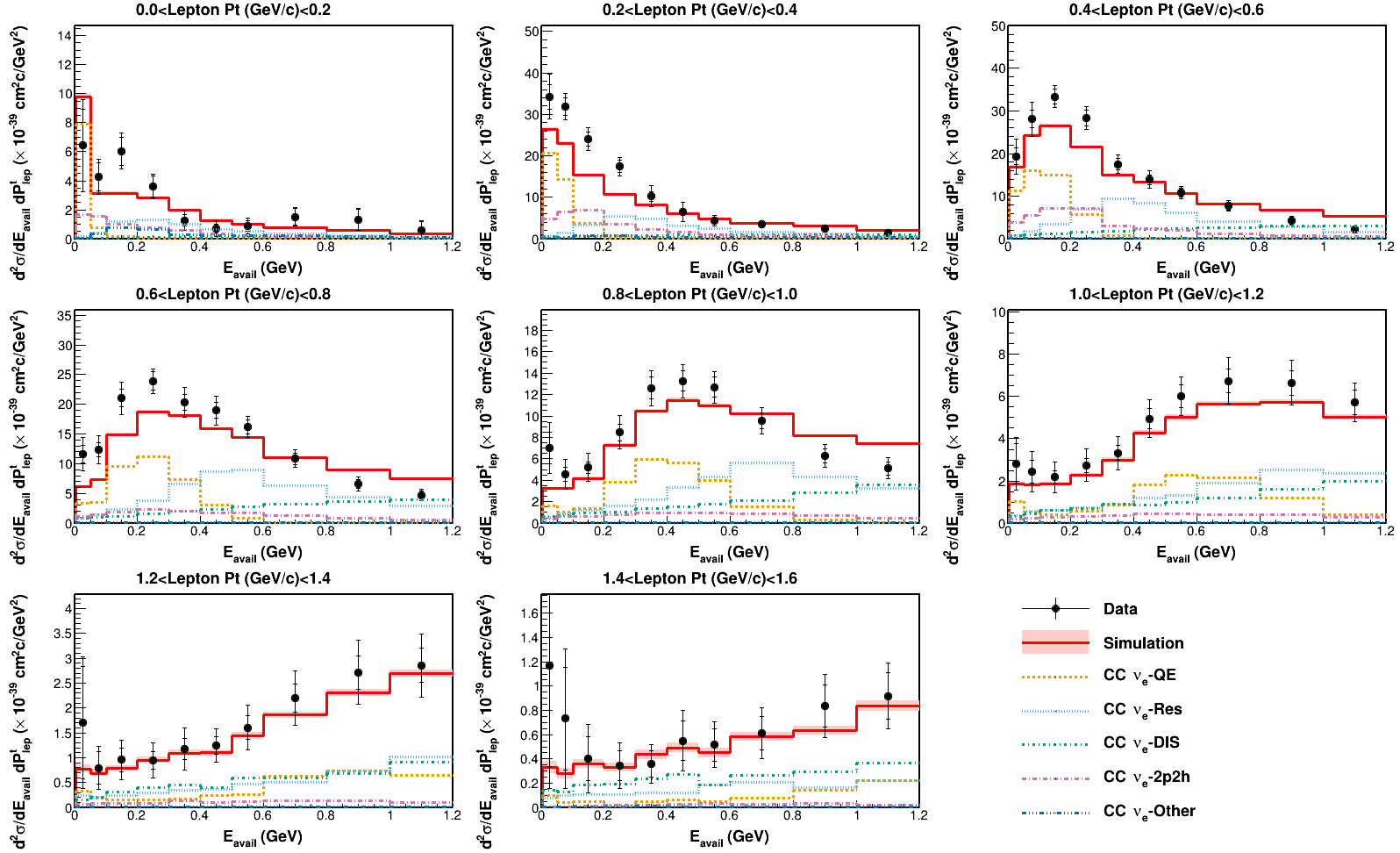}
    \caption{A decomposition of the $\nu_e$ cross section result into contributing interaction types in $E_{avail}$ vs $p_T$.}
    \label{fig:FHCPtSigdepXSec}
\end{figure*}

The normalization factors include 3.234$\times 10^{30}$ nucleon targets and the flux integral from 0 GeV to 100 GeV for a total integrated flux value of $\Phi=2.34 \times 10^{12}$ $\bar{\nu}/cm^{2}$ for the antineutrino analysis and $\Phi=6.7\pm0.2 \times 10^{11}$ $\nu/cm^2$ for the neutrino analysis. The double differential cross sections $d^2 \sigma/dE_{avail}dq_3$ and $d^2 \sigma/dE_{avail}dp_T$ are found in Figs. \ref{fig:RHCQ3SigdepLogXSec} and \ref{fig:RHCPtSigdepLogXSec} for the $\bar{\nu}_e$ analysis and Figs. \ref{fig:FHCQ3SigdepXSec} and \ref{fig:FHCPtSigdepXSec} for the $\nu_e$ analysis.

\FloatBarrier

\subsection{Systematic uncertainties}

\begin{figure*}[p]
    \centering
    \includegraphics[width=0.8\textwidth]{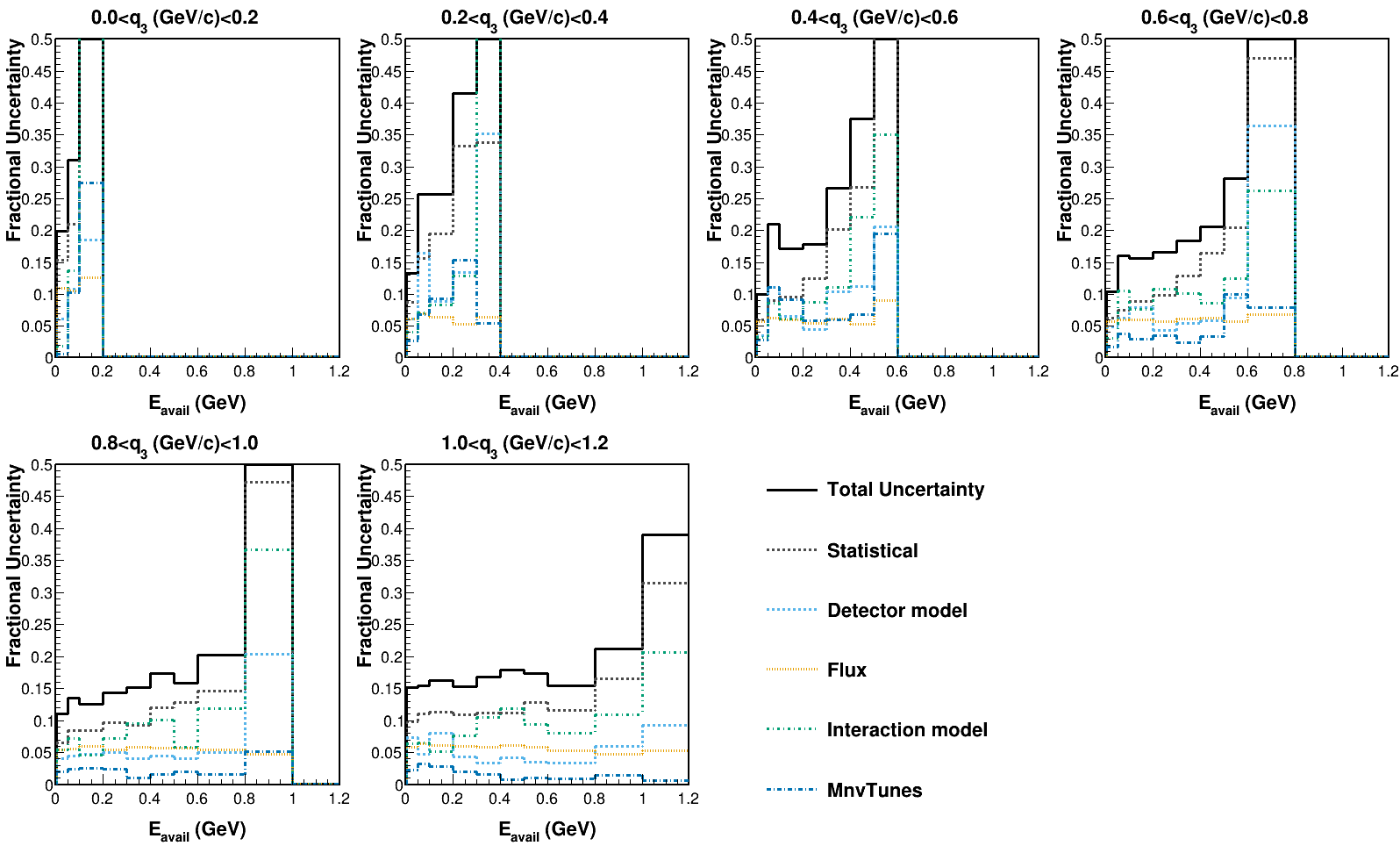}
    \caption{$\bar{\nu}_e$: Total cross section error summary broken down into four major subgroups for $E_{avail}$ vs $q_3$}
    \label{fig:rhc_xsec_q3_tot_error}
\end{figure*}

\begin{figure*}[p]
    \centering
    \includegraphics[width=0.8\textwidth]{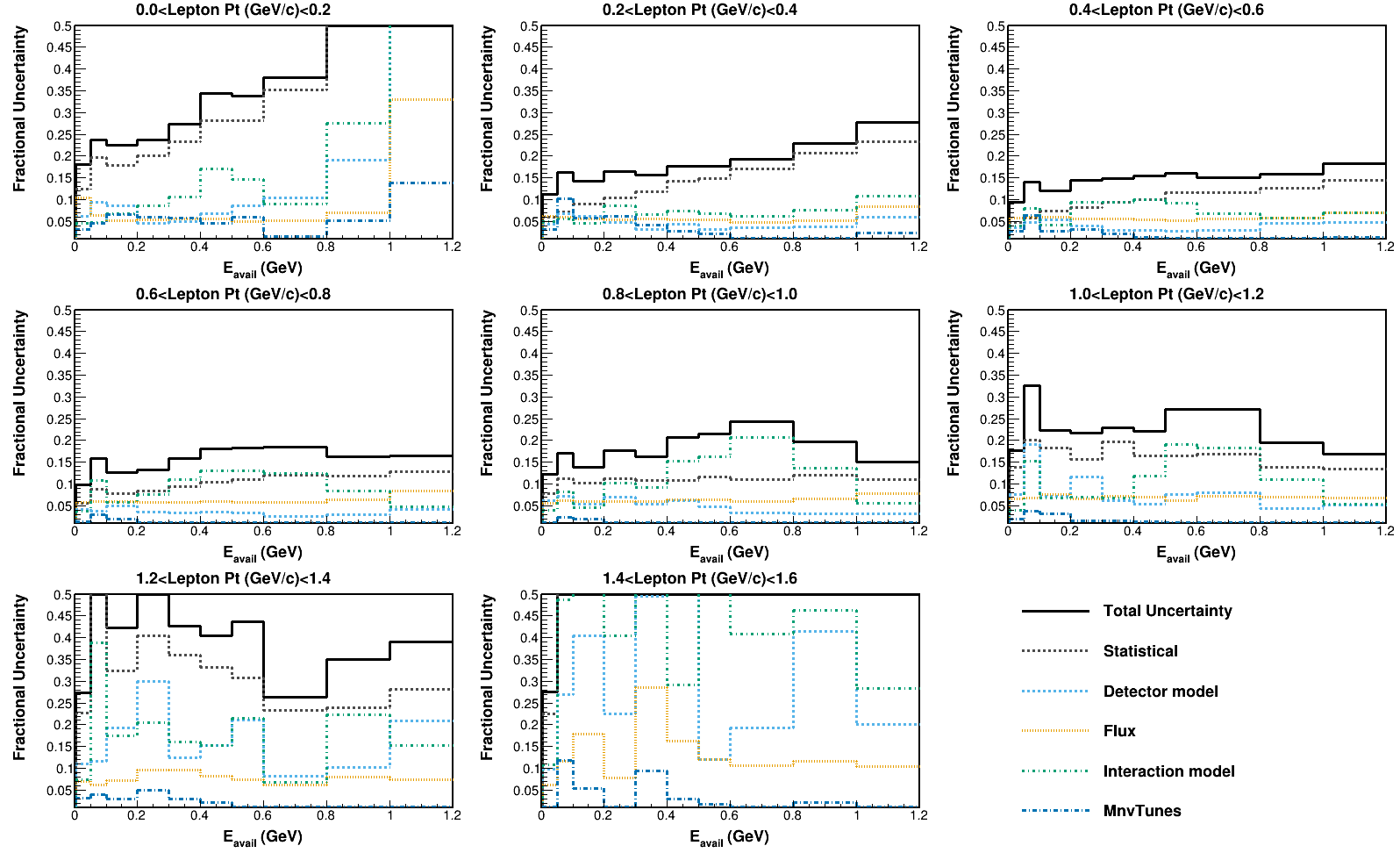}
    \caption{$\bar{\nu}_e$: Total cross section error summary broken down into four major subgroups for $E_{avail}$ vs $p_T$}
    \label{fig:rhc_xsec_pt_tot_error}
\end{figure*}

\begin{figure*}[p]
    \centering
    \includegraphics[width=0.8\textwidth]{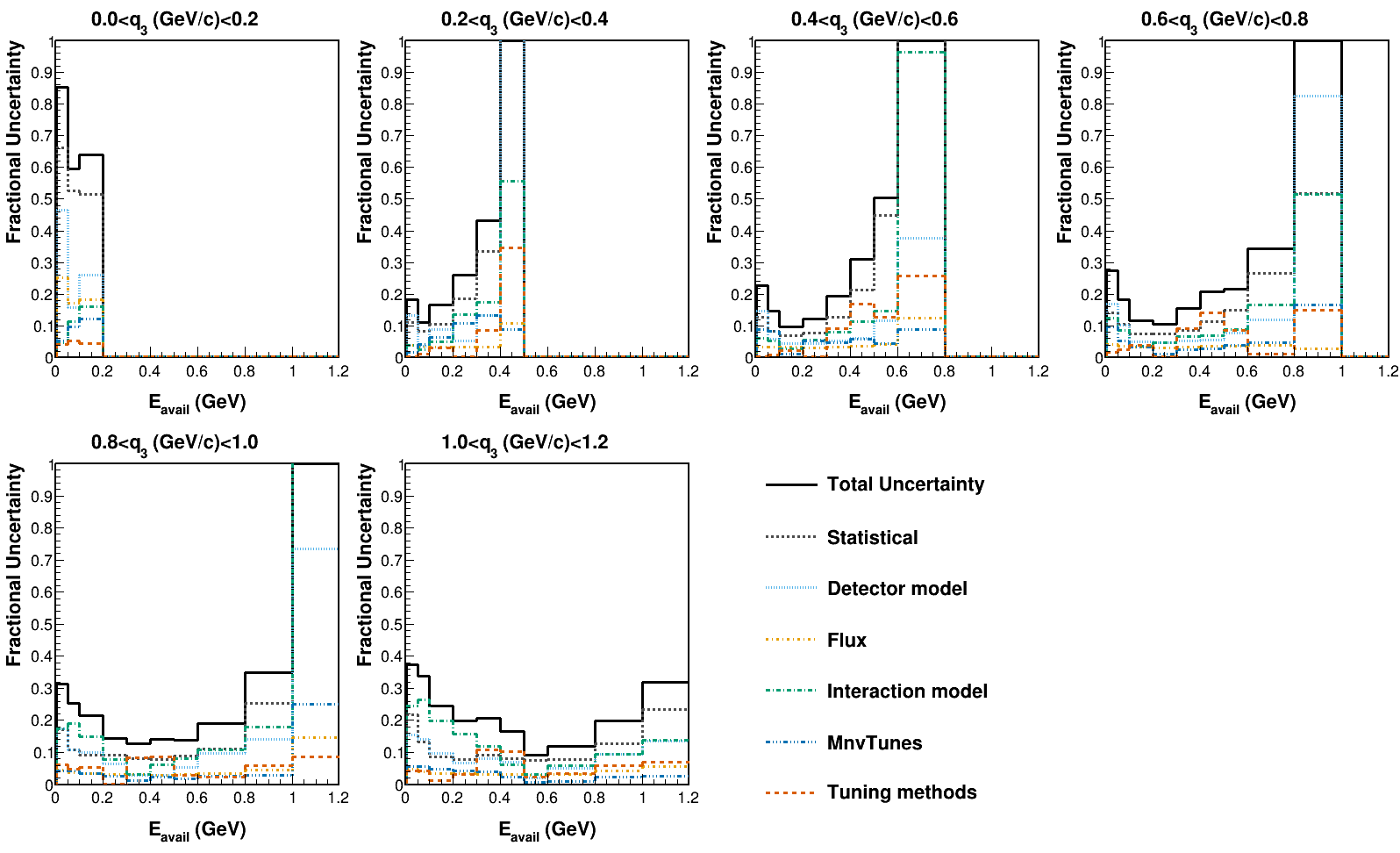}
    \caption{$\nu_e$: Total cross section error summary broken down into four major subgroups for $E_{avail}$ vs $q_3$}
    \label{fig:fhc_xsec_q3_tot_error}
\end{figure*}

\begin{figure*}[p]
    \centering
    \includegraphics[width=0.8\textwidth]{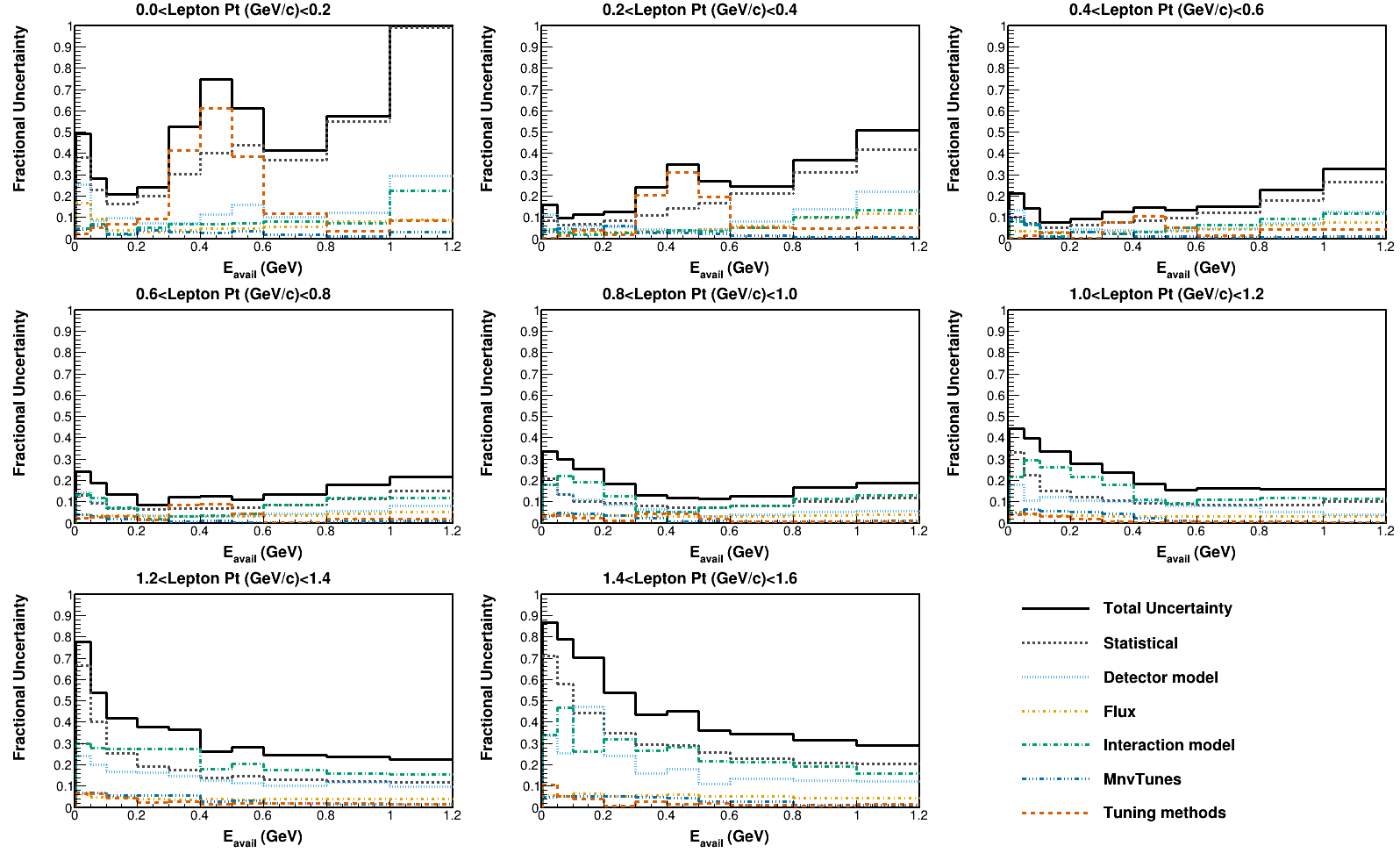}
    \caption{$\nu_e$: Total cross section error summary broken down into four major subgroups for $E_{avail}$ vs $p_T$}
    \label{fig:fhc_xsec_pt_tot_error}
\end{figure*}

Statistical uncertainties dominate systematic uncertainties in nearly every bin of these measurements.  Uncertainties on the measured cross sections can be categorized into four major groups: flux, detector model, interaction model, and MINERvA tunes. The breakdown of the $\bar{\nu}_e$ fractional systematic uncertainty for $d^2 \sigma/dE_{avail}dq_3$ is shown in Fig. \ref{fig:rhc_xsec_q3_tot_error} and $d^2 \sigma/dE_{avail}dp_T$ in Fig. \ref{fig:rhc_xsec_pt_tot_error}. The equivalent plots for $\nu_e$ are shown in Figs. \ref{fig:fhc_xsec_q3_tot_error} and \ref{fig:fhc_xsec_pt_tot_error} respectively. 

The uncertainties related to the flux can be broken down into two major categories: focusing uncertainties associated with all components related to the NuMI beam and hadron production uncertainties related to the uncertainty of hadron production from the proton beam incident on the graphite target. The flux uncertainty is fairly constant with $E_{avail}$ and $q_3$/$p_T$, around 4.7\%.

The detector model uncertainties consist of the uncertainties pertaining to the simulation of particle propagation through the detector, particle and kinematic reconstruction and the particle response of the detector. The detector model uncertainty can be broken into two groups: hadronic energy and electron reconstruction. A systematic uncertainty is assessed on the correction for leakage of electron energy outside of the electron cone. The energy leakage outside the cone leads to an overestimation of the available energy. The energy leakage was estimated to be 0.8\% of the electron energy. We estimate the energy leakage by simulating electron initiated showers with various energies and angles. 
By comparing this simulation to our sample of neutrino-electron elastic scattering ($\nu + e \rightarrow \nu + e$) events, we conclude that the simulation underestimates the energy leakage by $5\pm 2$~MeV, and the $2$~MeV uncertainty from this study is the assigned systematic uncertainty.  The leading uncertainty in $q_3$ bins with the highest $E_{avail}$ is the leakage uncertainty. The highest $p_T$ bin shows large systematic error values, similar to the GENIE error summary, due to the low number of events in that bin. The leakage uncertainty is the leading systematic uncertainty for the lower $p_T$ bins.

The interaction model uncertainties encompass GENIE interaction model uncertainties as well as GENIE final-state interaction uncertainties. For the $\bar{\nu}_e$ analysis, the leading systematic uncertainty for most bins is the axial mass $M_A$ resonance production  (MaRES) which adjusts the $M_A$ in the Rein-Sehgal cross section, affecting the shape and normalization. This next leading systematic is the $M_V$ resonance production (MvRES), which adjusts the axial vector mass $M_V$ in the Rein-Sehgal cross section, and the charged current resonance normalizaion (NormCCRES) that implements changes the normalization of CC Rein-Sehgal cross section. As shown in Figs. \ref{fig:RHCQ3SigdepLogXSec} and \ref{fig:RHCPtSigdepLogXSec}, the CC resonant pion production has a large contribution to the cross section measurement. Since the GENIE MaRES and MvRES parameters control resonant pion production it is not surprising that they are among the leading contributors to the uncertainty. For the $\bar{\nu}_e$ analysis, the probability for elastic scattering of nucleons while conserving the total rescattering probability (FrInelas\_N), contributes to the first bin of $q_3$ and highest $E_{avail}$ bin with less contribution for higher $q_3$ bins. Most likely this would include a neutron losing a large amount of energy in a collision, resulting with a proton in the final-state.

The systematic error breakdown for the MnvTunes shows the low $Q^2$ tune affects the highest bin of $E_{avail}$ for a given $q_3$ and $p_T$. This is expected because these are the regions in which the process is most dominant. The low-recoil 2p2h tune has a larger systematic uncertainty for values of $q_3$ compared to $p_T$. The shifting of the 2p2h model impacts the $E_{avail}$ distribution and the effects are seen more easily in $q_3$ due to the model dependency.

%\FloatBarrier

\subsection{Discussion and interpretation of antineutrino and neutrino results}

The measurement shows a larger
antineutrino cross section than predicted in the first bin of $E_{avail}$
%(\refcomComment{Replaced: "The measurement shows more rate than predicted for the antineutrino analysis in the first bin of $E_{avail}$}")
 in the cross section both for  $E_{avail}$ vs $q_3$ in Fig. \ref{fig:RHCQ3SigdepLogXSec} and for $E_{avail}$ vs $p_T$ as seen in Fig. \ref{fig:RHCPtSigdepLogXSec}. The events predicted to populate the first bin of available energy tend to be events where the final-state is neutral, typically composed by neutrons,
%(\refcomComment{Replaced: "typically final-state neutrons}"),
and in the first bin of $E_{avail}$ $\sim$ 60-90\% of the model prediction consists of
%(\refcomComment{added: "consists of"})
charged current quasielastic events. The quasielastic events are expected to be the dominant contributor to the first $E_{avail}$ bin because, in the absence of final-state interactions, there is only a lepton and neutron in the final-state.

 Looking more closely at the $q_3$ bin of 0.4-0.6 GeV in Fig. \ref{fig:RHCQ3SigdepLogXSec}, there is a population of inelastic events that leak into the first bin of $E_{avail}$. It is possible that some type of inelastic events with mostly neutrons in the final-state is not being correctly simulated. It also could involve events where the final-state pion does not have much energy and is absorbed within the nucleus, resulting in only final-state neutrons. The last proposal to explain the high cross section in the first $E_{avail}$ bin is that the MC simulation predicts too many quasielastic events at higher values of $E_{avail}$. Increasing the population of quasielastic MC events near zero $E_{avail}$ would improve this prediction.

In contrast to the antineutrino results there is a deficit of data events over the simulated prediction for the neutrino analysis found in the first bin of $E_{avail}$ seen in the cross section plots for both $E_{avail}$ vs $q_3$ as shown in Fig. \ref{fig:FHCQ3SigdepXSec} and $E_{avail}$ vs $p_T$ (Fig. \ref{fig:FHCPtSigdepXSec}) in lowest respective bins.

\subsection{Comparison to muon neutrino and antineutrino measurements}

These results can be compared with MINERvA's measurement of the analogous samples from muon neutrinos and antineutrinos.  However, there are differences in the measurements that make a direct comparison challenging.  In particular, all 
%(\refcomComment{removed: "there"}) 
of the measurements of muon neutrino and antineutrino processes are made with neutrino spectra which are substantially different than the ones measured in these results.

The $\bar{\nu}_{e}$ cross section result would be most appropriately compared with MINERvA's low-recoil LE $\bar{\nu}_{\mu}$ result \cite{MINERvA:2018nab}.   In addition to the flux differences, there are also selection differences between the two analyses, so the signal definitions are not identical. The $\bar{\nu}_{\mu}$ result requires a lepton momentum of greater than 1.5 GeV, while the $\bar{\nu}_{e}$ analysis requires lepton energy greater than 2.5 GeV to eliminate a large $\pi^0$ background at low electron energy. In addition, the $\bar{\nu}_{e}$ analysis has no scattering angle requirements while the $\bar{\nu}_{\mu}$ analysis requires the lepton scattering angle to be less than 20 degrees due to the difficulty of reconstructing high angle muons. The analyses are also reported using different binning. Table \ref{tab:BinningEavail} shows a binning comparison between the ME and LE results for $E_{avail}$ and Table \ref{tab:BinningQ3} shows the comparison for $q_3$.

\begin{table}[bpt]
\small
\centering
\sffamily
 \begin{tabular}{|c | c |} 
 \hline 
  ME $\bar{\nu}_e$ & LE $\bar{\nu}_\mu$\\
 \hline 
 $0.0 < E_{avail} (GeV) < 0.04 $ & $0.0 < E_{avail} (GeV) < 0.03 $\\ [0.5ex]
 \hline
  $0.04 < E_{avail} (GeV) < 0.08 $ & $0.03 < E_{avail} (GeV) < 0.07 $ \\ [0.5ex]
 \hline
  $0.8 < E_{avail} (GeV) < 0.12 $ & $0.07 < E_{avail} (GeV) < 0.17 $ \\ [0.5ex]
 \hline
  $0.12 < E_{avail} (GeV) < 0.16 $ & $0.17 < E_{avail} (GeV) < 0.27 $ \\ [0.5ex]
 \hline
  $0.16 < E_{avail} (GeV) < 0.24 $ & $0.27 < E_{avail} (GeV) < 0.35 $ \\ [0.5ex]
 \hline
  $0.24 < E_{avail} (GeV) < 0.32 $ & $0.35 < E_{avail} (GeV) < 0.5 $ \\ [0.5ex]
 \hline
  $0.32 < E_{avail} (GeV) < 0.4 $ & \\ [0.5ex]
 \hline
  $0.4 < E_{avail} (GeV) < 0.5 $ & \\ [0.5ex]
 \hline
 \end{tabular}
\caption{Comparison between the $E_{avail}$ binning used for the low-recoil ME  $\bar{\nu}_{e}$ analysis (left) and the low-recoil LE $\bar{\nu}_{\mu}$ analysis (right). The $\bar{\nu}_{e}$ binning is truncated at 0.5 GeV for the comparison to LE but the results are reported up to 1.2 GeV.}
\label{tab:BinningEavail}
\end{table}

\begin{table}[bpt]
\small
\centering
\sffamily
 \begin{tabular}{|c | c |} 
 \hline 
  ME $\bar{\nu}_e$ & LE $\bar{\nu}_\mu$\\
 \hline 
 $0.0 < q_3 (GeV) < 0.2 $ & $0.0 < q_3 (GeV) < 0.2 $\\ [0.5ex]
 \hline
  $0.2 < q_3 (GeV) < 0.4 $ & $0.2 < q_3 (GeV) < 0.3 $ \\ [0.5ex]
 \hline
  $0.4 < q_3 (GeV) < 0.6 $ & $0.3 < q_3 (GeV) < 0.4 $ \\ [0.5ex]
 \hline
  $0.6 < q_3(GeV) < 0.8 $ & $0.4 < q_3 (GeV) < 0.5 $ \\ [0.5ex]
 \hline
  $0.8 < q_3 (GeV) < 1.0 $ & $0.5 < q_3 (GeV) < 0.6 $ \\ [0.5ex]
 \hline
  $1.0 < q_3 (GeV) < 1.2 $ & $0.6 < q_3 (GeV) < 0.8 $ \\ [0.5ex]
 \hline
 \end{tabular}
\caption{Comparison between the $q_3$ binning used for the low-recoil ME  $\bar{\nu}_{e}$ analysis (left) and the low-recoil LE $\bar{\nu}_{\mu}$ analysis (right).}
\label{tab:BinningQ3}
\end{table}

\begin{figure*}[p]
    \centering
    \includegraphics[width=0.8\textwidth]{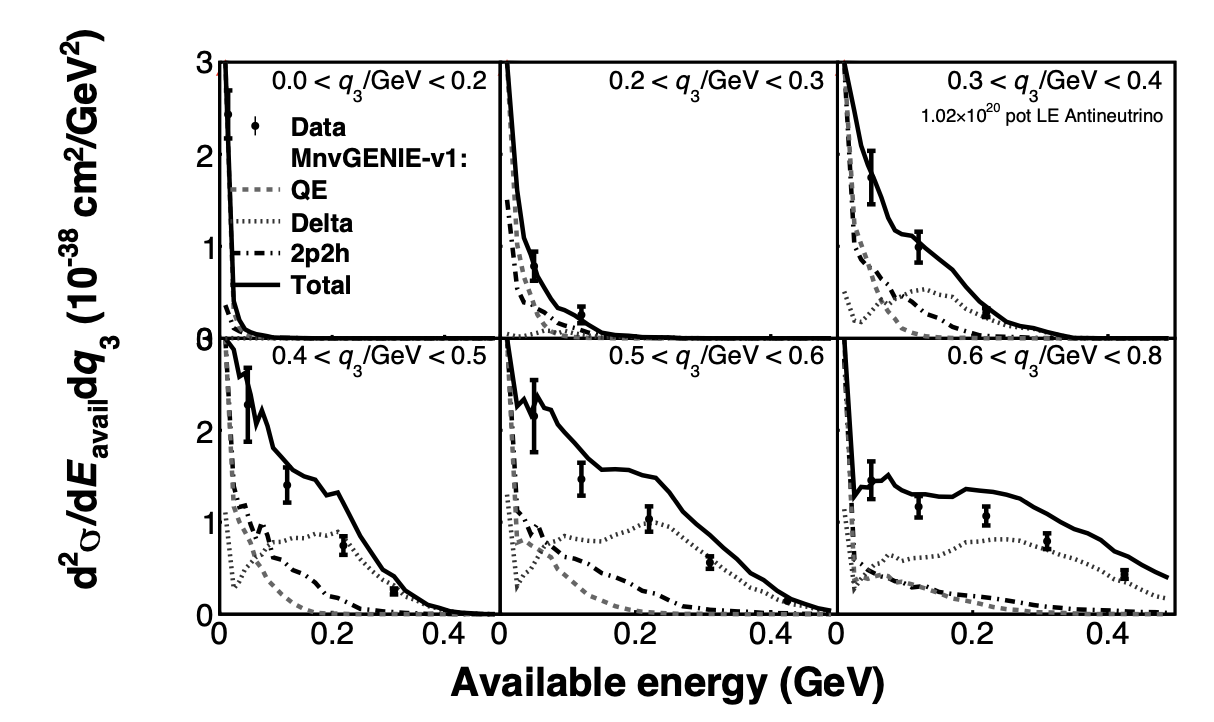}
    \caption{$d^2\sigma/dE_{avail}dq_3$ cross section per nucleon compared to the model with RPA and tune 2p2h components. Figure from Ref. \cite{MINERvA:2018nab}}
    \label{fig:LEAntiNuXSec}
\end{figure*}

\begin{figure*}[p]
    \centering
    \includegraphics[width=0.8\textwidth]{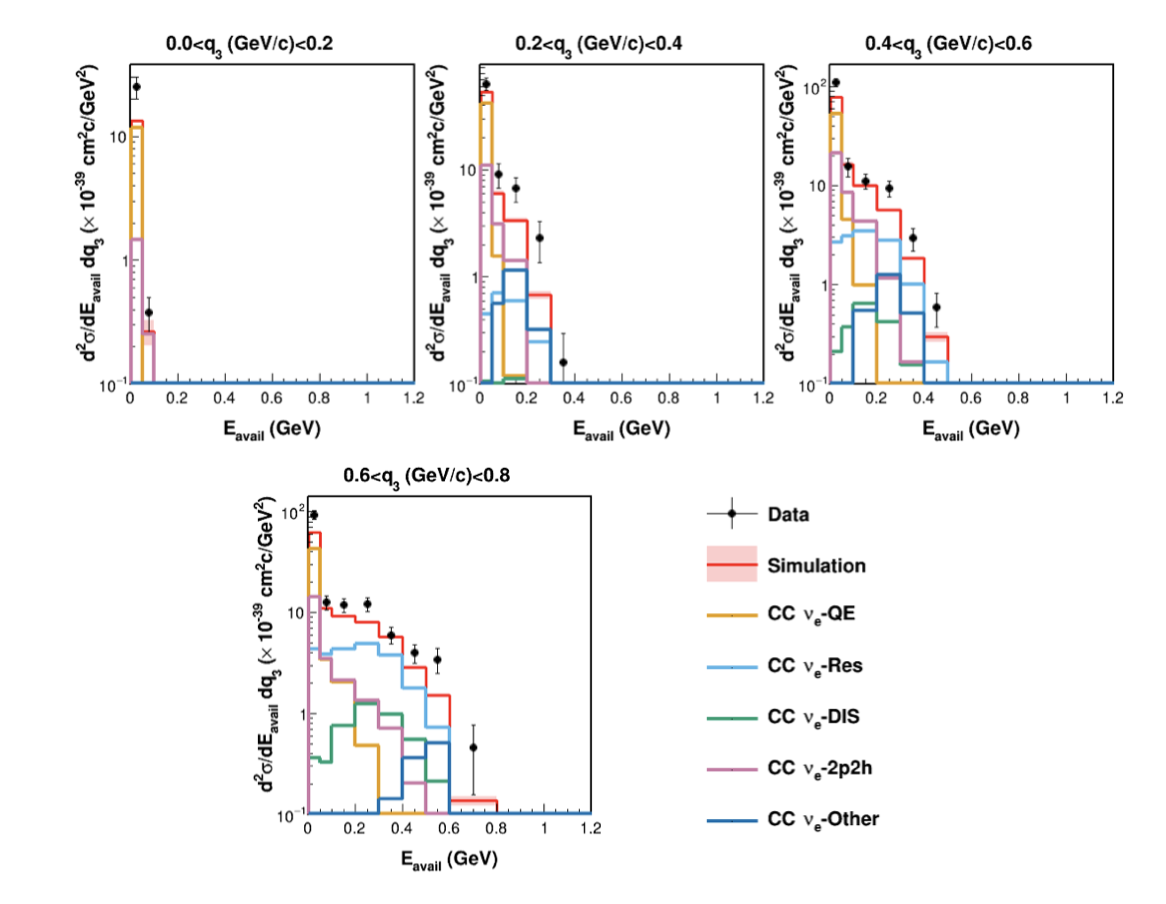}
    \caption{The $\bar{\nu}_{e}$ cross section result truncated at $0.8$ GeV on logy scale in $q_3$ for comparison to the LE result.}
    \label{fig:TruncatedSig}
\end{figure*}

Lastly, the two analyses took different approaches in unfolding. The $\bar{\nu}_{e}$ unfolds using coarse binning and a large number of iterations and the $\bar{\nu}_{\mu}$ analysis unfolds using fine binning and a small number of iterations. This is due in large part to the difference in observables. The ME $\bar{\nu}_{e}$ analysis has to account for the energy leakage outside the electron cone and into the available energy. Overall, the LE $\bar{\nu}_{\mu}$ has a much better energy resolution compared to the $\bar{\nu}_{e}$ analysis. With the consideration of the differences between the two analyses, the cross section result for the LE $\bar{\nu}_{\mu}$ is shown in Fig. \ref{fig:LEAntiNuXSec} and the relevant cross section bins for the $\bar{\nu}_{e}$ are shown in Fig. \ref{fig:TruncatedSig}.

\begin{figure*}[p]
    \centering
    \includegraphics[width=0.8\textwidth]{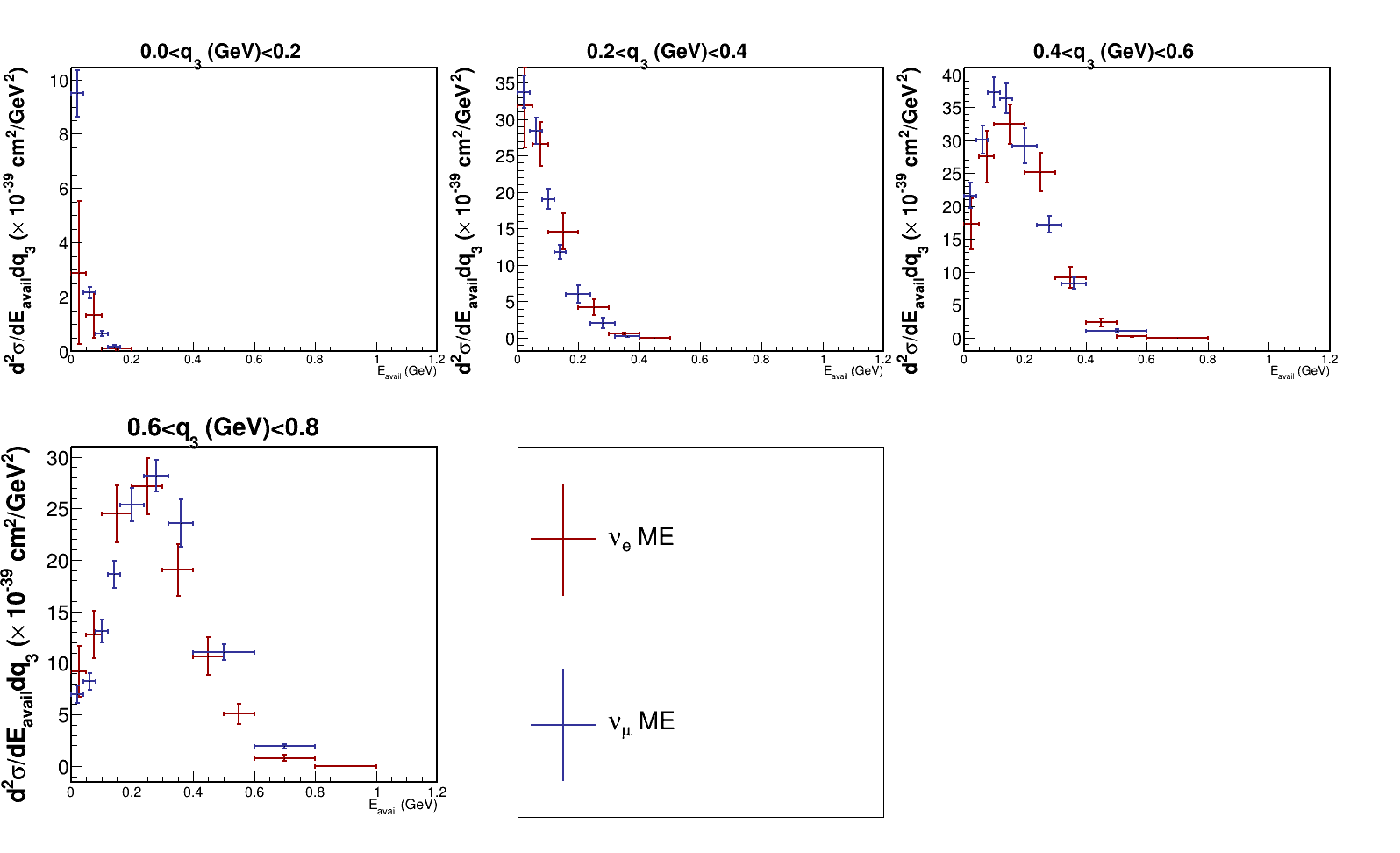}
    \caption{Comparison of published MINERvA $\nu_{\mu}$ ME  measurements with the present ME $\nu_{e}$ results.}
    \label{fig:nue2numuLE_ME}
\end{figure*}

There are similar features between the two results. As expected in the cross section model prediction, both cross section results have quasielastic events as the dominant contributor in the first bin of $E_{avail}$. There is a population of 2p2h events for values of low $E_{avail}$. The delta resonance becomes the dominant process at the higher values of $E_{avail}$ in both predicted cross sections. There is a noticeable difference between the data results for values of $\sim E_{avail} > 0.2$~GeV. The reference MC prediction for the $\bar{\nu}_{\mu}$ cross section consistently exceeds the data while the $\bar{\nu}_{e}$ prediction falls below the data.

Both cross section results contain many events that have no available energy. This creates a sharp peak at zero $E_{avail}$ followed by a cross section that falls slowly compared to the size of the peak in the first $E_{avail}$ bin. Therefore, to compare the first two $E_{avail}$ bins for both results we assume that the peak at zero $E_{avail}$ is a Kronecker delta function-like peak and the remaining cross section distribution is flat. To determine the magnitude of the delta function we subtract the second $E_{avail}$ bin from the first, or the flat distribution from the peak, leaving us with a $d\sigma/dE_{avail}$ value. We multiply $d\sigma/dE_{avail}$ by the bin width so that we end up with a cross section that is differential in each $q_3$ bin. This process is repeated for each result's data and MC values. The resultant bin combination for the two samples are 0.0 $< E_{avail}$(GeV) $<$ 0.08 for the  $\bar{\nu}_{e}$ cross section result and 0.0 $< E_{avail}$(GeV) $<$ 0.07 for the $\bar{\nu}_{\mu}$ cross section result. Tables \ref{table:SarahData} and \ref{table:LEData} summarize the results. Tables \ref{table:SarahCorr} and \ref{table:LECorr} are the correlation matrices for the reported bins with the correlation matrix ordering defined in Table \ref{table:CorrMapping}.

The conclusion drawn from the comparison between the data/MC peak at zero is that the ME $\bar{\nu}_e$ result is consistent with the LE $\bar{\nu}_\mu$ result in all $q_3$ bins except the first. The ME $\bar{\nu}_e$ result has a significant enhancement over the simulation. 

Similarly, we can in addition compare the $\nu_{e}$ cross section result with MINERvA's low-recoil ME $\nu_{\mu}$ result\cite{MINERvA:2021wjs}. 
As with the comparison above for the $\bar{\nu}_e$, there are significant differences between the $\nu_{e}$ and $\nu_{\mu}$ analyses including the flux and features of the reconstruction.  The cross section results of  ($\nu_{\mu}$ ME and $\nu_{e}$) are
compared in Fig. \ref{fig:nue2numuLE_ME}. We conclude the $\nu_{e}$ result is qualitatively consistent with the $\nu_{\mu}$ results, except in the lowest $q_{3}$ and $E_{avail}$ bin where there is some indication of a difference. In this bin the measured $\nu_{e}$ cross section is  smaller than the $\nu_{\mu}$ cross section, albeit with large uncertainties.

\begin{table*}[bpt]
\small
\centering
\sffamily
\begin{tabular}{|c|c|c|}
\hline
 & 0.0 $< q_3 (GeV) <$ 0.2  &  0.2 $< q_3 (GeV) <$ 0.4 \\
 \hline
 Estimated peak at zero & 1 & 3 \\
 Subtraction size from first bin & 2 & 4 \\
 \hline \hline
 & 0.4 $< q_3 (GeV) <$ 0.6  &  0.6 $< q_3 (GeV) <$ 0.8 \\
 \hline
 Estimated peak at zero & 5 & 7 \\
 Subtraction size from first bin & 6 & 8 \\
 \hline
\end{tabular}
\caption{Correlation matrix ordering for both results. Values in the table refer to bin numbers. Note that the correlation matrix value for the data/MC peak is equivalent to the estimated peak at zero value.}
\label{table:CorrMapping}
\vspace{-5mm}
\end{table*}

\begin{table*}[bpt]
\small
\centering
\sffamily
\begin{tabular}{|c | c | c|} 
\hline
 & 0.0 $< q_3 (GeV) <$ 0.2  & Diagonal Unc. \\ [0.5ex] 
\hline
Estimated data peak at zero & 0.99 & 0.20 \\
Data subtraction size from first bin & 0.015 &  0.005\\ 
Estimated MC peak at zero & 0.52 &n/a\\ 
MC subtraction size from first bin & 0.01 & n/a\\
Data/MC peak at zero & 1.90 & 0.38\\ 
\hline \hline
 & 0.2 $< q_3 (GeV) <$ 0.4  & Diagonal Unc. \\ [0.5ex] 
\hline
Estimated data peak at zero & 2.18 & 0.35 \\ 
Data subtraction size from first bin & 0.36 &  0.09\\ 
Estimated MC peak at zero & 1.87 &n/a\\ 
MC subtraction size from first bin & 0.29 & n/a\\
Data/MC peak at zero & 1.17 & 0.19\\ 
\hline \hline
 & 0.4 $< q_3 (GeV) <$ 0.6  & Diagonal Unc. \\ [0.5ex] 
\hline
Estimated data peak at zero & 3.12 & 0.45 \\ 
Data subtraction size from first bin & 0.67 &  0.13\\ 
Estimated MC peak at zero & 2.43 &n/a\\ 
MC subtraction size from first bin & 0.79 & n/a\\
Data/MC peak at zero & 1.53 & 0.18\\ 
\hline \hline
 & 0.6 $< q_3 (GeV) <$ 0.8  & Diagonal Unc. \\ [0.5ex] 
\hline
Estimated data peak at zero & 3.24 & 0.40\\ 
Data subtraction size from first bin & 0.51 &  0.08\\ 
Estimated MC peak at zero & 2.05 &n/a\\ 
MC subtraction size from first bin & 0.52 & n/a\\
Data/MC peak at zero & 1.58 & 0.19 \\ 
\hline
\end{tabular}
\caption{Summary of results for ME $\bar{\nu}_e$.}
\label{table:SarahData}
\vspace{-5mm}
\end{table*}

\begin{table*}[bpt]
\small
\centering
\sffamily
\begin{tabular}{|c | c | c|} 
\hline
 & 0.0 $< q_3 (GeV) <$ 0.2  & Diagonal Uncertainty \\ [0.5ex] 
\hline
Estimated data peak at zero & 0.73 & 0.078 \\ 
Data subtraction size from first bin & 0.00 &  0.000\\
Estimated MC peak at zero & 0.9 &n/a \\ 
MC subtraction size from first bin & 0.08 & n/a\\
Data/MC peak at zero & 0.74 & 0.08 \\ 
\hline \hline
 & 0.2 $< q_3 (GeV) <$ 0.4  & Diagonal Uncertainty \\ [0.5ex] 
\hline
Estimated data peak at zero & 3.1 & 0.33 \\ 
Data subtraction size from first bin & 0.38 &  0.06 \\ 
Estimated MC peak at zero & 2.55 &n/a \\ 
MC subtraction size from first bin & 0.36 & n/a\\
Data/MC peak at zero & 1.15 & 0.12\\ 
\hline \hline
 & 0.4 $< q_3 (GeV) <$ 0.6  & Diagonal Uncertainty \\ [0.5ex] 
\hline
Estimated data peak at zero & 4.1 & 0.51 \\ 
Data subtraction size from first bin & 0.66 &  0.12 \\ 
Estimated MC peak at zero & 3.46 &n/a \\ 
MC subtraction size from first bin & 0.51 & n/a\\
Data/MC peak at zero & 1.16 & 0.14 \\ 
\hline \hline
 & 0.6 $< q_3 (GeV) <$ 0.8  & Diagonal Uncertainty \\ [0.5ex] 
\hline
Estimated data peak at zero & 3.5 & 0.42 \\ 
Data subtraction size from first bin & 0.44 &  0.06 \\
Estimated MC peak at zero & 2.91 &n/a \\ 
MC subtraction size from first bin & 0.26 & n/a\\
Data/MC peak at zero & 1.20 & 0.15 \\ 
\hline
\end{tabular}
\caption{Summary of results for LE $\bar{\nu}_\mu$.}
\label{table:LEData}
\vspace{-5mm}
\end{table*}

\FloatBarrier

\begin{table}[bpt]
\small
\centering
\sffamily
\begin{tabular}{|c|c|c|c|c|c|c|c|}
\hline
1.00	&0.47	&0.16	&-0.03	&0.36	&0.04	&0.36	&0.15 \\
\hline
0.47	&1.00	&-0.06	&0.51	&0.07	&0.27	&0.13	&0.04 \\
\hline
0.16	&-0.06	&1.00	&-0.32	&0.49	&-0.30	&0.51	&-0.21 \\
\hline
-0.03	&0.51	&-0.32	&1.00	&-0.30	&0.60	&-0.30	&0.42 \\
\hline
0.36	&0.07	&0.49	&-0.30	&1.00	&-0.26	&0.53	&-0.26 \\
\hline
0.04	&0.27	&-0.30	&0.60	&-0.26	&1.00	&-0.30	&0.72 \\
\hline
0.36	&0.13	&0.51	&-0.30	&0.53	&-0.30	&1.00	&-0.14 \\
\hline
0.15	&0.04	&-0.21	&0.42	&-0.26	&0.72	&-0.14	&1.00 \\
\hline
\end{tabular}
\caption{Correlation matrix for ME $\bar{\nu}_e$ result for relevant bins.}
\label{table:SarahCorr}
\vspace{-5mm}
\end{table}

\begin{table}[bpt]
\small
\centering
\sffamily
\begin{tabular}{|c|c|c|c|c|c|c|c|}
\hline
1.00&0.00&0.82&0.37	&0.67	&0.38	&0.56	&0.42 \\
\hline
0.00&0.00&0.00&0.00	&0.00	&0.00	&0.00	&0.00 \\
\hline
0.82&0.00&1.00&0.19	&0.83	&0.30	&0.76	&0.38 \\
\hline
0.37&0.00&0.19&1.00&0.21	&0.75	&0.01	&0.70 \\
\hline
0.67&0.00&0.83&0.21	&1.00	&0.32	&0.86	&0.38 \\
\hline
0.38&0.00&0.30&0.75	&0.32	&1.00	&0.10	&0.85 \\
\hline
0.56&0.00&0.76&0.01	&0.86	&0.10	&1.00	&0.24 \\
\hline
0.42&0.00&0.38&0.70	&0.38	&0.85	&0.24	&1.00 \\
\hline
\end{tabular}
\caption{Correlation matrix for the LE $\bar{\nu}_\mu$ result for relevant bins.}
\label{table:LECorr}
\vspace{-5mm}
\end{table}

\begin{acknowledgements}
% KSM pulled version 15 frmo DocDB 13929 on June 2, 2023
% added D. Ruterbories and S. Henry
This document was prepared by members of the MINERvA Collaboration using the resources of the Fermi National Accelerator Laboratory (Fermilab), a U.S. Department of Energy, Office of Science, HEP User Facility. Fermilab is managed by Fermi Research Alliance, LLC (FRA), acting under Contract No. DE-AC02-07CH11359.
These resources included support for the MINERvA construction project, and support
for construction also
was granted by the United States National Science Foundation under
Award No. PHY-0619727 and by the University of Rochester. Support for
participating scientists was provided by NSF and DOE (USA); by CAPES
and CNPq (Brazil); by CoNaCyT (Mexico); by ANID PIA / APOYO AFB180002, CONICYT PIA ACT1413, and Fondecyt 3170845 and 11130133 (Chile); 
%by CONCYTEC, DGI-PUCP, and IDI/IGI-UNI (Peru); 
by CONCYTEC (Consejo Nacional de Ciencia, Tecnolog\'ia e Innovaci\'on Tecnol\'ogica), DGI-PUCP (Direcci\'on de Gesti\'on de la Investigaci\'on  - Pontificia Universidad Cat\'olica del Peru), and VRI-UNI (Vice-Rectorate for Research of National University of Engineering) (Peru); NCN Opus Grant No. 2016/21/B/ST2/01092 (Poland); by Science and Technology Facilities Council (UK); by EU Horizon 2020 Marie Skłodowska-Curie Action; by a Cottrell Postdoctoral Fellowship from the Research Corporation for Scientific Advancement; by an Imperial College London President's PhD Scholarship.
D. Ruterbories gratefully acknowledges support from a Cottrell Postdoctoral Fellowship, Research Corporation for Scientific Advancement Award No.\ 27467 and National Science Foundation Award CHE2039044. S.M.~Henry's work was supported in part by the National Science Foundation Graduate Research Fellowship under Grant No. DGE-1419118.  
We thank the MINOS Collaboration for use of its near detector data. Finally, we thank the staff of Fermilab for support of the beam line, the detector, and computing infrastructure.
\end{acknowledgements}

\clearpage

\bibliography{MINERvAHangSarah}

\end{document}